\pdfoutput=1

\documentclass[11pt]{article}
\usepackage{listings}
\usepackage[table]{xcolor} 

\usepackage[final]{acl}

\usepackage{times}
\usepackage{latexsym}

\usepackage{fontawesome5}

\usepackage{bm}
\usepackage{booktabs}

\usepackage{bbding}
\usepackage{amsmath}  
\definecolor{green}{RGB}{0,128,0}  
\definecolor{red}{RGB}{255,0,0}    
\usepackage{multirow}
\usepackage{amsmath}
\usepackage{amssymb}
\usepackage{array}
\usepackage{longtable}
\usepackage{tabularx}
\usepackage{supertabular}
\usepackage{makecell}
\usepackage{array}
\usepackage{hyperref}
\usepackage[capitalize,noabbrev]{cleveref}
\usepackage{enumitem}

\usepackage{microtype}
\usepackage{graphicx}

\usepackage{subcaption}

\usepackage{mathtools}
\usepackage{titlesec}

\usepackage{caption}
\usepackage{float} 
\usepackage{stfloats}

\usepackage{booktabs}
\usepackage{tabularx}
\usepackage{makecell}
\usepackage{threeparttable} 

\usepackage[normalem]{ulem}

\usepackage{lipsum} 

\usepackage[T1]{fontenc}

\usepackage[utf8]{inputenc}

\usepackage{microtype}

\usepackage{inconsolata}
\usepackage[most]{tcolorbox}
\newtcolorbox[auto counter]{response}[1][]{
    enhanced,
  top=15pt,
  bottom=15pt,
  left=25pt,
  right=25pt,
  left skip=30pt,
  right skip=30pt,
  colback=gray!5,
  colframe=black,
  fonttitle=\bfseries,
  coltitle=white,
  title=Prompt~\thetcbcounter: #1
}

\newcolumntype{C}[1]{>{\centering\arraybackslash}p{#1}}
\title{FedGUI: Benchmarking Federated GUI Agents 
across Heterogeneous Platforms, Devices, and Operating Systems}

\author{
Wenhao Wang\textsuperscript{1,3,4,*}, 
Haoting Shi\textsuperscript{2,4,*}, 
Mengying Yuan\textsuperscript{5},
Yiquan Lin\textsuperscript{1},
Panrong Tong\textsuperscript{3},\\ \bfseries
Hanzhang Zhou\textsuperscript{3},
Guangyi Liu\textsuperscript{1}, 
Pengxiang Zhao\textsuperscript{1}, 
Yue Wang\textsuperscript{3}\dag, 
Siheng Chen\textsuperscript{2,4}\dag,
\\ 
\textsuperscript{1} Zhejiang University \quad
\textsuperscript{2} Shanghai Jiao Tong University\quad
\textsuperscript{3} Tongyi Lab \\
\textsuperscript{4} Multi-Agent Governance \& Intelligence Crew (MAGIC) \quad
\textsuperscript{5} Wuhan University
\\
}

\lstdefinestyle{mystyle}{
    language=Python, 
    basicstyle=\scriptsize\ttfamily, 
}

\begin{document}
\maketitle

\begingroup
\renewcommand\thefootnote{$*$}
\footnotetext{Equal contributions. \textsuperscript{\dag}Corresponding authors.}
\endgroup


\begin{abstract}
Training GUI agents with traditional centralized methods faces significant cost and scalability challenges. Federated learning (FL) offers a promising solution, yet its potential is hindered by the lack of benchmarks that capture real-world, cross-platform heterogeneity. To bridge this gap, we introduce FedGUI, the first comprehensive benchmark for developing and evaluating federated GUI agents across mobile, web, and desktop platforms. FedGUI provides a suite of six curated datasets to systematically study four crucial types of heterogeneity: 
cross-platform, cross-device, cross-OS, and cross-source.
Extensive experiments reveal several key insights: First, we show that cross-platform collaboration improves performance, extending prior mobile-only federated learning to diverse GUI environments;
Second, we demonstrate the presence of distinct heterogeneity dimensions and identify platform and OS as the most influential factors.
FedGUI provides a vital foundation for the community to build more scalable and privacy-preserving GUI agents for real-world deployment.
Code and data are publicly available at \href{https://github.com/wwh0411/FedGUI}{https://github.com/wwh0411/FedGUI}.
\end{abstract}

\section{Introduction}
\label{sec:intro}
Recent advances in vision–language models (VLMs) have enabled the emergence of GUI agents that can perceive graphical user interfaces (GUI) and execute user instructions through sequential interactions \cite{zhou2025maiuitechnicalreportrealworld, liu2025llmpoweredguiagentsphone,lian2026uiagileadvancingguiagents}. 
Traditional approaches to GUI agents largely rely on centralized data collection and manual labeling. While effective, such paradigm suffers from high data collection costs and limited scalability \cite{sunOSGenesisAutomatingGUI2024}.
Meanwhile, the widespread and frequent use of GUI devices naturally generates abundant supervisory signals, which could serve as a low-cost data source for training GUI agents \cite{berkovitch2025identifying}. However, this real-world, large-scale data remains largely underutilized, as it cannot be publicly shared due to user privacy concerns \cite{zhangPrivacyAsstSafeguardingUser2024}.
This necessitates a distributed learning paradigm where each client trains locally on its own data without direct transmission \cite{heEmergedSecurityPrivacy2024}.


Initial research has explored this via federated learning (FL) for privacy-preserving collaborative training \cite{federatedscopellm}.
FedMABench \cite{wang-etal-2025-fedmabench} is the first benchmark designed for federated mobile agents, but it is limited to collaboration among Android users and overlooks the \uline{significant potential of incorporating users from web and desktop environments to further enhance performance.}
Moreover, FedMABench does not account for broader forms of heterogeneity across devices, operating systems (OSs), and data sources.

These limitations give rise to two fundamental challenges:
(1) \textbf{RQ1:} How to enable cross-platform collaboration of GUI agent training, and does the expanded collaboration from distinct platforms improve performance?
(2) \textbf{RQ2:} How to quantitatively characterize and measure the real-world heterogeneity spanning diverse platforms, OSs, devices, and data sources.


To address the challenges outlined above, we introduce \textbf{FedGUI}, a comprehensive benchmark designed for distributed GUI agents across diverse platforms and devices.
FedGUI is characterized by three key features: 
(1) \textbf{Diversity}. FedGUI covers a wide range of real-world GUI environments, including over 900 mobile applications, 40+ desktop applications, and 200+ websites. It supports both multi-step and cross-application tasks, enabling the evaluation of agents partitioned with different levels of complexity.
(2) \textbf{Comprehensiveness}. FedGUI integrates seven representative FL algorithms and supports more than 20 base models, including state-of-the-art open-source VLMs and proprietary models. In addition, FedGUI provides a comprehensive set of evaluation metrics that jointly assess task performance and system efficiency.
(3) \textbf{Heterogeneity}. FedGUI models four representative real-world heterogeneity settings, simulating the complexity of user collaboration across different platforms, devices, and operating systems, thereby reflecting realistic deployment scenarios.

\begin{figure}
    \centering
    \includegraphics[width=1\linewidth]{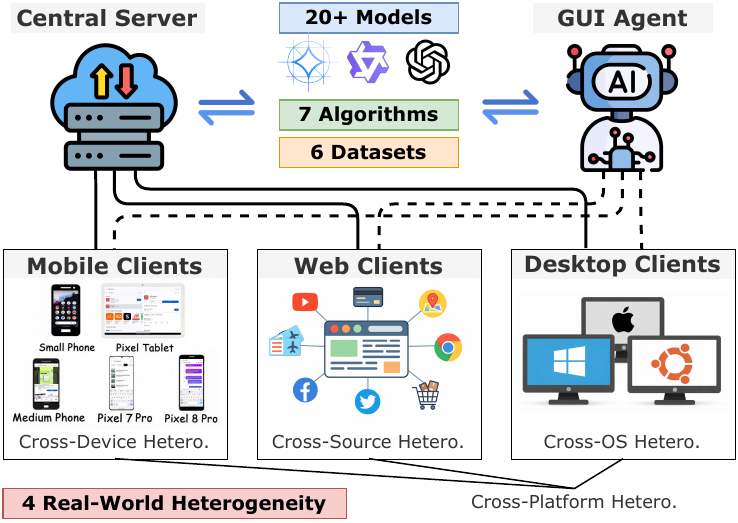}
    \caption{ 
    Overview of FedGUI.
    FedGUI provides a federated framework that coordinates a central server and heterogeneous clients across mobile, web, and desktop platforms to train a generalized cross-platform GUI agent.
    Hetero. is short for heterogeneity.
    }
    \label{fig:FedGUI_overview}
    \vspace{-1mm}
\end{figure}

Specifically, FedGUI constructs six datasets from eight data sources to study four different heterogeneities. (1) \textit{FedGUI-Platform} is constructed using samples collected from three platforms (i.e., mobile, web and desktop), aiming to model \textbf{cross-platform heterogeneity}.
(2) We further evaluate \textbf{cross-device heterogeneity} within the same mobile platform. Samples are collected from five different devices, forming the \textit{FedGUI-Device} dataset.
(3) To study \textbf{cross-OS heterogeneity}, we construct \textit{FedGUI-OS}, which comprises samples collected from desktop environments on Ubuntu, macOS, and Windows.
(4) To isolate and analyze the impact of data sources, we additionally construct more controlled datasets within each platform. \textit{FedGUI-Web} and \textit{FedGUI-Mobile} are designed to measure \textbf{cross-source heterogeneity}, where data originate from different underlying datasets.
(5) Finally, \textit{FedGUI-Full} is introduced to jointly capture both cross-platform and cross-source heterogeneity, providing a comprehensive benchmark for complex realistic distributed settings.

Based on FedGUI, we conduct an exhaustive empirical study to investigate and compare performance across three key axes: federated algorithm, heterogeneity level, and base model. Our extensive experiments yield several key observations:
(1) Increasing user participation in federated learning boosts model performance, even when these users contribute from highly diverse platforms and devices.
(2) Platform-level heterogeneity poses a more significant challenge to model performance than within-platform heterogeneity.
(3) Adaptive FL algorithms (e.g., FedYogi \cite{fedopt}) outperform other baselines, demonstrating particular robustness in cross-platform settings.
To summarize, our contributions are:
\begin{enumerate}[nosep, left=0em]
    \item We present FedGUI, a comprehensive and unified benchmark for training cross-platform GUI agents through federated learning. 
    \item We construct six datasets targeted for four type of real-world heterogeneity. The datasets covers diverse applications and websites.
    \item Extensive experiments thoroughly investigate federated GUI agents across platforms and devices, revealing insightful discoveries.
\end{enumerate}

\newcommand{\cmark}{\textcolor{green}{\Checkmark}}
\newcommand{\xmark}{\textcolor{red}{\XSolidBrush}}

\begin{table}[t]
\centering
\small
\setlength{\tabcolsep}{2pt}
\begin{tabular}{p{2.2cm}cccc}
\toprule
\multirow{2}{*}{\textbf{Benchmark}} & \textbf{Distributed} & \multicolumn{3}{l}{\textbf{Cross-Platform Evaluation}} \\
& \textbf{Training}    & Mobile & Web & Desktop \\
\midrule
Mobile-Bench    & \xmark & \cmark & \xmark & \xmark \\
MMBench-GUI     & \xmark & \cmark & \cmark & \cmark \\
FedMABench      & \cmark & \cmark & \xmark & \xmark \\
FedGUI (Ours)   & \cmark & \cmark & \cmark & \cmark \\
\bottomrule
\end{tabular}
\caption{Comparison of related benchmarks.}
\label{tab:comparison_benchmark}
\end{table}

\begin{table}[t]
\centering
\small
\setlength{\tabcolsep}{1.9pt}
\begin{tabular}{lcc}
\toprule
\textbf{Dimension} & \textbf{FedMABench} & \textbf{FedGUI (Ours)} \\
\midrule
Platform           & Mobile Only             & Mobile, Web, Desktop \\
Supported OS        & Android             & Android, Ubuntu, mac, Win \\
Data Sources       &  AC, AitW	  & 9 Diverse Datasets\\
Heterogeneity      & User-based       & 4 Types (Plat., Dev., OS, Src.) \\
Eval Metric             & TF-IDF Sim.         & Action Type, Grounding, SR \\
\bottomrule
\end{tabular}
\caption{Detailed benchmark comparison between FedGUI and FedMABench \cite{wang-etal-2025-fedmabench}.}
\label{tab:fedgui_vs_fedmabench}
\vspace{-1mm}
\end{table}

\section{Related Work (\textit{Summarized in Table \ref{tab:comparison_benchmark}})}

\label{sec:related_work}
\subsection{Centralized GUI Agents}


\paragraph{Single-Platform Agents.}
Previous research has explored GUI agents for individual platforms, with mobile agents \cite{ye2025mobile, zhang2026dontactblindlyrobust} focusing on app navigation, web agents \cite{ning2025survey, chen2025guicourse} specializing in browser-based tasks, and desktop agents \cite{OSWorld} targeting workflow automation on computing environments. 
However, these single-platform approaches exhibit limited cross-platform generalization, as their specialized capabilities are confined to corresponding environments and fail to transfer effectively across different interface modalities.

\paragraph{Cross-Platform Agents.}
To address the generalization limitations, recent work has shifted toward cross-platform GUI agents capable of operating across diverse interface environments. 
Representative efforts span both models \cite{qin2025uitars} and benchmarks \cite{wang2025mmbenchgui}, demonstrating unified reasoning across mobile, web, and desktop platforms.
Despite their promising performance, these approaches remain constrained by centralized data collection pipelines that are costly to scale and fail to capture the full spectrum of data diversity encountered in real-world distributed scenarios.



%
\subsection{Distributed GUI Agents}
\paragraph{Federated Learning.}
FL \cite{wen2023survey} offers a decentralized paradigm in which models are trained collaboratively without transmitting raw data.
By performing local optimization and periodic server-side aggregation, FL enables large-scale learning under privacy constraints while naturally capturing user-specific behaviors.


\paragraph{Federated GUI Agents.}
FedMABench \cite{wang-etal-2025-fedmabench} represents the first benchmark for federated mobile agents. However, as detailed in Table \ref{tab:fedgui_vs_fedmabench}, its scope is fundamentally limited: it only considers collaboration among mobile device users, ignoring the broader context where users from different platforms can contribute to a unified agent system.
In contrast, FedGUI presents the first comprehensive benchmark that addresses heterogeneous data sources, device types, and operating systems in federated agent training, benchmarking across a substantially broader spectrum of scenarios.



\begin{table*}[t]
\centering
\small
\begin{threeparttable}
\setlength{\tabcolsep}{5pt}
\begin{tabular}{l l l c c c c}
\toprule
\textbf{Dataset Name} & \textbf{Heterogeneity} & \textbf{Source Datasets} & \textbf{N\# Clients} & \textbf{N\# Subsets} & \textbf{N\# Epi.} & \textbf{N\# Steps} \\ \midrule
FedGUI-Platform & Cross-Platform & AC, GA, AS & 15 & 3 & 3,000 & 23,157 \\
FedGUI-Device   & Cross-Device   & GO & 5 & 4 & 2,500 & 38,064 \\
FedGUI-OS       & Cross-OS       & AS, OA-Mac, OA-Win & 3 & 5 & 1,800 & 5,813 \\
FedGUI-Web      & Cross-Source   & M2W, GA-W, OA-W & 3 & 5 & 1,800 & 9,975 \\
FedGUI-Mobile   & Cross-Source   & AC, AitW, GO & 3 & 5 & 6,000 & 59,328 \\
FedGUI-Full     & Cross-Platform \& Source & All Nine Sources\tnote{*} & 9-36 & 7 & 5,400 & 33,341 \\ 
\bottomrule
\end{tabular}
\begin{tablenotes}
    \footnotesize
    \item[*] \textbf{Source Datasets:} \textbf{AC}: AndroidControl; \textbf{GA}: GUIAct; \textbf{AS}: AgentSynth; \textbf{GO}: GUI Odyssey; \textbf{OA-W/Mac/Win}: OmniAct-Web/MacOS/Windows; \textbf{M2W}: Mind2Web; \textbf{AitW}: Android-in-the-Wild.
    \item[*] \textbf{Evaluation Datasets:} 
    Prior to constructing the six training datasets, we create a dedicated test set of 100 episodes for each of the nine sources.
    These test sets serve as a fixed benchmark for all subsequent evaluations.
\end{tablenotes}
\end{threeparttable}
\vspace{-1mm}
\caption{Statistic summary of FedGUI's datasets. The table details the heterogeneity type, source composition, and scale for each of the six sets. N\# is short for "the number of" and Epi. represents "episode"
}
\label{tab:fedgui_stats}
\vspace{-1mm}
\end{table*}

\section{FedGUI}
\subsection{System Overview.}
FedGUI presents a comprehensive benchmark accompanied by six datasets that emphasize multi-dimensional heterogeneity and diversity in real-world mobile, web, and desktop GUI interactions. 
As illustrated in Figure \ref{fig:FedGUI_overview}, FedGUI follows the typical federated learning protocol and offers a research-friendly infrastructure that integrates 7 representative federated learning algorithms and supports over 20 mainstream VLMs. 

To systematically characterize the complex non-IID patterns inherent in GUI interaction data, we categorize heterogeneity along four key dimensions: cross-platform, cross-device, cross-source, and cross-OS. Based on these dimensions, we construct six datasets with 29 subsets that span varying degrees of data skew and client scales, with the statistics summarized in Table \ref{tab:fedgui_stats}.

\subsection{Data Collection (\textit{Details in Appendix \ref{sec:dataset_details}})}


\paragraph{Data Source.}
We collect a diverse set of GUI interaction datasets covering mobile, web, and desktop platforms,
including AndroidControl (AC) \cite{li2024effects}, Android-in-the-Wild (AitW) \cite{rawlesAndroidWildLargeScale2023}, GUI Odyssey (GO) \cite{lu2025guiodyssey}, GUIAct (GA) \cite{chen2025guicourse}, Mind2Web (M2W) \cite{deng2023mind2web}, OmniAct (OA) \cite{kapoor2024omniact}, and AgentSynth \cite{xie2025agentsynth}.
These datasets differ in task complexity, platform, device and sources, enabling the study of heterogeneous data in federated settings.


\paragraph{Unified Action Space.}
We design a unified action space to standardize user interactions across mobile, web, and desktop platforms. Specifically, we identify six basic actions shared across all platforms (e.g., $\mathtt{CLICK}$, $\mathtt{TYPE}$), while mapping platform-specific actions into two distinct domains defined in the system prompt. This unified action space enables consistent policy learning and parameter aggregation across platforms; details are provided in Appendix, Table~\ref{tab:action_space}.

\begin{figure*}[t]
    \centering
    \begin{subfigure}{0.24\linewidth}
        \centering
        \includegraphics[width=\linewidth]{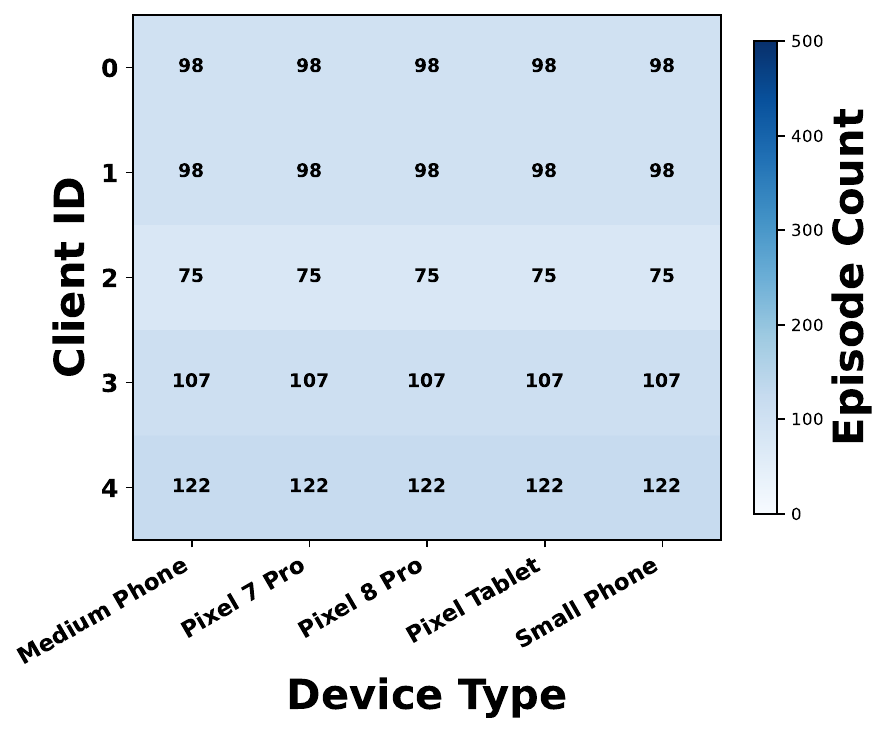}
        \caption{Device IID}
    \end{subfigure}
    \hfill
    \begin{subfigure}{0.24\linewidth}
        \centering
        \includegraphics[width=\linewidth]{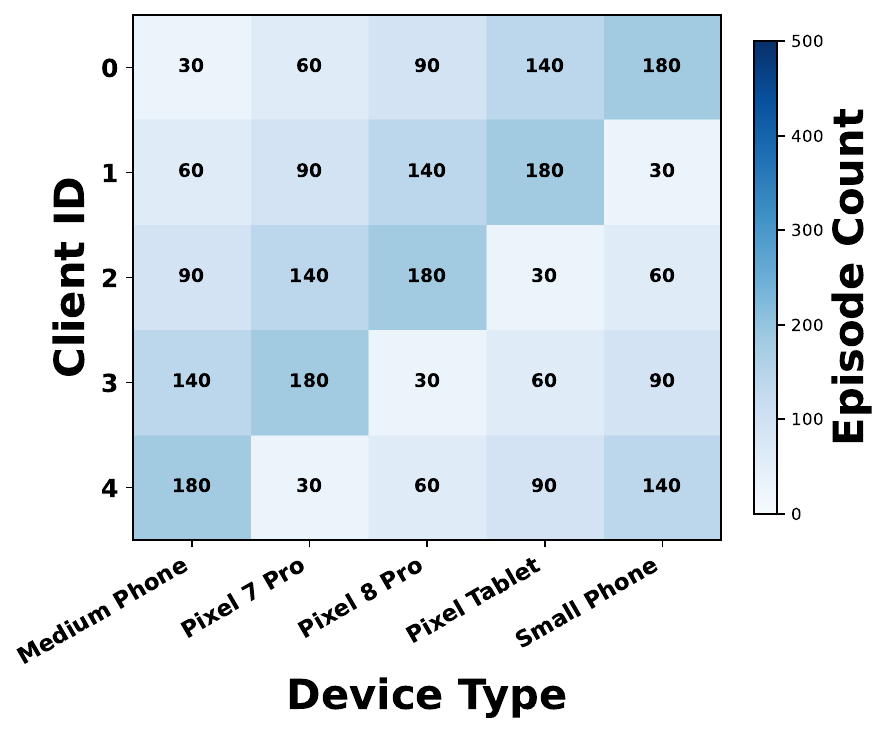}
        \caption{Device Non-Uniform}  
    \end{subfigure}
    \hfill 
    \begin{subfigure}{0.24\linewidth}
        \centering
        \includegraphics[width=\linewidth]{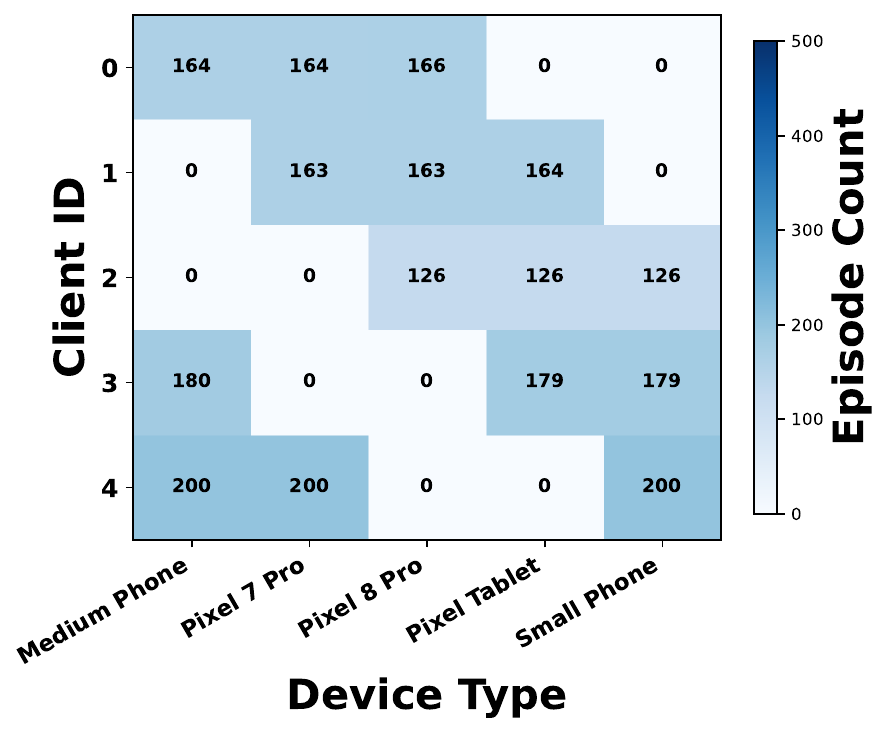}
        \caption{Device Partial}
    \end{subfigure}
    \hfill
    \begin{subfigure}{0.24\linewidth}
        \centering
        \includegraphics[width=\linewidth]{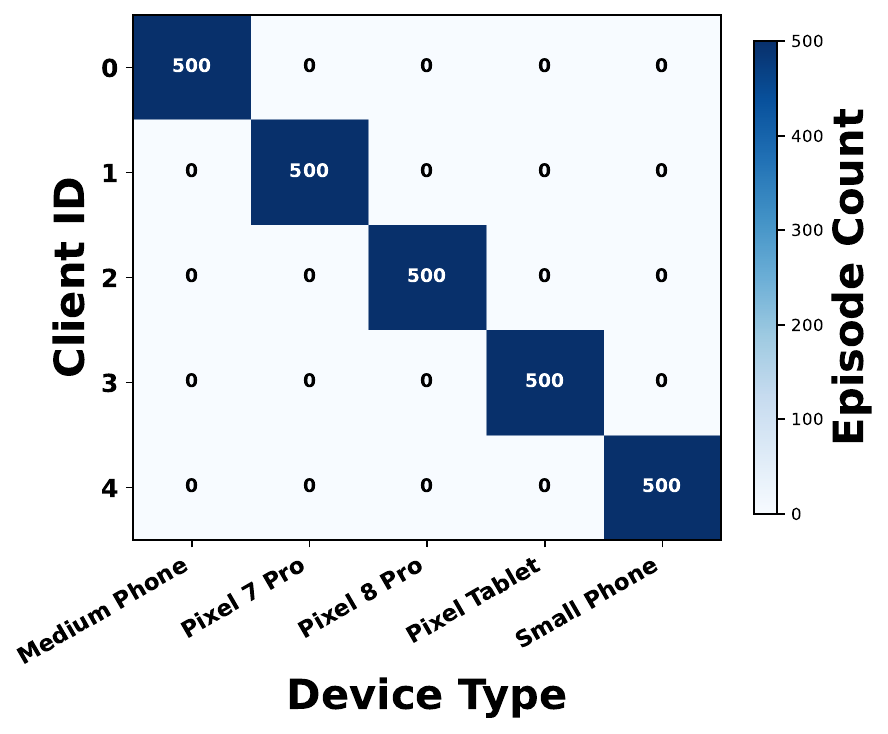}
        \caption{Device Skew}
    \end{subfigure}
    \caption{Distributions of the 5 different Android device across five clients in \textit{FedGUI-Device}. Through 4 types of cross-device heterogeneity subsets, diverse patterns across clients are simulated.}
    \label{fig:fedgui_device}
\end{figure*}

\paragraph{Data Processing.}
We implement systematic data cleaning to remove episodes with missing images, redundant or semantically unnecessary actions, and ambiguous action parameters (e.g., undefined spatial ranges). 
To model realistic non-IID scenarios, we construct various federated partitioning configurations spanning platform-specific settings, varying client counts, and extreme distribution skew.

\subsection{Data Description \& Visualization}
\label{sec:data_description}

\subsubsection{FedGUI-Platform}

\paragraph{Cross-Platform Heterogeneity.} 
Data collected from different platforms, such as mobile devices and web browsers, inherently exhibit significant distributional shifts, as characterized by differences in visual attributes, application ecosystems, task structures, and action spaces.
This phenomenon, termed cross-platform heterogeneity, poses a substantial challenge to federated learning, as clients contributing platform-specific data can degrade the performance and convergence of the global model.
Therefore, a systematic study and quantification of this heterogeneity is of paramount importance.

\paragraph{Description.}
%


To enable a systematic investigation of cross-platform heterogeneity, we construct the \textit{FedGUI-Platform} dataset.
To systematically control the degree of heterogeneity, we propose three carefully curated data partitions:
(1) In the Platform IID subset, each client holds a uniform mixture of data from all platforms.
(2) A moderate degree of heterogeneity is introduced in Platform Partial, where each client is denied access to data from one particular platform.
(3) In stark contrast, Platform Skew exclusively assign each client data from a single platform, thereby maximizing statistical heterogeneity across the network.


\subsubsection{FedGUI-Device}
\label{sec:fedgui_device}
\paragraph{Cross-Device Heterogeneity.}
In real-world scenarios, participating clients provide data collected from heterogeneous devices, including smartphones, laptops, and tablets.
Even among clients operating on identical Android OS, substantial device-level variations persist, such as distinct visual patterns in image resolution, screen size, aspect ratio, and background rendering.
To minimize the confounding effects introduced by platform and OS differences, we restrict our analysis to Android devices and define cross-device heterogeneity as the fine-grained variability that arises solely from device-specific characteristics.

\paragraph{Description.}
Leveraging the data collection methodology of GUI Odyssey \cite{lu2025guiodyssey}, which employs multiple device simulators, we construct \textit{FedGUI-Device} by labeling each episode with its device type to model cross-device heterogeneity.
Specifically, we consider the following settings with increasing level of heterogeneity:
(1) Device IID, where all clients observe data from all devices with identical distributions;
(2) Device Non-Uniform, where all clients observe all devices but with heterogeneous proportions.
(3) Device Partial, where each client observes data from a subset of devices and
(4) Device Skew, where each client receives data from only one device.

As visualized in Figure \ref{fig:fedgui_device}, these variants induce progressively stronger cross-source heterogeneity while keeping platform and device factors fixed.

\subsubsection{FedGUI-OS}
\paragraph{Cross-OS Heterogeneity.}

Even when underlying hardware and application functionalities are comparable, the diversity of client operating systems, such as macOS and Windows, introduces substantial discrepancies in system architecture, GUI design, window management, and interaction conventions. 
These differences lead to distinct visual layouts, system behaviors, and event-handling mechanisms, resulting in heterogeneous data distributions across clients.
We define cross-OS heterogeneity as the variability arising from differences among operating systems, independent of device-level or platform-level factors.

\paragraph{Description.}
We construct \textit{FedGUI-OS} to facilitate the exploration of cross-OS heterogeneity.
For Ubuntu, we collect data using AgentSynth, which is sourced from desktop simulations built upon OS-World \cite{OSWorld}.
For macOS and Windows, we leverage data from OmniAct and manually categorize the samples according to their operating systems.
The client participation protocols follow those of \textit{FedGUI-Device}, including IID, non-uniform, skewed, and partial-observation settings.
In addition, we introduce a fully random variant to provide an additional homogeneous evaluation.

\begin{figure*}[t]
    \centering
    \begin{subfigure}[t]{0.23\textwidth}
        \centering
        \includegraphics[width=\linewidth]{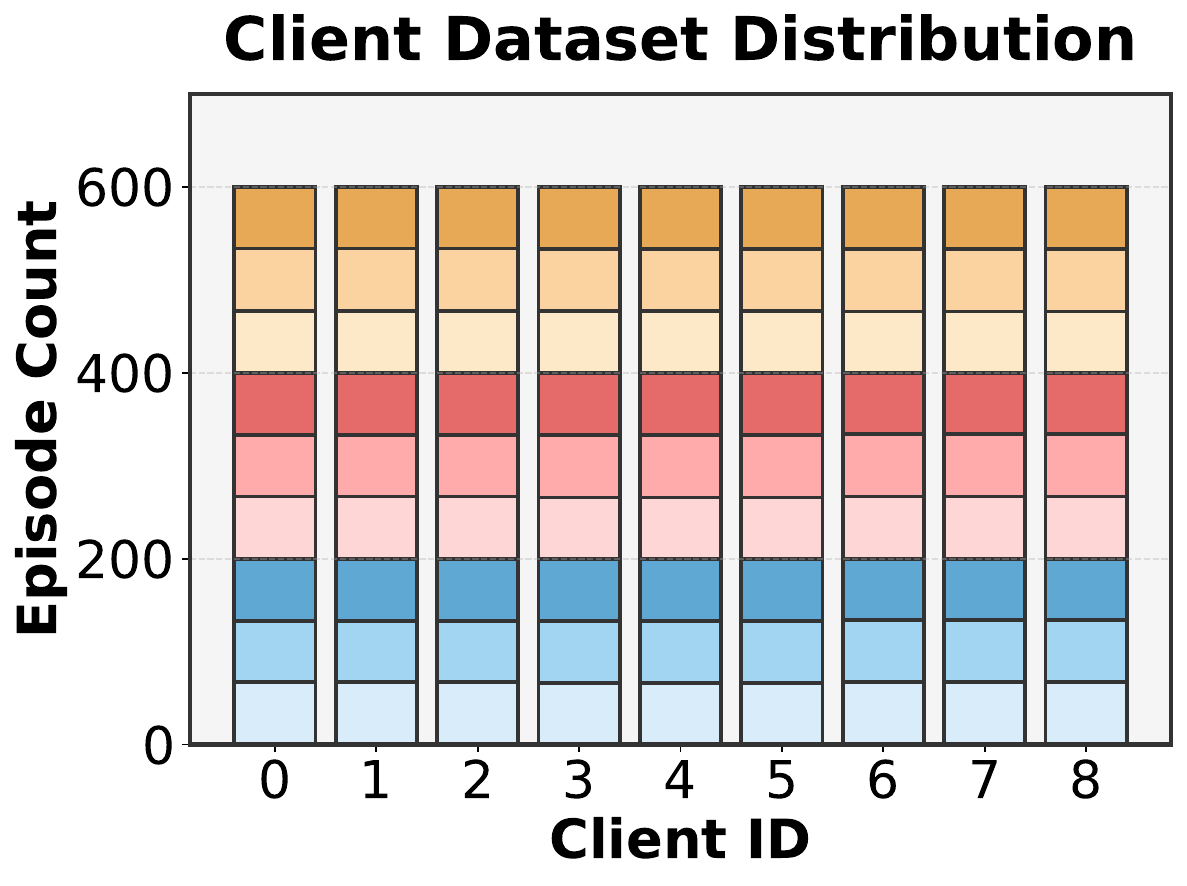}
        \caption{Full IID}
        \label{fig:iid}
    \end{subfigure}
    \hfill
    \begin{subfigure}[t]{0.23\textwidth}
        \centering
        \includegraphics[width=\linewidth]{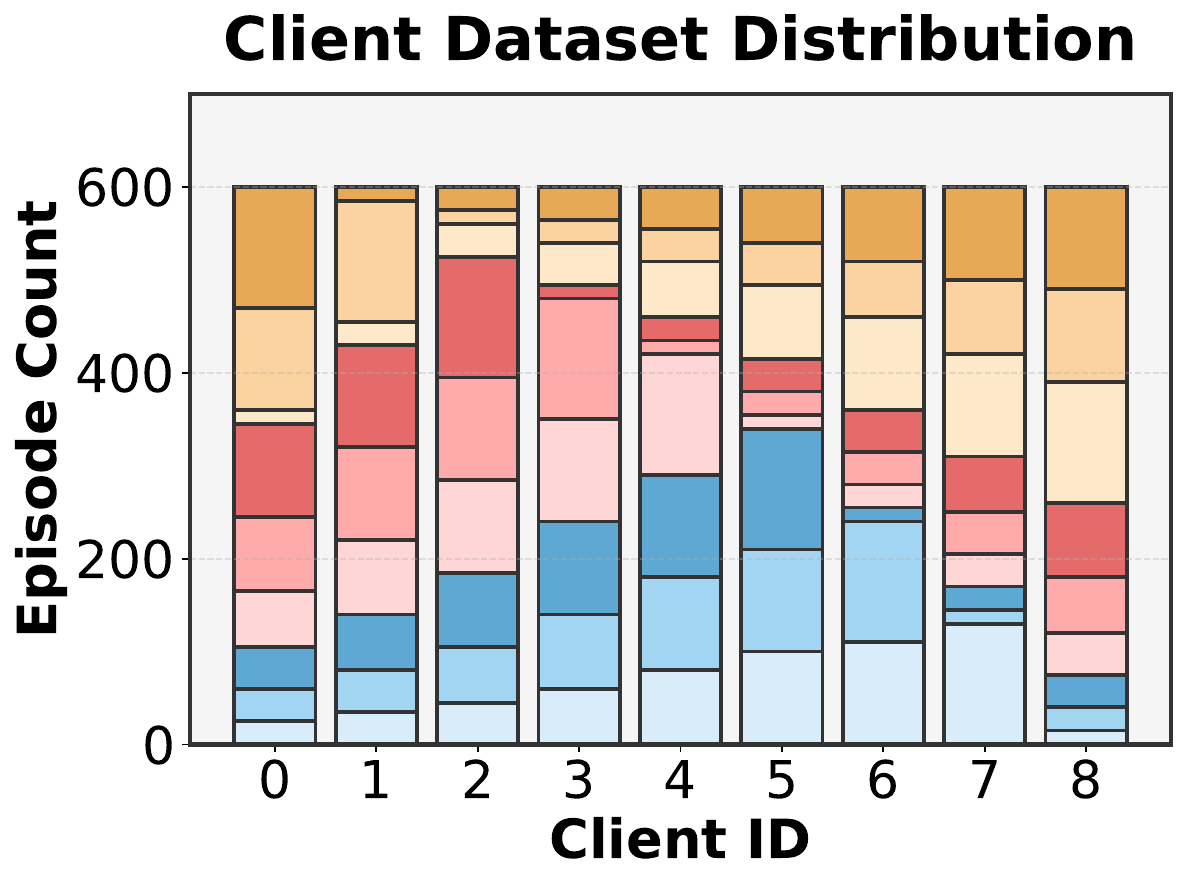}
        \caption{Full Non-Uniform}
        \label{fig:iid_random}
    \end{subfigure}
    \hfill
    \hfill
        \begin{subfigure}[t]{0.23\textwidth}
        \centering
        \includegraphics[width=\linewidth]{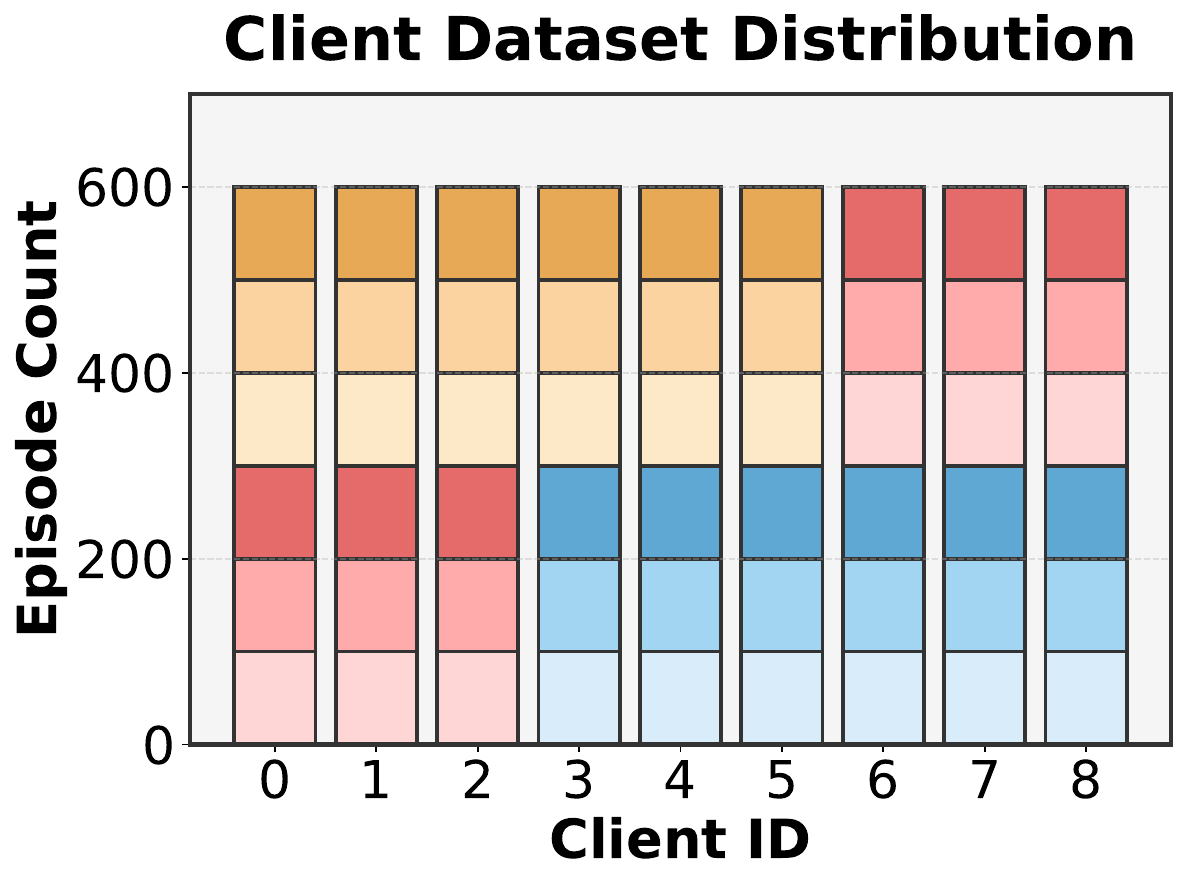}
        \caption{Platform Partial}
        \label{fig:partial_device_skew}
    \end{subfigure}
    \hfill
    \begin{subfigure}[t]{0.23\textwidth}
        \centering
        \includegraphics[width=\linewidth]{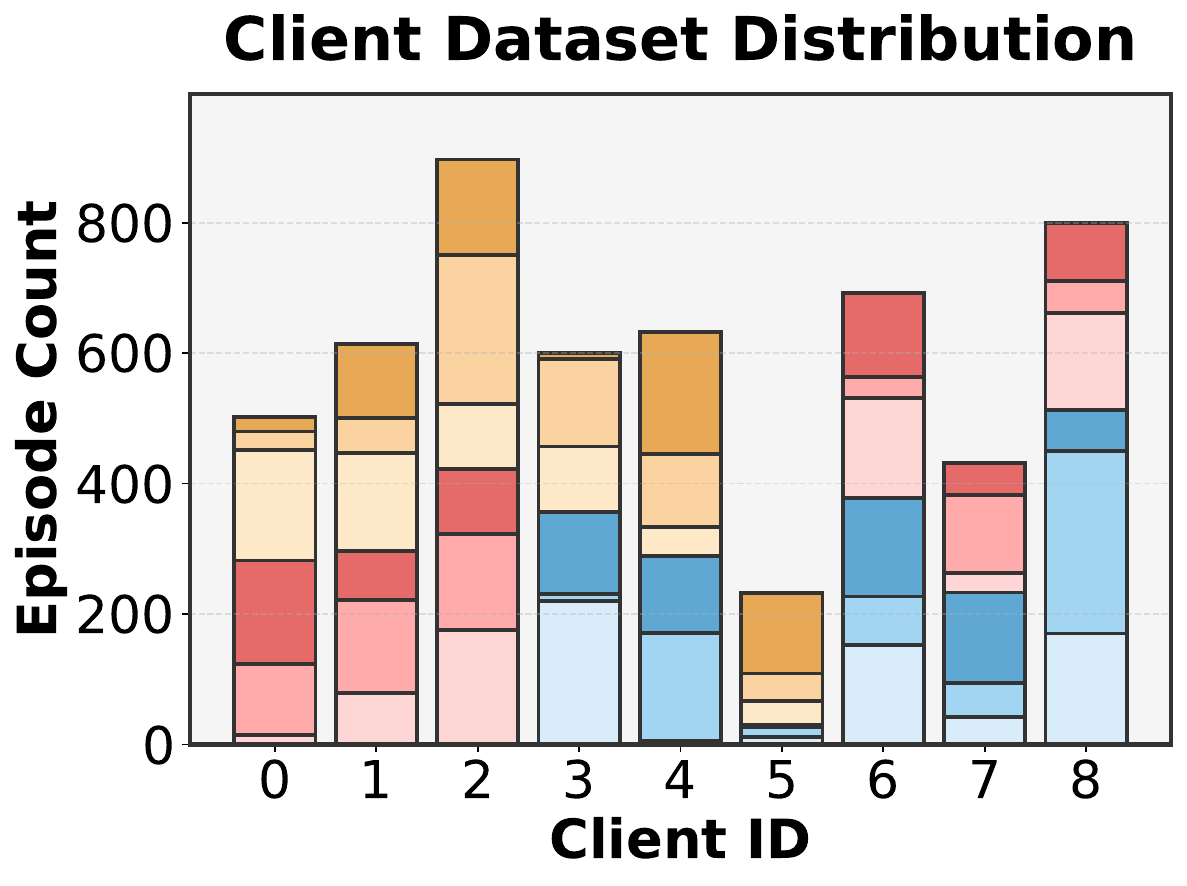}
        \caption{Platform Non-Uniform}
        \label{fig:partial_device_skew_random}
    \end{subfigure}
    
    \begin{subfigure}[t]{0.23\textwidth}
        \centering
        \includegraphics[width=\linewidth]{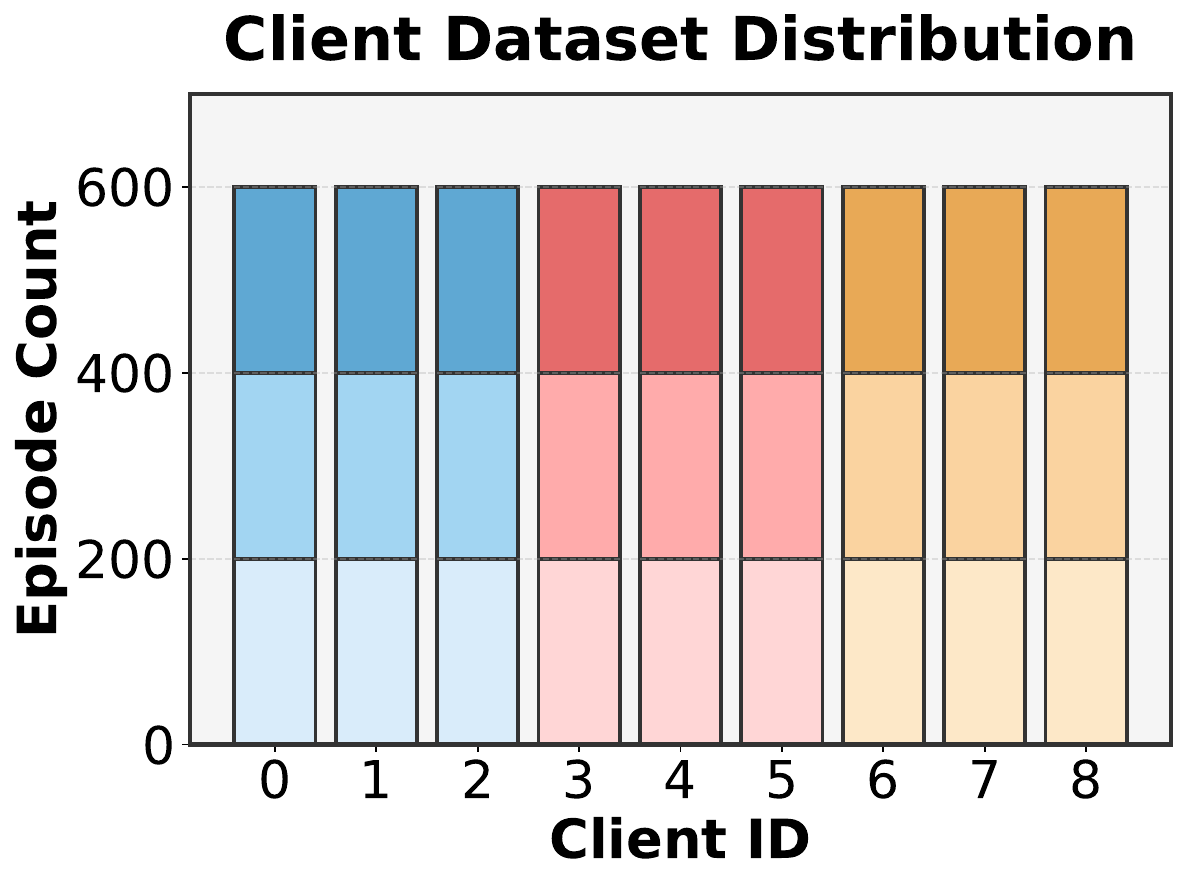}
        \caption{Platform Skew}
        \label{fig:device_skew}
    \end{subfigure}
    \hfill
    \begin{subfigure}[t]{0.23\textwidth}
        \centering
        \includegraphics[width=\linewidth]{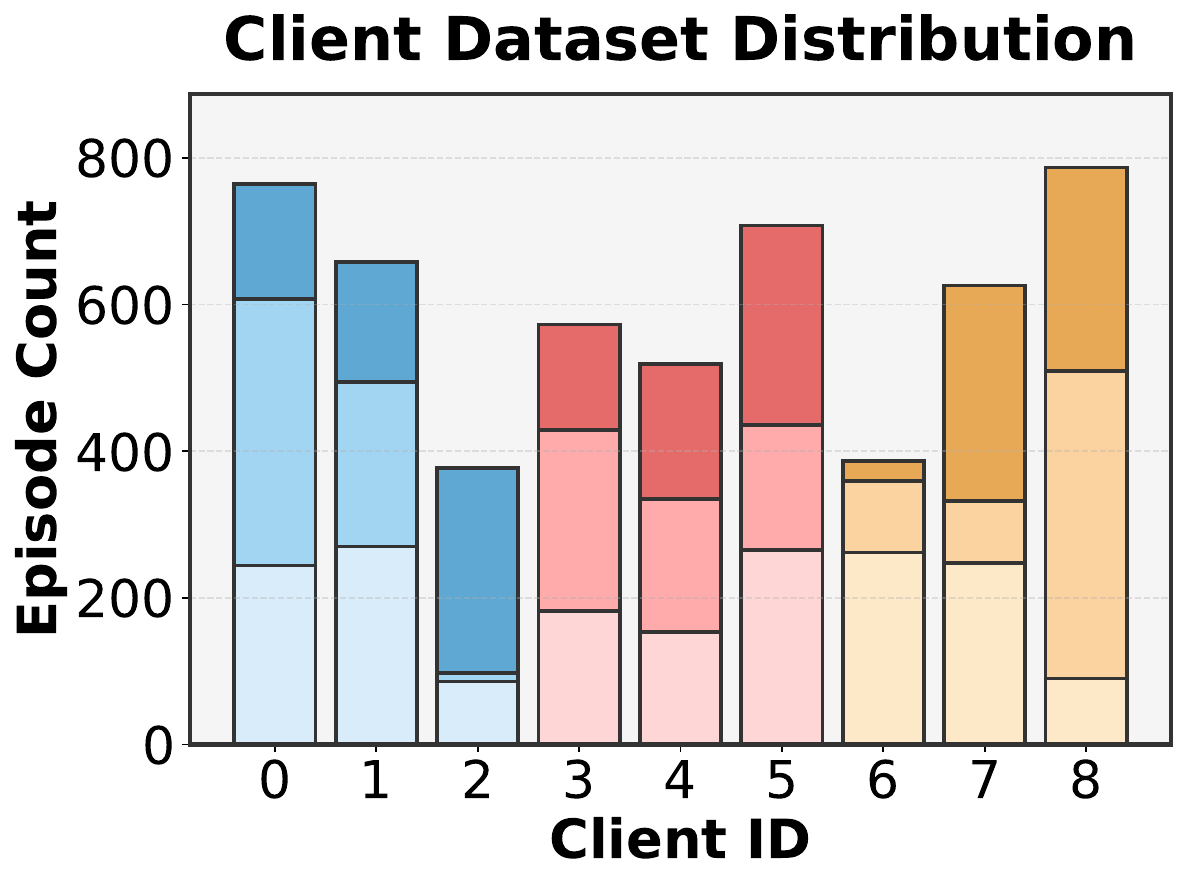}
        \caption{Source Non-Uniform}
        \label{fig:device_skew_random}
    \end{subfigure}
    \hfill
    \begin{subfigure}[t]{0.23\textwidth}
        \centering
        \includegraphics[width=\linewidth]{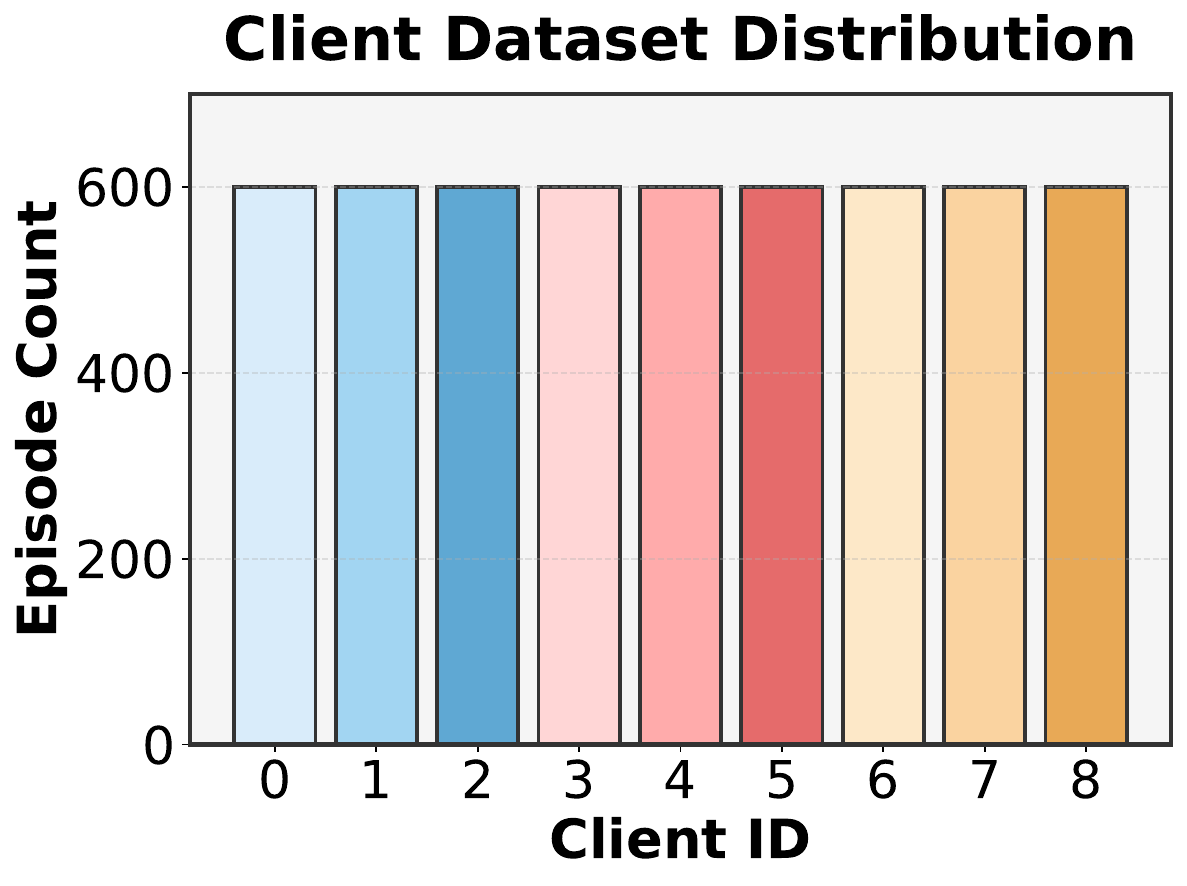}
        \caption{Source Skew}
        \label{fig:single_dataset}
    \end{subfigure}
    \hfill
    \begin{subfigure}[t]{0.23\textwidth}
        \centering
        \includegraphics[width=\linewidth]{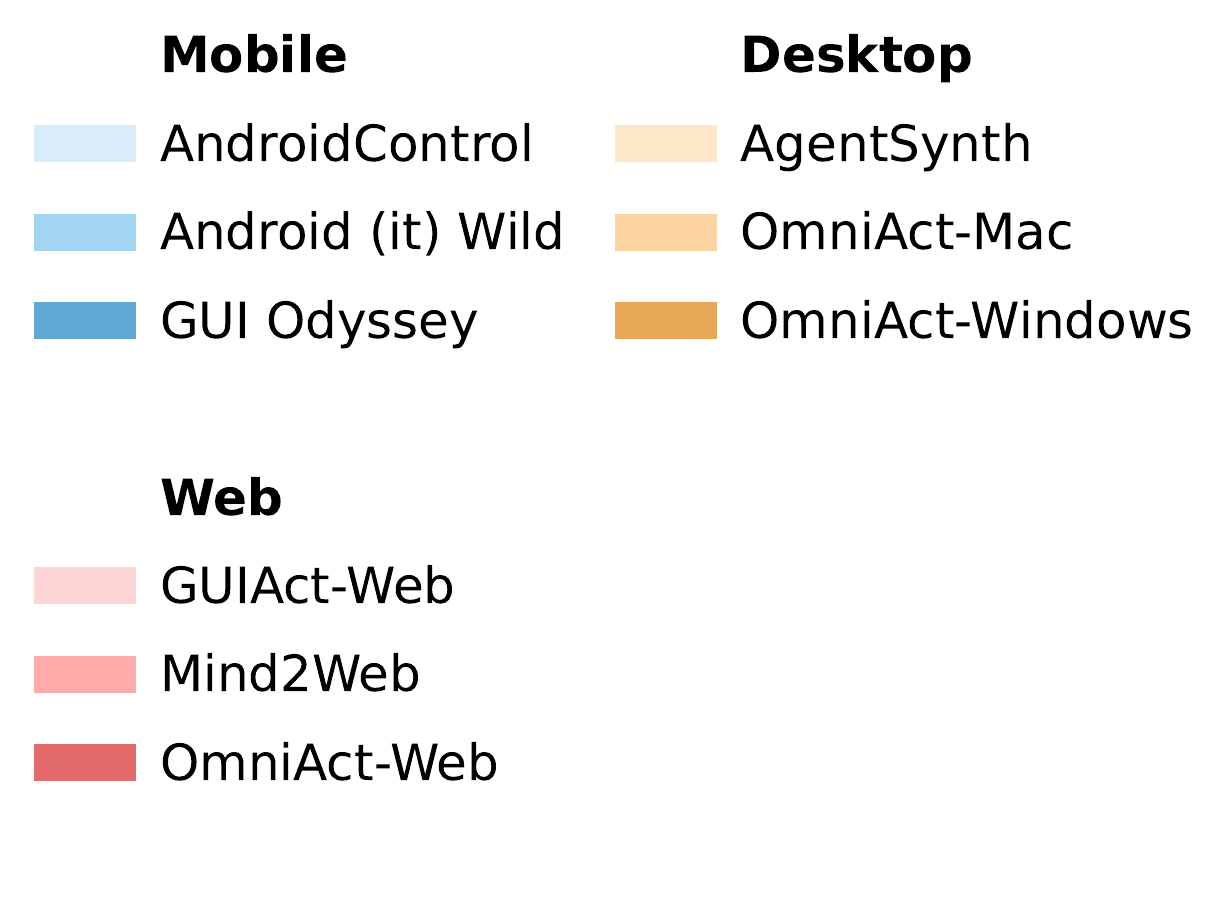}
        \caption{Source Legend}
        \label{fig:dataset_legend}
    \end{subfigure}
    
    \caption{Distributions of episode counts across clients in \textit{FedGUI-Full}. The 7 subsets demonstrate various degrees of cross-source and cross-platform heterogeneity, ranging from IID coverage to highly skewed distributions.}
    \label{fig:hetero_partitions}
\end{figure*}

\subsubsection{FedGUI-Mobile \& FedGUI-Web}
\paragraph{Cross-Source Heterogeneity.}

Cross-source heterogeneity refers to the statistical discrepancies in data that arise from different acquisition methods and sources. Such heterogeneity can be pronounced even when data is collected on the same platform and operating system. 
For instance, the trajectory data from AC (collected via large-scale outsourcing) and GO (generated by automated synthesis on simulators) exhibit different distributions.


\paragraph{Description.}
To investigate cross-source heterogeneity at a granular level, we constructed two platform-specific datasets: \textit{FedGUI-Mobile} and \textit{FedGUI-Web}. These datasets are designed to highlight the nuanced discrepancies stemming from their distinct data acquisition methods.
\textit{FedGUI-Mobile} comprises three sources, AC, AitW, and GO, originating from different task distributions and collection pipelines. In contrast, \textit{FedGUI-Web} is built from OA-W, GA, and M2W, covering diverse web interaction tasks and website domains. 
Both datasets follow the same construction principle as \textit{FedGUI-OS}, while focusing specifically on source-level differences in GUI environments.

\subsubsection{FedGUI-Full}
\textit{FedGUI-Full} represents the most realistic and challenging setting in our benchmark. It constitutes all data sources and simultaneously incorporates cross-platform and cross-source heterogeneity, while further scaling the number of participating clients.
To systematically characterize diverse degrees and forms of heterogeneity, we construct seven partition variants by jointly controlling source coverage, platform purity, and quantity balance across clients. 


As visualized in Figure~\ref{fig:hetero_partitions}, we consider:
(1) Full IID, where each client observes all sources and platforms with equal proportions;
(2) Full Non-Uniform, where all sources are observed but with uneven data distributions;
(3) Platform Partial, where each client observes only two platforms;
(4) Platform Non-Uniform, which further introduces quantity imbalance across platforms;
(5) Platform Skew, where each client is restricted to a single platform;
(6) Source Non-Uniform, combining platform skew with source-level imbalance; and
(7) Source Skew, where each client receives data from only one source and one platform.
Together, these variants span a broad spectrum of realistic federated heterogeneity patterns.


\begin{table}[t]
\centering
\small
\setlength{\tabcolsep}{6pt}
\begin{tabular}{ll}
\toprule
\multicolumn{2}{l}{\textbf{Supported Base Models}, \;Details in Appendix \ref{sec:model_details}} \\
OpenAI: 
& GPT-5, GPT-4o, GPT-4o-mini \\

Tongyi: 
& Qwen3-VL-2B/4B/8B, \\
& Qwen2.5-VL-3B/7B, Qwen2-VL-2B/7B \\

Google: 
& Gemma-3-4B/12B, Gemini-Pro-2.5 \\

Intern: 
& InternVL2-1B/2B/4B/8B \\

DeepSeek: 
& DeepSeekVL2, DeepSeekVL2-tiny/small \\

Seed: 
& UI-TARS-1.5-7B, Doubao-1.5-Vision-pro \\

\midrule
\multicolumn{2}{l}{\textbf{Integrated FL Algorithms}, \;Details in Appendix \ref{sec:training_details}} \\
\multicolumn{2}{l}{FedAvg, FedProx, SCAFFOLD, FedAvgM,} \\
\multicolumn{2}{l}{FedAdam, FedYogi, FedAdagrad} \\
\bottomrule
\end{tabular}
\caption{Summary of supported base models and integrated federated learning algorithms in FedGUI.}

\label{tab:models_algs}
\end{table}

\subsection{Framework Description}
FedGUI is built upon the prevalent $\texttt{ms-swift}$ framework~\cite{zhao2024swiftascalablelightweightinfrastructure}, which we extend for federated learning by modularizing the core pipeline to seamlessly integrate federated components. This architecture establishes FedGUI as a unified and extensible benchmark for evaluating GUI agents across diverse platforms, supporting a wide array of VLMs and federated algorithms while enabling efficient experimentation.


\paragraph{Model Support \& Algorithm Integration.}
As summarized in Table~\ref{tab:models_algs}, FedGUI accommodates diverse GUI-capable foundation models including both proprietary and open-source VLMs, enabling evaluation across different model scales and architectural designs.
In addition, we integrate commonly used federated optimization algorithms, ensuring fair and reproducible comparisons under identical training configurations.

\begin{table*}[t]
\centering
\small
\setlength\tabcolsep{3.8pt}
\begin{tabular}{
l|
c c c|
c c c|
c c c|
ccc
}
\toprule
\textbf{Distribution} &
\multicolumn{3}{c|}{\textbf{\faMobile\; AndroidControl}} &
\multicolumn{3}{c|}{\textbf{\faGlobeEurope\; GUIAct-Web}} &
\multicolumn{3}{c|}{\textbf{\faLaptop\; AgentSynth}} &
\multicolumn{3}{c}{\textbf{\faCalculator\; Average}} \\
\cmidrule(lr){2-4} \cmidrule(lr){5-7}
\cmidrule(lr){8-10} \cmidrule(lr){11-13}
Metric & Type & Ground & SR & Type & Ground & SR & Type & Ground & SR & Type & Ground & SR \\
\midrule
\rowcolor{gray!10}
Central &
48.41 & 45.47 & 59.89 &
68.48 & 51.48 & 48.73 &
70.54 & 76.74 & 54.91 &
\cellcolor{gray!20}71.28 & \cellcolor{gray!20}55.17 & \cellcolor{gray!20}49.32 \\

\midrule

\rowcolor{yellow!10}
Only Mobile &
72.53 & 29.14 & 40.21 &
46.74 & 18.79 & 23.91 &
43.98 & 19.55 & 8.58 &
\cellcolor{yellow!20}54.42 & \cellcolor{yellow!20}22.49 & \cellcolor{yellow!20}24.23 \\

\rowcolor{yellow!10}
Only Web &
54.78 & 4.49 & 5.16 &
66.12 & 45.09 & 43.66 &
48.96 & 22.08 & 11.89 &
\cellcolor{yellow!20}56.62 & \cellcolor{yellow!20}23.89 & \cellcolor{yellow!20}20.24 \\

\rowcolor{yellow!10}
Only Desktop &
53.57 & 6.46 & 10.17 &
34.78 & 7.89 & 2.90 &
72.06 & 74.77 & 54.63 &
\cellcolor{yellow!20}53.47 & \cellcolor{yellow!20}29.71 & \cellcolor{yellow!20}22.57 \\

\midrule
\cellcolor{red!10}Full IID 
& \cellcolor{red!10}67.83 & \cellcolor{red!10}22.79 & \cellcolor{red!10}30.86 
& \cellcolor{red!10}64.13 & \cellcolor{red!10}44.79 & \cellcolor{red!10}43.84 
& \cellcolor{red!10}69.85 & \cellcolor{red!10}55.77 & \cellcolor{red!10}40.53 
& \cellcolor{red!20}67.27 & \cellcolor{red!20}41.12 & \cellcolor{red!20}38.39 \\

\cellcolor{red!10}Full Non-Uniform 
& \cellcolor{red!10}67.07 & \cellcolor{red!10}19.57 & \cellcolor{red!10}30.96 
& \cellcolor{red!10}59.78 & \cellcolor{red!10}41.86 & \cellcolor{red!10}37.50 
& \cellcolor{red!10}68.74 & \cellcolor{red!10}57.55 & \cellcolor{red!10}41.91 
& \cellcolor{red!20}65.20 & \cellcolor{red!20}39.66 & \cellcolor{red!20}36.79 \\

\midrule

\cellcolor{blue!10}Platform Partial 
& \cellcolor{blue!10}68.44 & \cellcolor{blue!10}18.71 & \cellcolor{blue!10}30.50 
& \cellcolor{blue!10}65.76 & \cellcolor{blue!10}38.55 & \cellcolor{blue!10}40.12 
& \cellcolor{blue!10}70.40 & \cellcolor{blue!10}51.71 & \cellcolor{blue!10}39.54 
& \cellcolor{blue!20}68.20 & \cellcolor{blue!20}36.32 & \cellcolor{blue!20}36.72 \\

\cellcolor{blue!10}Platform Non-Uniform
& \cellcolor{blue!10}70.11 & \cellcolor{blue!10}21.20 & \cellcolor{blue!10}30.02 
& \cellcolor{blue!10}66.49 & \cellcolor{blue!10}39.87 & \cellcolor{blue!10}39.64 
& \cellcolor{blue!10}69.02 & \cellcolor{blue!10}56.85 & \cellcolor{blue!10}40.17 
& \cellcolor{blue!20}68.54 & \cellcolor{blue!20}39.31 & \cellcolor{blue!20}36.61 \\

\cellcolor{blue!10}Platform Skew 
& \cellcolor{blue!10}70.86 & \cellcolor{blue!10}18.44 & \cellcolor{blue!10}33.84 
& \cellcolor{blue!10}60.87 & \cellcolor{blue!10}39.75 & \cellcolor{blue!10}40.58 
& \cellcolor{blue!10}64.87 & \cellcolor{blue!10}37.18 & \cellcolor{blue!10}26.14 
& \cellcolor{blue!20}65.53 & \cellcolor{blue!20}31.79 & \cellcolor{blue!20}33.52 \\

\cellcolor{blue!10}Source Non-Uniform 
& \cellcolor{blue!10}72.08 & \cellcolor{blue!10}20.67 & \cellcolor{blue!10}34.29 
& \cellcolor{blue!10}59.60 & \cellcolor{blue!10}32.76 & \cellcolor{blue!10}36.05 
& \cellcolor{blue!10}66.80 & \cellcolor{blue!10}40.22 & \cellcolor{blue!10}28.63 
& \cellcolor{blue!20}66.16 & \cellcolor{blue!20}31.22 & \cellcolor{blue!20}32.99 \\

\cellcolor{blue!10}Source Skew 
& \cellcolor{blue!10}61.91 & \cellcolor{blue!10}6.91 & \cellcolor{blue!10}16.24 
& \cellcolor{blue!10}56.52 & \cellcolor{blue!10}42.08 & \cellcolor{blue!10}34.24 
& \cellcolor{blue!10}69.29 & \cellcolor{blue!10}58.07 & \cellcolor{blue!10}42.05 
& \cellcolor{blue!20}62.57 & \cellcolor{blue!20}35.69 & \cellcolor{blue!20}30.84 \\

\bottomrule
\end{tabular}
\vspace{-1mm}
\caption{
Results of FedAvg on \textit{FedGUI-Full} under different data distributions. 
\colorbox{yellow!10}{Only X} denotes that the participating clients are restricted to the X domain in the \textit{Source Skew} subset. 
Overall, the performance of the GUI agent degrades as the level of heterogeneity increases, while contributions from other platforms can help improve performance.
}
\label{tab:fedgui_full_distribution}
\vspace{-1mm}
\end{table*}

\section{Experiments}
\label{sec:experiment}

To address the key research questions, we first conduct experiments on \textit{FedGUI-Full}, the most comprehensive and realistic dataset, in Section~\ref{sec:exp_fedgui_full}.
Subsequently, Section~\ref{sec:experiment_fedgui_platform} compares the performance of various FL algorithms, while Section~\ref{sec:exp_fedgui_source} explores additional forms of heterogeneity across their corresponding datasets.
We also analyze the impact of backbone models in Section~\ref{sec:exp_base_model} and assess system efficiency in Section~\ref{sec:exp_efficiency}. 
Supplementary results are detailed in Appendix~\ref{sec:additional_experiment}.

\subsection{Basic Setups (\textit{Details in Appendix \ref{app:ExperimentalDetails}})}
\label{sec:exp_basic_setup}
\paragraph{Base Model.}
We adopt Qwen2-VL-7B \cite{Qwen2VL} as the main base model in our experiments.
To enable efficient adaptation under resource constraints on client-side devices, we apply Low-Rank Adaptation (LoRA) \cite{hu2021lora}.

\paragraph{Training Configuration.}
All models are trained for 30 federated rounds.
At each round, we uniformly sample 10\% of the training data and randomly select 3 clients to participate, simulating realistic user availability and client dropout scenarios \cite{dropout}. 

\paragraph{Evaluation Metrics.}
We evaluate model performance at the action level using three metrics.
\textit{Type} measures whether the predicted action type matches the ground truth.
\textit{Grounding Accuracy (Ground)} evaluates spatial correctness for grounding actions (e.g., \texttt{CLICK}),which measures whether the predicted target GUI element matches the gold element; 
\textit{Success Rate (SR)} requires both correct action type and correct content grounding, including spatial alignment for coordinate-based actions and semantic similarity for text-based actions.

\begin{table*}[t]
\centering
\small

\setlength{\tabcolsep}{6.4pt} 
\begin{tabular}{
l | 
c c c c | 
c c c c | 
c c c c
}
\toprule
\textbf{Algorithm} &
\multicolumn{4}{c|}{\textbf{Platform IID}} &
\multicolumn{4}{c|}{\textbf{Platform Partial}} &
\multicolumn{4}{c}{\textbf{Platform Skew}} \\
\cmidrule(lr){2-5} \cmidrule(lr){6-9} \cmidrule(lr){10-13}
SR & \textbf{\faMobile} & \textbf{\faGlobeEurope} & \textbf{\faLaptop} & \textbf{\faCalculator}
& \textbf{\faMobile} & \textbf{\faGlobeEurope} & \textbf{\faLaptop} & \textbf{\faCalculator}
& \textbf{\faMobile} & \textbf{\faGlobeEurope} & \textbf{\faLaptop} & \textbf{\faCalculator} \\
\midrule

Central 
& \cellcolor{gray!10}48.10 & \cellcolor{gray!10}53.26 & \cellcolor{gray!10}60.72 & \cellcolor{gray!20}54.03
& \cellcolor{gray!10}- & \cellcolor{gray!10}- & \cellcolor{gray!10}- & \cellcolor{gray!10}-
& \cellcolor{gray!10}- & \cellcolor{gray!10}- & \cellcolor{gray!10}- & \cellcolor{gray!10}- \\

\midrule
Local
& \cellcolor{red!10}27.77 & \cellcolor{red!10}35.51 & \cellcolor{red!10}28.63 & \cellcolor{red!20}30.64
& \cellcolor{blue!10}28.07 & \cellcolor{blue!10}33.15 & \cellcolor{blue!10}24.62 & \cellcolor{blue!20}28.61
& \cellcolor{blue!10}33.84 & \cellcolor{blue!10}10.51 & \cellcolor{blue!10}5.53 & \cellcolor{blue!20}16.63
\\
FedAvg
& \cellcolor{red!10}35.05 & \cellcolor{red!10}43.12 & \cellcolor{red!10}33.06 & \cellcolor{red!20}37.08
& \cellcolor{blue!10}32.78 & \cellcolor{blue!10}42.21 & \cellcolor{blue!10}35.27 & \cellcolor{blue!20}36.75
& \cellcolor{blue!10}27.16 & \cellcolor{blue!10}43.84 & \cellcolor{blue!10}34.16 & \cellcolor{blue!20}35.05
\\
FedProx
& \cellcolor{red!10}32.63 & \cellcolor{red!10}42.93 & \cellcolor{red!10}37.22 & \cellcolor{red!20}37.59
& \cellcolor{blue!10}31.63 & \cellcolor{blue!10}41.85 & \cellcolor{blue!10}36.21 & \cellcolor{blue!20}36.56
& \cellcolor{blue!10}27.77 & \cellcolor{blue!10}43.84 & \cellcolor{blue!10}34.72 & \cellcolor{blue!20}35.44
\\
SCAFFOLD
& \cellcolor{red!10}33.69 & \cellcolor{red!10}43.48 & \cellcolor{red!10}34.02 & \cellcolor{red!20}37.06
& \cellcolor{blue!10}32.12 & \cellcolor{blue!10}44.20 & \cellcolor{blue!10}33.33 & \cellcolor{blue!20}36.55
& \cellcolor{blue!10}26.86 & \cellcolor{blue!10}43.48 & \cellcolor{blue!10}33.89 & \cellcolor{blue!20}34.74
\\
FedYogi
& \cellcolor{red!10}35.81 & \cellcolor{red!10}44.38 & \cellcolor{red!10}52.28 & \cellcolor{red!20}44.16
& \cellcolor{blue!10}35.81 & \cellcolor{blue!10}43.12 & \cellcolor{blue!10}48.13 & \cellcolor{blue!20}42.35
& \cellcolor{blue!10}34.75 & \cellcolor{blue!10}48.55 & \cellcolor{blue!10}50.90 & \cellcolor{blue!20}44.73
\\
FedAdam
& \cellcolor{red!10}37.94 & \cellcolor{red!10}43.84 & \cellcolor{red!10}53.53 & \cellcolor{red!20}45.10
& \cellcolor{blue!10}35.05 & \cellcolor{blue!10}42.93 & \cellcolor{blue!10}48.82 & \cellcolor{blue!20}42.27
& \cellcolor{blue!10}33.84 & \cellcolor{blue!10}44.38 & \cellcolor{blue!10}50.62 & \cellcolor{blue!20}42.95
\\
FedAvgM
& \cellcolor{red!10}33.54 & \cellcolor{red!10}42.93 & \cellcolor{red!10}34.30 & \cellcolor{red!20}36.92
& \cellcolor{blue!10}34.33 & \cellcolor{blue!10}42.03 & \cellcolor{blue!10}33.59 & \cellcolor{blue!20}36.65
& \cellcolor{blue!10}28.22 & \cellcolor{blue!10}44.93 & \cellcolor{blue!10}34.85 & \cellcolor{blue!20}36.00
\\
FedAdagrad
& \cellcolor{red!10}35.81 & \cellcolor{red!10}43.12 & \cellcolor{red!10}46.75 & \cellcolor{red!20}41.89
& \cellcolor{blue!10}35.36 & \cellcolor{blue!10}41.30 & \cellcolor{blue!10}33.61 & \cellcolor{blue!20}36.76
& \cellcolor{blue!10}26.71 & \cellcolor{blue!10}26.99 & \cellcolor{blue!10}35.13 & \cellcolor{blue!20}29.61
\\

\bottomrule
\end{tabular}
\vspace{-1mm}
\caption{
Success Rate (SR, \%) results on \textit{FedGUI-Platform}. Each algorithm is evaluated under \colorbox{red!10}{IID} and \colorbox{blue!10}{non-IID} settings. Results are shown for three benchmarks (AC: \faMobile, GA-W: \faGlobeEurope, AS: \faLaptop) and their average (\faCalculator).
Algorithms exhibit diverse performance patterns, while optimizer-based methods are more robust under skewed distributions.
}
\label{tab:fedgui_platform}
\vspace{-1mm}
\end{table*}

\subsection{Verification of Cross-Platform Federated Collaboration}
\label{sec:exp_fedgui_full}
\paragraph{Setups.}

To verify whether cross-platform collaboration improves federated learning performance, we first conduct experiments on \textit{FedGUI-Full} across different distribution subsets using FedAvg \cite{fedavg}.
The comparisons are organized along two dimensions: 
(1) comparison between collaboration on \colorbox{red!10}{homogeneous} and \colorbox{blue!10}{heterogeneous} distributions with increasing heterogeneity levels;
(2) comparison of collaboration among clients from \colorbox{yellow!10}{a single platform} versus clients from \colorbox{blue!10}{all platforms (\textit{Source Skew})}.



\paragraph{Results. }
From Table~\ref{tab:fedgui_full_distribution}, we draw the following key conclusions:
(1) Single-domain models suffer from catastrophic failure on unseen platforms, but federated learning (even under \textit{Source Skew}) significantly restores baseline utility. This underscores that federated collaboration is essential to bridge the domain isolation that renders localized GUI agents non-functional across platforms.
(2) Overall performance degrades consistently as the data distribution shifts from IID coverage to increasingly heterogeneous settings, confirming that cross-platform heterogeneity fundamentally challenges federated GUI learning.
(3) A clear platform sensitivity hierarchy emerges: desktop exhibits the highest sensitivity, followed by web, then mobile. This suggests asymmetric knowledge transfer, where mobile interactions transfer more effectively to other platforms than desktop does.

\begin{figure*}[t]
    \centering
    \begin{minipage}[t]{0.325\linewidth}
        \centering
        \includegraphics[width=0.9\linewidth]{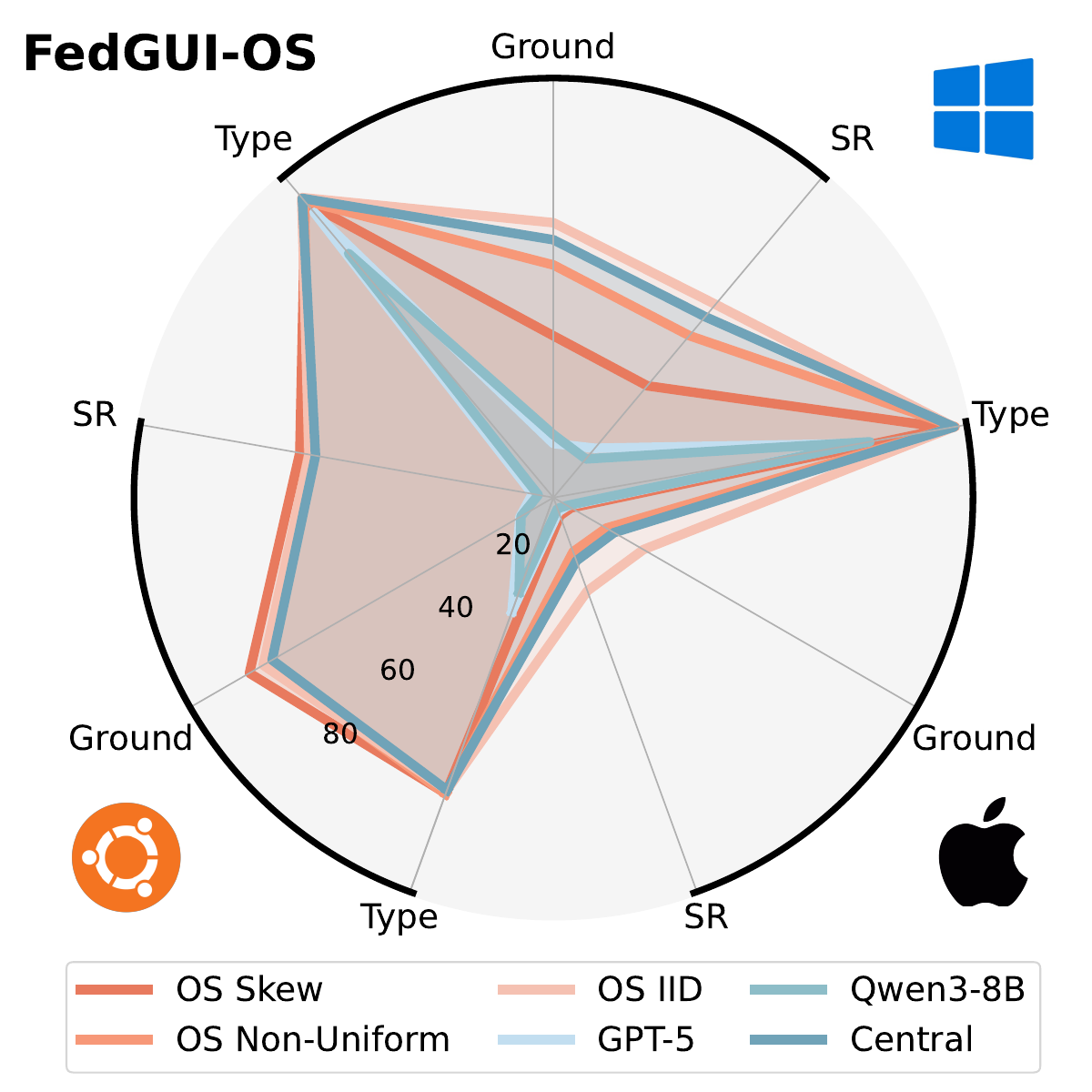}
        \caption{
Results on \textit{FedGUI-OS}, with evaluations conducted on Windows, Ubuntu, and macOS. 
Our Federated GUI agents demonstrate superior cross-OS generalization compared to strong VLMs like GPT-5.
}
        \label{fig:fedgui_os}
    \end{minipage}
    \hfill
    \begin{minipage}[t]{0.65\linewidth} 
        \centering
        \begin{subfigure}{0.45\linewidth} 
            \centering
            \includegraphics[width=\linewidth]{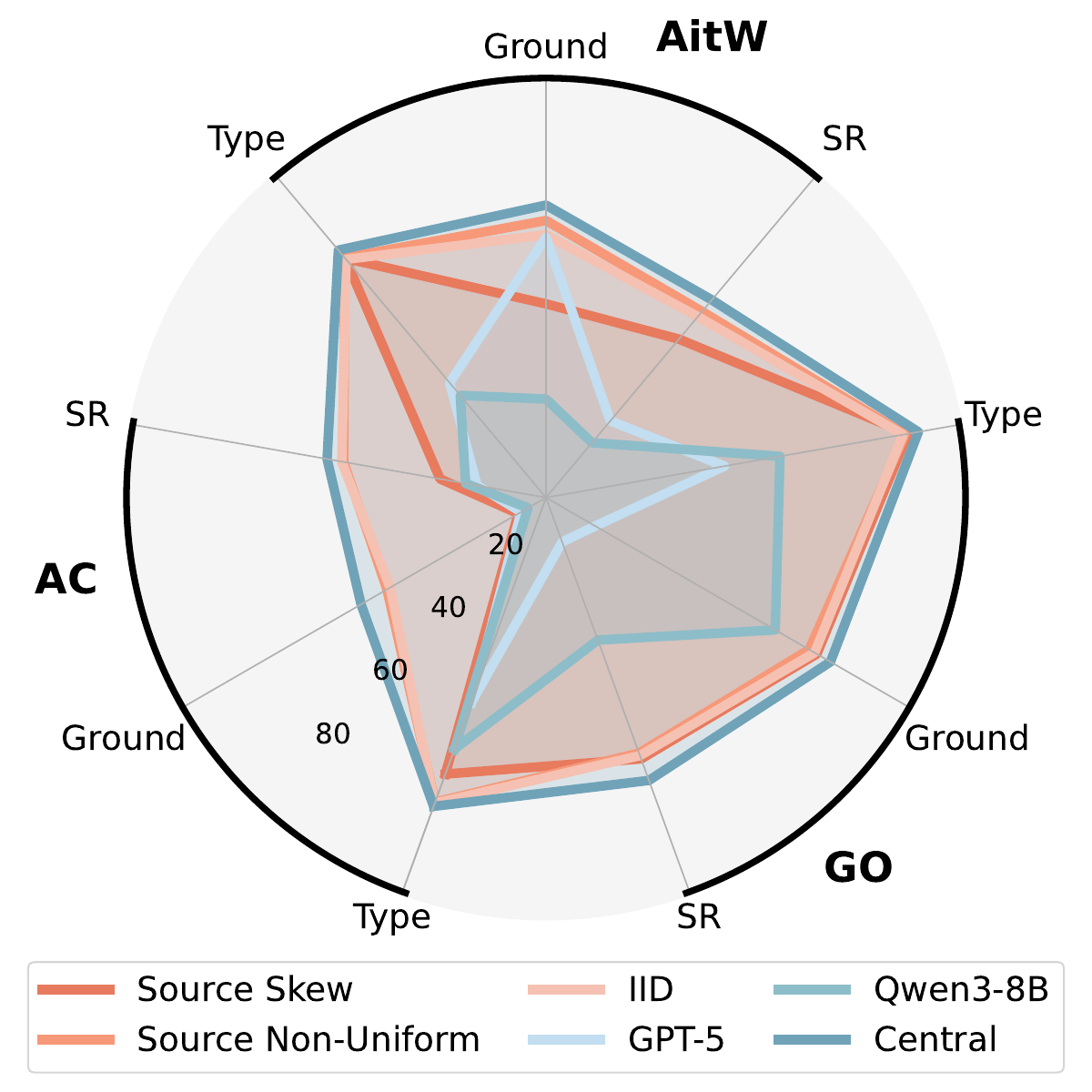}
            \caption{FedGUI-Mobile.}
            \label{fig:radar_mobile}
        \end{subfigure}
        \hfill
        \begin{subfigure}{0.45\linewidth}
            \centering
            \includegraphics[width=\linewidth]{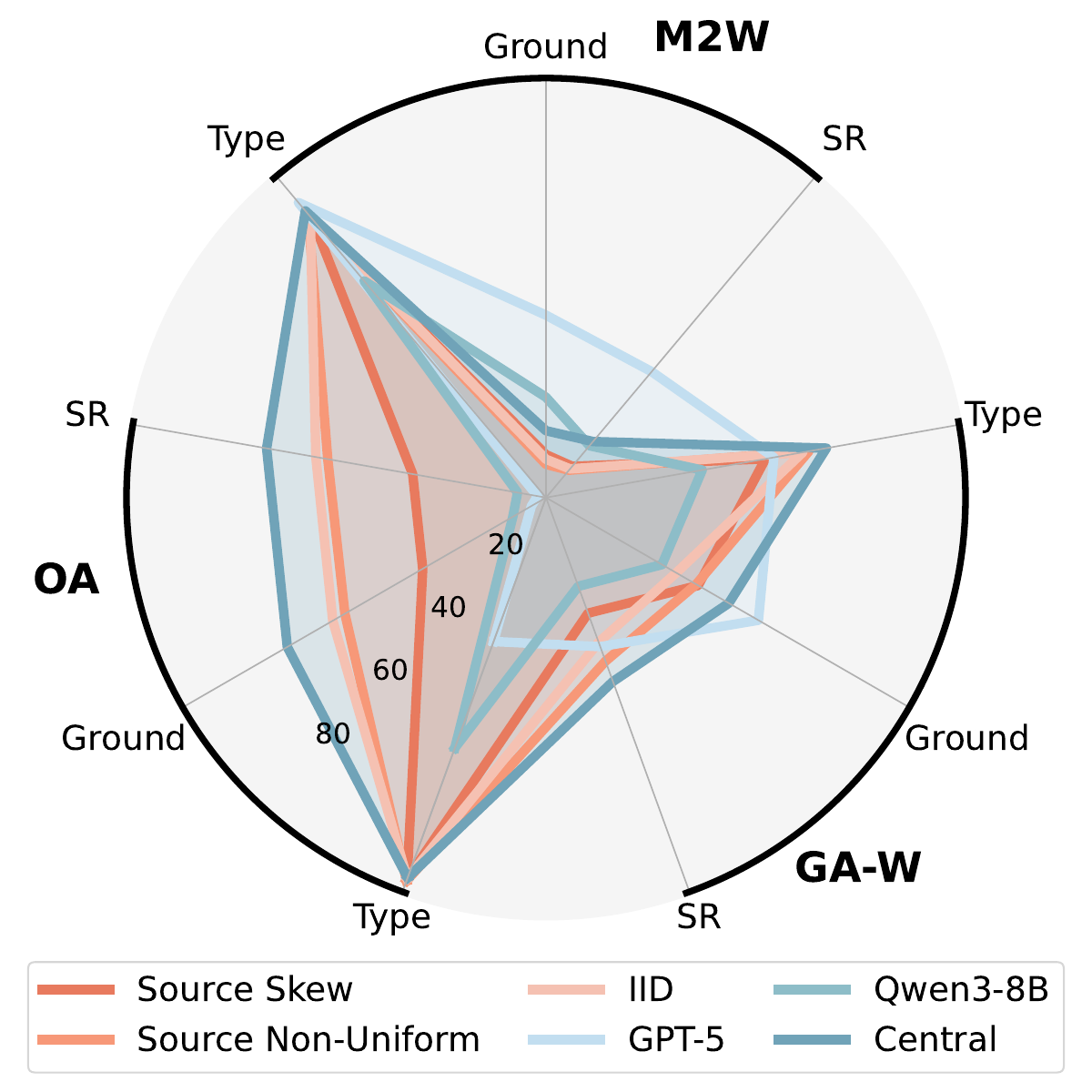}
            \caption{FedGUI-Web.}
            \label{fig:radar_web}
        \end{subfigure}

        \vspace{-1.5mm}
        \caption{Radar-chart comparison on \textit{FedGUI-Mobile} and \textit{FedGUI-Web}.
        Results for models trained with FedAvg on their respective source-specific subsets are plotted in red.
        The diverse data distributions in our dataset effectively reveal varied performance patterns, highlighting its capability to measure robustness against real-world distribution shifts.
        } 
        \label{fig:radars_combined}
    \end{minipage}
\end{figure*}

\subsection{Comparison of FL Algorithms}
\label{sec:experiment_fedgui_platform}
\paragraph{Setups.}
To examine their performance under diverse data distributions, we further compare different FL algorithms in three representative subsets on both \textit{FedGUI-Platform} and \textit{FedGUI-Full}, with results shown in Table~\ref{tab:fedgui_platform} and Table~\ref{tab:fedgui_full_alg}, respectively. 
We also provide centralized and local baselines under identical configurations for reference.

We conclude the following findings:
(1) Federated learning exhibits a strong salvage effect under cross-platform heterogeneity: while local models trained on a single platform fail to generalize to others, FL algorithms substantially recover cross-platform usability, elevating unusable models to a functional level.
(2) Adaptive optimizers (e.g., FedAdam) achieve higher average performance under platform-skewed distributions, suggesting stronger robustness to platform-induced interference compared to correction-based methods (e.g., FedProx).
(3) Despite these gains, a substantial gap to centralized training remains, indicating considerable headroom for future work and exposing an inherent trade-off in federated settings between global generalization and platform-specific expertise under extreme heterogeneity.



\subsection{Investigation of Cross-Device, Cross-OS, and Cross-Source Heterogeneity}

\label{sec:exp_fedgui_source}
\paragraph{Setups.}

Beyond cross-platform heterogeneity, we further investigate more fine-grained types of heterogeneity: cross-device, cross-OS, and cross-source variations on the corresponding datasets described in Section \ref{sec:data_description}.
For cross-device heterogeneity, we evaluate all algorithms across all five devices on \textit{FedGUI-Device}, with complete results provided in Table \ref{tab:fedgui_device}.
For cross-OS and cross-source experiments, we evaluate models on the corresponding source test sets. For example, on \textit{FedGUI-OS} shown in Figure \ref{fig:fedgui_os}, we test on samples from AS (Ubuntu), OA-Mac (macOS), and OA-Win (Windows).
All models are fine-tuned for 30 rounds under three data distribution settings: IID, non-uniform, and skewed. 

\begin{table*}[t]
\centering
\small
\setlength\tabcolsep{8pt}
\begin{tabular}{l | ccccccccc | c}
\toprule
\textbf{Setting} & \textbf{Edu} & \textbf{Fin} & \textbf{Game} & \textbf{Travel} & \textbf{Shop} & \textbf{Ent} & \textbf{Lit} & \textbf{Ref} & \textbf{Trans} & \textbf{ Avg.} \\ 
\midrule

\rowcolor{red!10}
App IID 
& 71.54 & 60.71 & 48.48 & 43.75 
& 53.66 & 34.33 & 51.61 & 39.58 & 25.45 & \cellcolor{red!20}\textbf{47.68} \\

\midrule

\rowcolor{blue!10}
App Partial 
& 70.59 & 62.12 & 43.55 & 41.26 
& 53.12 & 32.89 & 50.98 & 39.12 & 23.56 & \cellcolor{blue!20}46.35 \\

\rowcolor{blue!10}
App Skew 
& 69.23 & 64.28 & 42.42 & 40.00 
& 52.65 & 34.33 & 50.00 & 38.52 & 21.81 & \cellcolor{blue!20}45.92 \\

\bottomrule
\end{tabular}
\vspace{-1mm}
\caption{Cross-application heterogeneity results on web environments using {Qwen2-VL-7B}. Performances across \colorbox{red!10}{homogeneous} and \colorbox{blue!10}{heterogeneous} distributions show the model's robustness across diverse application domains.}
\label{tab:cross_app}
\vspace{-1mm}
\end{table*}

\paragraph{Results.} 
From Figure \ref{fig:fedgui_os} and \ref{fig:radars_combined}, we observe that:
(1) The gap between skewed and IID settings (where the light-red curves diverge from and cover the dark-red curves) further evidences the presence of the identified heterogeneities in FedGUI.
(2) Federated fine-tuning consistently outperforms strong VLM baselines (e.g., GPT-5), highlighting the value of aggregating in-domain GUI interaction knowledge beyond zero-shot generalization.
(3) In some cases (e.g., \textit{FedGUI-OS}), FedAvg achieve results close to centralized learning, demonstrating the practical value of federated GUI agents.
(4) Cross-OS heterogeneity appears more critical than source or device heterogeneity, as OS Skew causes substantially larger performance drops compared to the modest variations observed for the same algorithm across device distributions in Table \ref{tab:fedgui_device}.

\subsection{Cross-Application Heterogeneity.}
\paragraph{Setups.}
To explore how cross-application heterogeneity affects federated training performance, we conduct experiments on \textit{FedGUI-Web}, covering 10 application categories ranging from education to transportation. 
Heterogeneity increases from balanced App IID, to subset-based App Partial, and finally single-app dominated App Skew.

\paragraph{Results.}
Key observations from Table~\ref{tab:cross_app} are as follows: 
(1) Although VLM agents can handle a wide variety of web tasks, performance is still affected by extreme skew in website applications.
(2) Compared to the larger degradation under cross-platform heterogeneity (Table~\ref{tab:fedgui_full_distribution}), cross-application shifts have a much smaller impact, suggesting that platform-level GUI differences are more challenging than application-level variations.

\begin{figure*}[t]
    \centering
    \begin{subfigure}{0.24\linewidth}
        \centering
        \includegraphics[width=\linewidth]{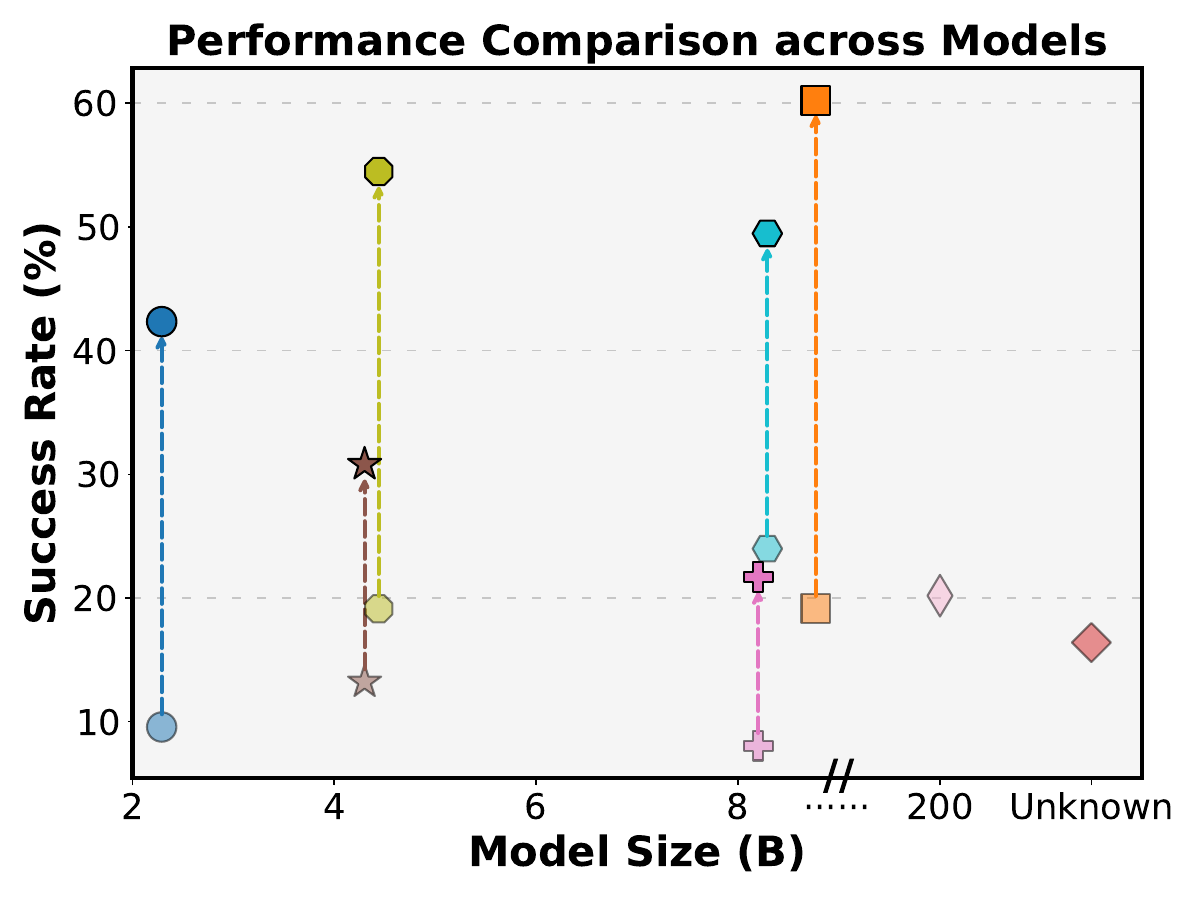}
        \caption{Moblie Domain}
    \end{subfigure}
    \hfill
    \begin{subfigure}{0.24\linewidth}
        \centering
        \includegraphics[width=\linewidth]{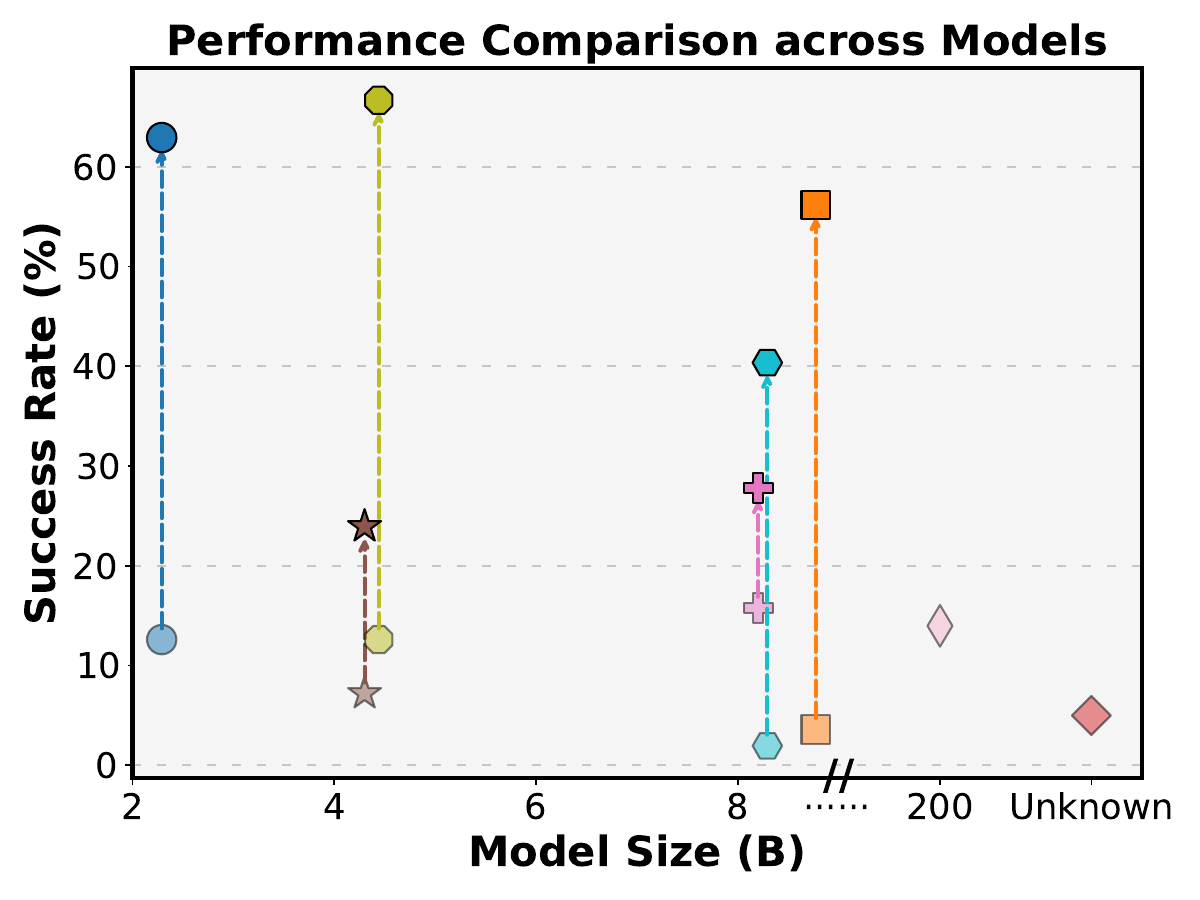}
        \caption{Web Domain}  
    \end{subfigure}
    \hfill 
    \begin{subfigure}{0.24\linewidth}
        \centering
        \includegraphics[width=\linewidth]{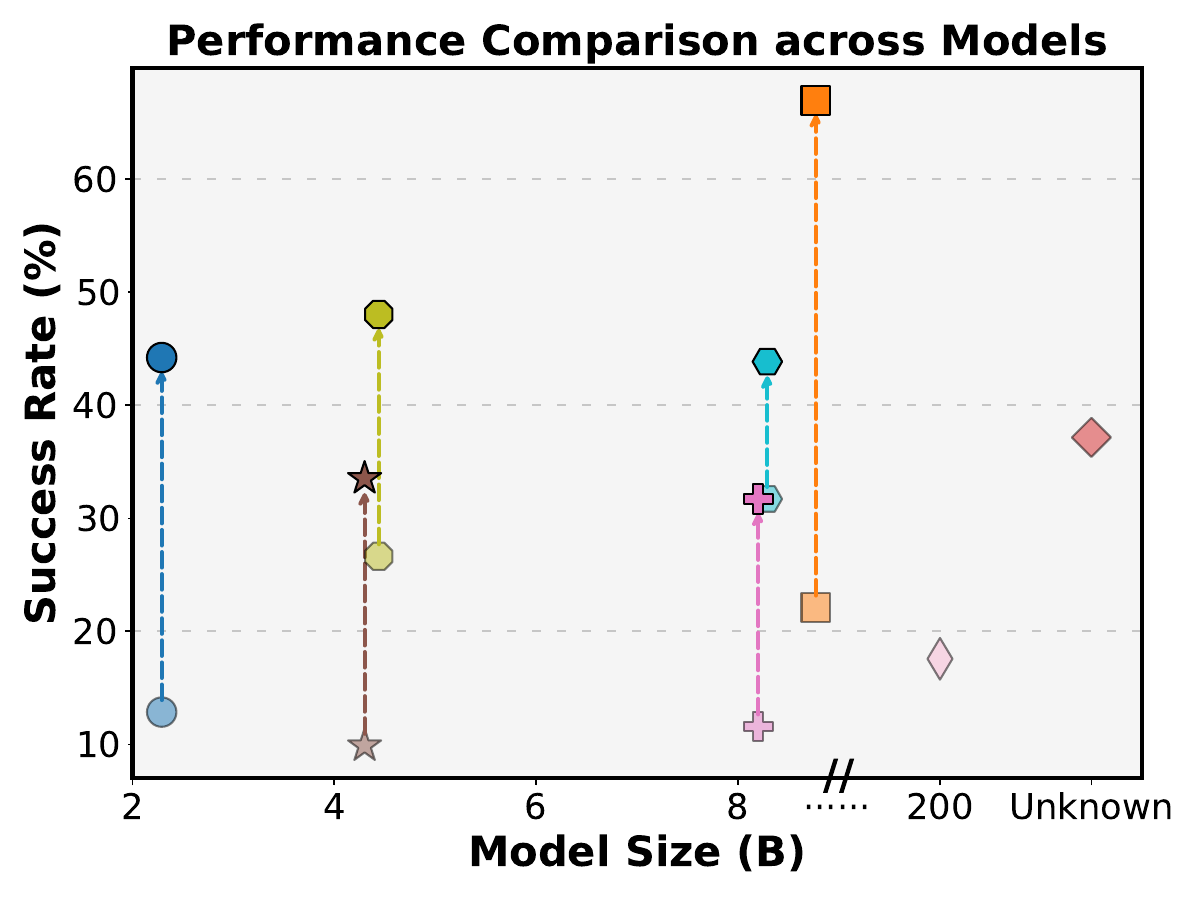}
        \caption{Desktop Domain}
    \end{subfigure}
    \begin{subfigure}{0.24\linewidth}
        \centering
        \includegraphics[width=\linewidth]{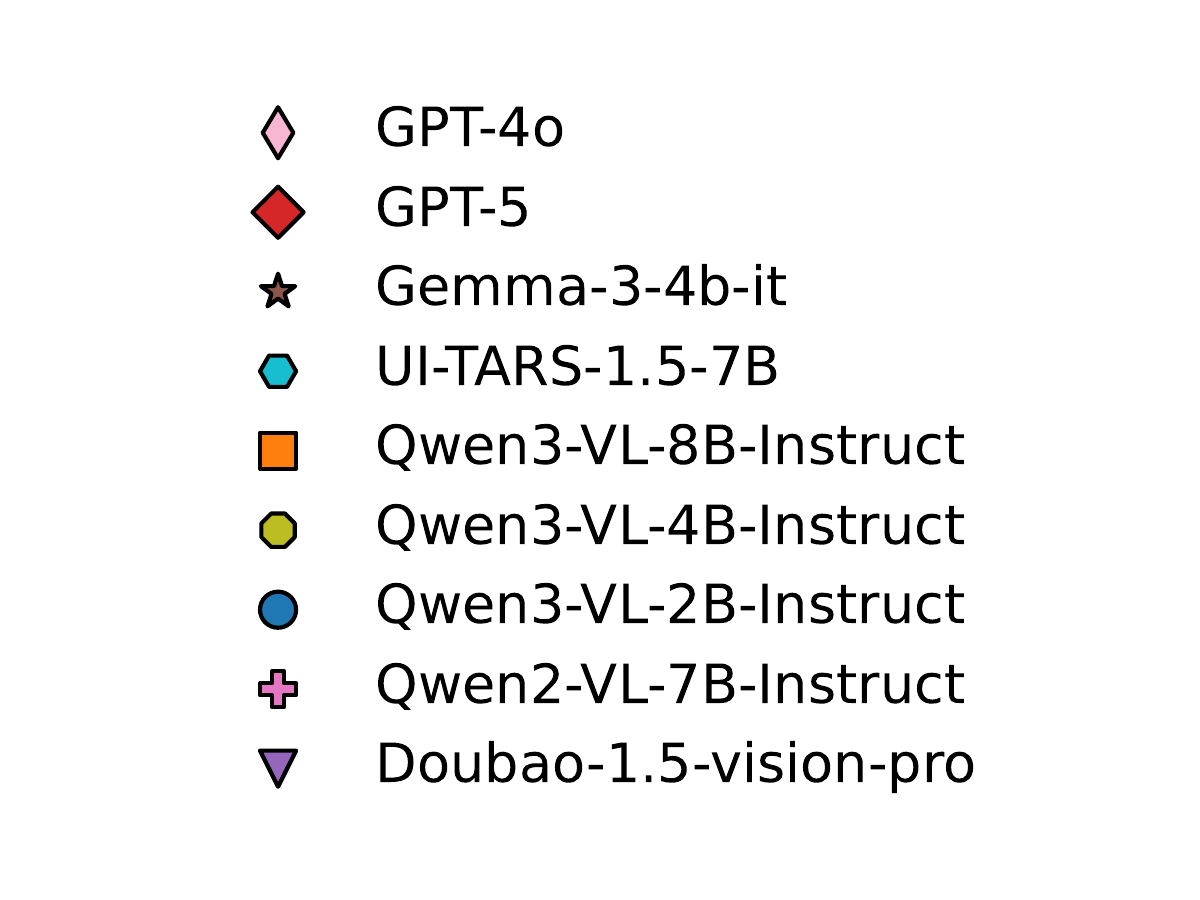}
        \caption{Model Legend}

    \end{subfigure}
    \vspace{-1mm}
    \caption{Model performance comparison across mobile, web, and desktop domains.
Semi-transparent and solid markers denote base and FedAvg fine-tuned models, respectively, with dashed arrows indicating federated gains. Notably, after federated fine-tuning, several compact open-source models outperform large proprietary models.}
    \label{fig:fedgui_model}
    \vspace{-1mm}
\end{figure*}

\subsection{Comparison of Model Backbones}
\label{sec:exp_base_model}
\paragraph{Setups.}
To analyze the impact of model backbones on cross-platform federated GUI training, we conduct experiments on over 20 VLMs, covering both open-source and proprietary models, including Qwen, Gemma, and GPT~\cite{gpt4}. 
For open-source models, we evaluate both their base performance and federated fine-tuned counterparts on \textit{FedGUI-Full} as a representative setting.

\paragraph{Results.}
As visualized in Figure \ref{fig:fedgui_model} and \ref{fig:fedgui_model_supplementary}: 
(1) The inconsistent performance of base VLMs across domains necessitates in-domain adaptation for executable GUI behavior.
(2) After federated training, performance becomes more structured and shows a clearer positive correlation with model scale, particularly within the Qwen3 family \cite{Qwen3-VL}.
(3) Collaboration on distributed GUI data narrows the gap between small open-source models and large proprietary models, enabling compact VLMs to achieve competitive GUI performance.

\subsection{Efficiency Evaluation}
\label{sec:exp_efficiency}


\begin{table}[t]
\centering
\small
\setlength{\tabcolsep}{4pt}
\begin{tabular}{l c}

\toprule
\textbf{Approach} & \textbf{Overhead} \\
\midrule
Central + Unpacked data (5k Epi.)   & $\approx$ 20 GB \\   
Central + Unpacked data (All)   & $\approx$ 159 GB \\
\midrule
FL + Full model (Qwen3-VL-8B)         & 16.50 GB $\times$ round \\
FL + LoRA adapter (rank=8,$\alpha$=32)          & 84.32 MB $\times$ round \\
\bottomrule
\end{tabular}
\caption{Communication overhead of FedGUI using different approaches.
Transmitting only LoRA adapters achieves the highest communication efficiency.}
\label{tab:communication}
\end{table}

\begin{table}[t]
\centering
\setlength{\tabcolsep}{2pt}
\small
\begin{tabular}{l c c}
\toprule
{\textbf{Base Model}} & \textbf{GPU Memory}(MB) & \textbf{Time per Round}(s) \\
\midrule
SmolVLM-500M   & \phantom{0}6,056  & 04:47 \\
Intern2-VL-1B  & 10,498 & 06:56 \\
Qwen3-VL-2B    & 14,536 & 10:24 \\
Gemma-4B       & 17,650 & 18:10 \\
Qwen2-VL-7B    & 21,610 & 18:24 \\
Qwen3-VL-8B    & 30,490 & 31:03 \\
\bottomrule
\end{tabular}
\caption{Computational statistics for training VLMs using \textit{FedGUI-Full}, showing its efficiency in both GPU memory usage and training time.
}
\label{tab:computation}
\vspace{-1mm}
\end{table}

\paragraph{Communication.}
Assuming equal episode sizes, we compare three settings: 
centralized training, federated full fine-tuning, and federated training with LoRA.
As shown in Table~\ref{tab:communication}, LoRA-based FL achieves the lowest communication cost.
\paragraph{Computation.}
Table~\ref{tab:computation} reports the computational cost of \textit{FedGUI-Full} across a diverse set of VLMs, spanning from lightweight to medium-scale models. We measure the peak GPU memory usage and the average wall-clock time per federated round.  These results reinforce the training efficiency of FedGUI, which can be fully executed on a single RTX 4090 GPU. Notably, 2B models are compact enough to be deployed on standard mobile devices with 16GB of RAM.

\paragraph{Deployability on Edge Devices.}
To better reflect realistic mobile and web scenarios, we evaluate the performance of sub-2B models that are deployable on high-end smartphones or edge devices. We specifically test Qwen3-VL-2B-Instruct and SmolVLM-500M-Instruct \cite{marafioti2025smolvlm}. As shown in Table~\ref{tab:deployable_performance}, federated training consistently improves sub-2B models across all distributions, making them hardware-feasible for real-world mobile and edge deployment.


\begin{table}[t]

\centering
\small
\setlength\tabcolsep{3pt}
\begin{tabular}{c l c c c c}
\toprule
\textbf{Model} & \textbf{Distribution} & 
\textbf{\faMobile} & \textbf{\faGlobeEurope} & \textbf{\faLaptop} & \textbf{\faCalculator} \\
\midrule

\multirow{3}{*}{\makecell{Qwen3-VL-\\2B-Instruct} }
& \cellcolor{gray!10} Base 
& \cellcolor{gray!10} 9.56 
& \cellcolor{gray!10} 12.59 
& \cellcolor{gray!10} 12.86 
& \cellcolor{gray!20} 11.67 \\

& \cellcolor{red!10} Full IID 
& \cellcolor{red!10} 42.34 
& \cellcolor{red!10} 62.93 
& \cellcolor{red!10} 44.20 
& \cellcolor{red!20} 49.82 \\

& \cellcolor{blue!10} Source Skew 
& \cellcolor{blue!10} 32.92 
& \cellcolor{blue!10} 58.32 
& \cellcolor{blue!10} 43.32 
& \cellcolor{blue!20} 44.85 \\

\midrule

\multirow{3}{*}{\makecell{SmolVLM-\\500M}}
& \cellcolor{gray!10} Base 
& \cellcolor{gray!10} 0.95 
& \cellcolor{gray!10} 1.02 
& \cellcolor{gray!10} 0.85 
& \cellcolor{gray!20} 0.94 \\

& \cellcolor{red!10} Full IID 
& \cellcolor{red!10} 15.15 
& \cellcolor{red!10} 22.28 
& \cellcolor{red!10} 20.60 
& \cellcolor{red!20} 19.34 \\

& \cellcolor{blue!10} Source Skew 
& \cellcolor{blue!10} 5.31 
& \cellcolor{blue!10} 7.42 
& \cellcolor{blue!10} 20.47 
& \cellcolor{blue!20} 11.07 \\

\bottomrule
\end{tabular}

\caption{Success Rate (\%) results using mobile-trainable models show that even models deployable on-device remain effective under FedGUI, demonstrating the feasibility of practical edge-side agents.}
\label{tab:deployable_performance}
\end{table}

\section{Conclusion}

In this paper, we introduce \textbf{FedGUI}, the first unified benchmark for cross-platform federated GUI agents that systematically models heterogeneity across mobile, web, and desktop environments. 
Through six carefully curated datasets, FedGUI exposes realistic non-IID challenges largely overlooked by existing benchmarks.
Extensive experiments across 7 algorithms and 20+ models show that such heterogeneity fundamentally degrades performance, while cross-platform collaboration exhibits a strong salvage effect even under severe isolation. 

Overall, FedGUI bridges federated learning research and practical GUI agent evaluation, providing a solid foundation for robust and privacy-preserving agent development.

\section*{Limitations}

Despite its comprehensive design, coverage of cross-platform heterogeneity, and the informative insights provided, FedGUI still faces one major challenge.
Our benchmark relies on publicly available GUI datasets rather than real user-owned private data collected in live federated deployments. While real user data would better reflect authentic interaction patterns, privacy, ethical, and reproducibility constraints make such data difficult to obtain and release at scale.
Therefore, to better benefit the research community and facilitate more accessible experimentation, we incorporate diverse publicly available data sources, which, by virtue of their scale and heterogeneity, can reasonably simulate real-world distributed user data.




\bibliography{custom}

\appendix

\section{Future Directions}
\subsection{Cross-Platform Heterogeneity-Aware Optimization}
While our experiments demonstrate that cross-platform collaboration can improve overall performance, the substantial heterogeneity across mobile, web, and desktop environments remains a significant challenge for federated GUI agents. Future work should explore specialized aggregation mechanisms that can dynamically weight client contributions based on platform-specific characteristics and task relevance. For instance, developing platform-aware federated optimization algorithms that can identify and leverage complementary knowledge across different GUI modalities (e.g., touch-based interactions on mobile vs. keyboard-mouse interactions on desktop) could substantially improve convergence speed and final performance. 

Additionally, investigating hierarchical federated learning architectures that first aggregate updates within the same platform before cross-platform aggregation may help mitigate the negative impact of extreme heterogeneity. 
The promising method based on our empirical results are design optimizer based algorithms based on FedYogi \cite{fedopt}.


\subsection{Real-World Data Collection and Continuous Learning}
The current benchmark relies on curated datasets from existing sources, which may not fully capture the complexity and diversity of real-world user interactions across different geographical regions, application versions, and usage patterns. Future research should investigate scalable frameworks for continuous data collection from real users in production environments \cite{kong2025mobileworldbenchmarkingautonomousmobile, zhang2026silobenchscalableenvironmentevaluating}, while addressing the inherent challenges of data quality control, annotation efficiency, and temporal distribution shifts. Specifically, developing semi-supervised or self-supervised learning techniques that can leverage abundant unlabeled GUI interaction traces would significantly reduce the annotation burden. 
\subsection{Privacy-Preserving Mechanisms for Federated GUI Agents}
Although federated learning provides a foundation for privacy preservation by keeping raw data local, GUI interaction traces often contain highly sensitive information including personal identities, financial transactions, health records, and behavioral patterns that could be exposed through model updates or inferred from agent actions. Future work should integrate advanced privacy-enhancing technologies specifically designed for GUI agent training. Differential privacy (DP) \cite{dwork2006differential} mechanisms and secure aggregation \cite{secureagg} could be adapted to account for the sequential and high-dimensional nature of GUI interaction data, ensuring formal privacy guarantees without severely degrading model utility. 

\begin{table*}[t]
\centering
\setlength\tabcolsep{1.5pt}
\small
\resizebox{1\textwidth}{!}{
\begin{tabular}{l|C{1cm}C{1cm}C{1cm}C{1cm}C{1cm}C{1cm}|l|C{1cm}C{1cm}C{1cm}C{1cm}C{1cm}C{1cm}}
\toprule
\textbf{Algorithm} & P7P & P8P & Sm. & Med. & Tab & \textbf{Avg.} & \textbf{Algorithm} & P7P & P8P & Sm. & Med. & Tab & \textbf{Avg.} \\ 
\midrule
Central & \cellcolor{gray!10}78.95 & \cellcolor{gray!10}76.74 & \cellcolor{gray!10}77.71 & \cellcolor{gray!10}78.81 & \cellcolor{gray!10}63.64 & \cellcolor{gray!20}75.17 & Central & \cellcolor{gray!10}78.95 & \cellcolor{gray!10}76.74 & \cellcolor{gray!10}77.71 & \cellcolor{gray!10}78.81 & \cellcolor{gray!10}63.64 & \cellcolor{gray!20}75.17 \\
\midrule
\textbf{Homo.} & \multicolumn{6}{c|}{\cellcolor{red!10}\textbf{Device IID}} & \textbf{Hetero.} & \multicolumn{6}{c}{\cellcolor{blue!10}\textbf{Device Non-Uniform}}\\
Local & \cellcolor{red!10}56.15 & \cellcolor{red!10}60.91 & \cellcolor{red!10}54.29 & \cellcolor{red!10}57.25 & \cellcolor{red!10}46.46 & \cellcolor{red!20}55.01 & Local & \cellcolor{blue!10}53.85 & \cellcolor{blue!10}63.64 & \cellcolor{blue!10}45.14 & \cellcolor{blue!10}61.83 & \cellcolor{blue!10}49.49 & \cellcolor{blue!20}54.79 \\
FedAvg & \cellcolor{red!10}66.15 & \cellcolor{red!10}70.00 & \cellcolor{red!10}63.71 & \cellcolor{red!10}70.99 & \cellcolor{red!10}60.61 & \cellcolor{red!20}66.29 & FedAvg & \cellcolor{blue!10}64.62 & \cellcolor{blue!10}67.27 & \cellcolor{blue!10}60.76 & \cellcolor{blue!10}64.12 & \cellcolor{blue!10}54.55 & \cellcolor{blue!20}62.26 \\
FedProx & \cellcolor{red!10}67.69 & \cellcolor{red!10}68.18 & \cellcolor{red!10}62.86 & \cellcolor{red!10}69.47 & \cellcolor{red!10}61.62 & \cellcolor{red!20}65.96 & FedProx & \cellcolor{blue!10}63.08 & \cellcolor{blue!10}66.36 & \cellcolor{blue!10}62.29 & \cellcolor{blue!10}64.12 & \cellcolor{blue!10}57.58 & \cellcolor{blue!20}62.69 \\
FedAvgM & \cellcolor{red!10}67.69 & \cellcolor{red!10}70.91 & \cellcolor{red!10}61.71 & \cellcolor{red!10}67.18 & \cellcolor{red!10}60.61 & \cellcolor{red!20}65.62 & FedAvgM & \cellcolor{blue!10}66.15 & \cellcolor{blue!10}66.36 & \cellcolor{blue!10}58.86 & \cellcolor{blue!10}64.89 & \cellcolor{blue!10}56.57 & \cellcolor{blue!20}62.57 \\
SCAFFOLD & \cellcolor{red!10}66.15 & \cellcolor{red!10}69.09 & \cellcolor{red!10}61.14 & \cellcolor{red!10}70.23 & \cellcolor{red!10}56.57 & \cellcolor{red!20}64.64 & SCAFFOLD & \cellcolor{blue!10}64.62 & \cellcolor{blue!10}67.27 & \cellcolor{blue!10}61.71 & \cellcolor{blue!10}66.41 & \cellcolor{blue!10}54.55 & \cellcolor{blue!20}62.91 \\
FedAdagrad & \cellcolor{red!10}66.15 & \cellcolor{red!10}73.64 & \cellcolor{red!10}65.71 & \cellcolor{red!10}72.52 & \cellcolor{red!10}64.65 & \cellcolor{red!20}68.53 & FedAdagrad & \cellcolor{blue!10}68.46 & \cellcolor{blue!10}71.82 & \cellcolor{blue!10}68.57 & \cellcolor{blue!10}68.70 & \cellcolor{blue!10}59.60 & \cellcolor{blue!20}67.43 \\
FedYogi & \cellcolor{red!10}\textbf{71.54} & \cellcolor{red!10}75.45 & \cellcolor{red!10}\textbf{76.57} & \cellcolor{red!10}\textbf{75.57} & \cellcolor{red!10}\textbf{68.69} & \cellcolor{red!20}\textbf{73.56} & FedYogi & \cellcolor{blue!10}70.77 & \cellcolor{blue!10}\textbf{76.36} & \cellcolor{blue!10}\textbf{70.86} & \cellcolor{blue!10}74.05 & \cellcolor{blue!10}61.62 & \cellcolor{blue!20}70.73 \\
FedAdam & \cellcolor{red!10}66.92 & \cellcolor{red!10}\textbf{77.27} & \cellcolor{red!10}70.86 & \cellcolor{red!10}71.76 & \cellcolor{red!10}60.61 & \cellcolor{red!20}69.48 & FedAdam & \cellcolor{blue!10}\textbf{73.08} & \cellcolor{blue!10}72.73 & \cellcolor{blue!10}69.71 & \cellcolor{blue!10}\textbf{75.57} & \cellcolor{blue!10}\textbf{63.64} & \cellcolor{blue!20}\textbf{70.95} \\
\midrule
\textbf{Hetero.} & \multicolumn{6}{c|}{\cellcolor{blue!10}\textbf{Device Partial}} & \textbf{Hetero.} & \multicolumn{6}{c}{\cellcolor{blue!10}\textbf{Device Skew}} \\
Local & \cellcolor{blue!10}58.46 & \cellcolor{blue!10}60.00 & \cellcolor{blue!10}45.14 & \cellcolor{blue!10}58.02 & \cellcolor{blue!10}45.45 & \cellcolor{blue!20}53.41 & Local & \cellcolor{blue!10}59.23 & \cellcolor{blue!10}67.27 & \cellcolor{blue!10}46.29 & \cellcolor{blue!10}51.91 & \cellcolor{blue!10}42.42 & \cellcolor{blue!20}53.42 \\ 
FedAvg & \cellcolor{blue!10}59.88 & \cellcolor{blue!10}69.09 & \cellcolor{blue!10}61.14 & \cellcolor{blue!10}62.60 & \cellcolor{blue!10}56.57 & \cellcolor{blue!20}61.86 & FedAvg & \cellcolor{blue!10}52.38 & \cellcolor{blue!10}66.36 & \cellcolor{blue!10}55.43 & \cellcolor{blue!10}60.31 & \cellcolor{blue!10}54.55 & \cellcolor{blue!20}57.81 \\ 
FedProx & \cellcolor{blue!10}64.62 & \cellcolor{blue!10}68.18 & \cellcolor{blue!10}61.71 & \cellcolor{blue!10}64.89 & \cellcolor{blue!10}52.53 & \cellcolor{blue!20}62.39 & FedProx & \cellcolor{blue!10}68.46 & \cellcolor{blue!10}68.18 & \cellcolor{blue!10}57.14 & \cellcolor{blue!10}64.12 & \cellcolor{blue!10}52.53 & \cellcolor{blue!20}62.09 \\ 
FedAvgM & \cellcolor{blue!10}64.62 & \cellcolor{blue!10}66.36 & \cellcolor{blue!10}64.57 & \cellcolor{blue!10}64.12 & \cellcolor{blue!10}52.53 & \cellcolor{blue!20}62.44 & FedAvgM & \cellcolor{blue!10}66.92 & \cellcolor{blue!10}68.18 & \cellcolor{blue!10}57.14 & \cellcolor{blue!10}63.36 & \cellcolor{blue!10}55.56 & \cellcolor{blue!20}62.23 \\ 
SCAFFOLD & \cellcolor{blue!10}66.15 & \cellcolor{blue!10}70.91 & \cellcolor{blue!10}61.71 & \cellcolor{blue!10}67.18 & \cellcolor{blue!10}54.55 & \cellcolor{blue!20}64.10 & SCAFFOLD & \cellcolor{blue!10}68.46 & \cellcolor{blue!10}67.27 & \cellcolor{blue!10}55.43 & \cellcolor{blue!10}62.60 & \cellcolor{blue!10}54.55 & \cellcolor{blue!20}61.66 \\ 
FedAdagrad & \cellcolor{blue!10}67.69 & \cellcolor{blue!10}\textbf{71.82} & \cellcolor{blue!10}70.29 & \cellcolor{blue!10}65.65 & \cellcolor{blue!10}57.58 & \cellcolor{blue!20}66.61 & FedAdagrad & \cellcolor{blue!10}66.15 & \cellcolor{blue!10}71.82 & \cellcolor{blue!10}61.14 & \cellcolor{blue!10}66.41 & \cellcolor{blue!10}60.61 & \cellcolor{blue!20}65.23 \\ 
FedYogi & \cellcolor{blue!10}74.62 & \cellcolor{blue!10}70.91 & \cellcolor{blue!10}73.14 & \cellcolor{blue!10}\textbf{74.81} & \cellcolor{blue!10}60.61 & \cellcolor{blue!20}70.82 & FedYogi & \cellcolor{blue!10}73.08 & \cellcolor{blue!10}\textbf{79.09} & \cellcolor{blue!10}\textbf{70.29} & \cellcolor{blue!10}\textbf{71.76} & \cellcolor{blue!10}\textbf{66.67} & \cellcolor{blue!20}\textbf{72.18} \\ 
FedAdam & \cellcolor{blue!10}\textbf{75.38} & \cellcolor{blue!10}\textbf{71.82} & \cellcolor{blue!10}\textbf{73.71} & \cellcolor{blue!10}72.52 & \cellcolor{blue!10}\textbf{64.65} & \cellcolor{blue!20}\textbf{71.62} & FedAdam & \cellcolor{blue!10}\textbf{74.62} & \cellcolor{blue!10}78.18 & \cellcolor{blue!10}66.86 & \cellcolor{blue!10}\textbf{71.76} & \cellcolor{blue!10}65.66 & \cellcolor{blue!20}71.42 \\ 
\bottomrule
\end{tabular}
}
\caption{Experimental results on \textit{FedGUI-Device}. We evaluate four types of cross-device distributions across five devices: Pixel 7 Pro (P7P), Pixel 8 Pro (P8P), Small Phone (Sm.), Medium Phone (Med.), Pixel Tablet (Tab) and their average. 
Accuracy is generally higher in \colorbox{red!10}{Device IID} settings, while heterogeneous scenarios (\colorbox{blue!10}{Device Non-Uniform}, \colorbox{blue!10}{Device Partial}, and \colorbox{blue!10}{Device Skew}) demonstrate the varying robustness of FL algorithms to cross-device variability.}
\label{tab:fedgui_device}
\end{table*}

\begin{table*}[t]
\centering
\small
\setlength{\tabcolsep}{6.4pt} 
\begin{tabular}{
l | 
c c c c | 
c c c c | 
c c c c
}
\toprule
\multirow{2}{*}{\textbf{\makecell{Client\\ Count}}} &
\multicolumn{4}{c|}{\textbf{Full IID}} &
\multicolumn{4}{c|}{\textbf{Platform Skew}} &
\multicolumn{4}{c}{\textbf{Source Skew}} \\
\cmidrule(lr){2-5} \cmidrule(lr){6-9} \cmidrule(lr){10-13}
 & \textbf{\faMobile} & \textbf{\faGlobeEurope} & \textbf{\faLaptop} & \textbf{\faCalculator}
& \textbf{\faMobile} & \textbf{\faGlobeEurope} & \textbf{\faLaptop} & \textbf{\faCalculator}
& \textbf{\faMobile} & \textbf{\faGlobeEurope} & \textbf{\faLaptop} & \textbf{\faCalculator} \\
\midrule

Client 9
& \cellcolor{red!10}30.86 & \cellcolor{red!10}43.84 & \cellcolor{red!10}40.53 & \cellcolor{red!20}38.39
& \cellcolor{blue!10}16.24 & \cellcolor{blue!10}34.24 & \cellcolor{blue!10}42.05 & \cellcolor{blue!20}30.84
& \cellcolor{blue!10}33.84 & \cellcolor{blue!10}40.58 & \cellcolor{blue!10}26.14 & \cellcolor{blue!20}33.52
\\

Client 18
& \cellcolor{red!10}21.40 & \cellcolor{red!10}32.79 & \cellcolor{red!10}32.79 & \cellcolor{red!20}28.99
& \cellcolor{blue!10}22.46 & \cellcolor{blue!10}31.16 & \cellcolor{blue!10}31.16 & \cellcolor{blue!20}28.26
& \cellcolor{blue!10}11.53 & \cellcolor{blue!10}15.58 & \cellcolor{blue!10}15.58 & \cellcolor{blue!20}14.23
\\

Client 27
& \cellcolor{red!10}22.61 & \cellcolor{red!10}32.07 & \cellcolor{red!10}32.07 & \cellcolor{red!20}28.92
& \cellcolor{blue!10}23.67 & \cellcolor{blue!10}33.51 & \cellcolor{blue!10}33.51 & \cellcolor{blue!20}30.23
& \cellcolor{blue!10}14.11 & \cellcolor{blue!10}20.65 & \cellcolor{blue!10}20.65 & \cellcolor{blue!20}18.47
\\

Client 36
& \cellcolor{red!10}18.97 & \cellcolor{red!10}29.71 & \cellcolor{red!10}29.71 & \cellcolor{red!20}26.13
& \cellcolor{blue!10}14.87 & \cellcolor{blue!10}24.82 & \cellcolor{blue!10}24.82 & \cellcolor{blue!20}21.50
& \cellcolor{blue!10}13.35 & \cellcolor{blue!10}24.82 & \cellcolor{blue!10}24.82 & \cellcolor{blue!20}21.00
\\

\bottomrule
\end{tabular}
\caption{Client scalability evaluation across different client numbers (9, 18, 27, 36). 
Each setting is evaluated under \colorbox{red!10}{Full IID}, \colorbox{blue!10}{Platform Skew} and \colorbox{blue!10}{Source Skew} distributions. 
Results are shown for three benchmarks (AndroidControl: \faMobile, GUIAct-Web: \faGlobeEurope, AgentSynth: \faLaptop) and their average (\faCalculator).
Model performance degrades with a larger number of clients, owing to the escalating difficulty of collaboration.
}
\label{tab:scalability-results}
\end{table*}

\begin{table*}[t]
\centering
\small
\setlength{\tabcolsep}{6.4pt} 
\begin{tabular}{
l | 
c c c c | 
c c c c 
}
\toprule
\multirow{2}{*}{\textbf{\makecell{Sample\\Count}}} &
\multicolumn{4}{c|}{\textbf{Full IID}} &
\multicolumn{4}{c}{\textbf{Source Skew}} \\
\cmidrule(lr){2-5} \cmidrule(lr){6-9}
 & \textbf{\faMobile} & \textbf{\faGlobeEurope} & \textbf{\faLaptop} & \textbf{\faCalculator}
& \textbf{\faMobile} & \textbf{\faGlobeEurope} & \textbf{\faLaptop} & \textbf{\faCalculator} \\
\midrule

Sample 3
& \cellcolor{red!10}22.61 & \cellcolor{red!10}28.62 & \cellcolor{red!10}32.07 & \cellcolor{red!20}27.77
& \cellcolor{blue!10}18.61 & \cellcolor{blue!10}26.62 & \cellcolor{blue!10}26.32 & \cellcolor{blue!20}23.85
\\

Sample 6
& \cellcolor{red!10}22.91 & \cellcolor{red!10}28.72 & \cellcolor{red!10}32.68 & \cellcolor{red!20}28.10
& \cellcolor{blue!10}19.52 & \cellcolor{blue!10}27.27 & \cellcolor{blue!10}28.65 & \cellcolor{blue!20}25.15
\\

Sample 9
& \cellcolor{red!10}23.51 & \cellcolor{red!10}28.98 & \cellcolor{red!10}32.32 & \cellcolor{red!20}28.27
& \cellcolor{blue!10}21.51 & \cellcolor{blue!10}28.98 & \cellcolor{blue!10}32.32 & \cellcolor{blue!20}27.60
\\

Sample 12
& \cellcolor{red!10}23.32 & \cellcolor{red!10}29.34 & \cellcolor{red!10}33.38 & \cellcolor{red!20}28.68
& \cellcolor{blue!10}22.22 & \cellcolor{blue!10}28.34 & \cellcolor{blue!10}33.68 & \cellcolor{blue!20}28.08
\\

\bottomrule
\end{tabular}
\caption{Client sampling scalability evaluation. Success rate (\%) across different number of sampled clients (3, 6, 9, 12) under \colorbox{red!10}{Full IID} and \colorbox{blue!10}{Source Skew} distributions. Results cover Mobile, Web, and OS platforms with their average performance.}
\label{tab:fedgui_client_sample}  
\end{table*}

\begin{table*}[t]
\centering
\small
\setlength{\tabcolsep}{6.4pt} 
\begin{tabular}{
l | 
c c c c | 
c c c c | 
c c c c
}
\toprule
\textbf{Algorithm} &
\multicolumn{4}{c|}{\textbf{Full IID}} &
\multicolumn{4}{c|}{\textbf{Platform Skew}} &
\multicolumn{4}{c}{\textbf{Source Skew}} \\
\cmidrule(lr){2-5} \cmidrule(lr){6-9} \cmidrule(lr){10-13}
SR & \textbf{\faMobile} & \textbf{\faGlobeEurope} & \textbf{\faLaptop} & \textbf{\faCalculator}
& \textbf{\faMobile} & \textbf{\faGlobeEurope} & \textbf{\faLaptop} & \textbf{\faCalculator}
& \textbf{\faMobile} & \textbf{\faGlobeEurope} & \textbf{\faLaptop} & \textbf{\faCalculator} \\
\midrule
Central 
& \cellcolor{gray!10}44.31 & \cellcolor{gray!10}48.73 & \cellcolor{gray!10}54.91 & \cellcolor{gray!20}49.32
& \cellcolor{gray!10}- & \cellcolor{gray!10}- & \cellcolor{gray!10}- & \cellcolor{gray!10}-
& \cellcolor{gray!10}- & \cellcolor{gray!10}- & \cellcolor{gray!10}- & \cellcolor{gray!10}- \\

\midrule
Local
& \cellcolor{red!10}31.11 & \cellcolor{red!10}67.22 & \cellcolor{red!10}33.20 & \cellcolor{red!20}33.33
& \cellcolor{blue!10}37.10 & \cellcolor{blue!10}45.15 & \cellcolor{blue!10}5.97 & \cellcolor{blue!20}23.96
& \cellcolor{blue!10}30.81 & \cellcolor{blue!10}26.46 & \cellcolor{blue!10}5.29 & \cellcolor{blue!20}19.94
\\
FedAvg
& \cellcolor{red!10}30.80 & \cellcolor{red!10}69.85 & \cellcolor{red!10}40.53 & \cellcolor{red!20}38.39
& \cellcolor{blue!10}16.24 & \cellcolor{blue!10}69.29 & \cellcolor{blue!10}42.05 & \cellcolor{blue!20}30.84
& \cellcolor{blue!10}33.84 & \cellcolor{blue!10}64.87 & \cellcolor{blue!10}26.14 & \cellcolor{blue!20}33.52
\\
FedProx
& \cellcolor{red!10}31.26 & \cellcolor{red!10}70.12 & \cellcolor{red!10}41.63 & \cellcolor{red!20}38.97
& \cellcolor{blue!10}35.66 & \cellcolor{blue!10}64.73 & \cellcolor{blue!10}26.69 & \cellcolor{blue!20}34.37
& \cellcolor{blue!10}16.24 & \cellcolor{blue!10}68.74 & \cellcolor{blue!10}41.22 & \cellcolor{blue!20}29.48
\\
SCAFFOLD
& \cellcolor{red!10}30.50 & \cellcolor{red!10}69.85 & \cellcolor{red!10}41.49 & \cellcolor{red!20}38.43
& \cellcolor{blue!10}34.90 & \cellcolor{blue!10}64.87 & \cellcolor{blue!10}26.28 & \cellcolor{blue!20}33.86
& \cellcolor{blue!10}16.39 & \cellcolor{blue!10}69.85 & \cellcolor{blue!10}41.36 & \cellcolor{blue!20}30.06
\\
FedYogi
& \cellcolor{red!10}37.78 & \cellcolor{red!10}73.31 & \cellcolor{red!10}58.64 & \cellcolor{red!20}48.14
& \cellcolor{blue!10}43.40 & \cellcolor{blue!10}69.85 & \cellcolor{blue!10}39.28 & \cellcolor{blue!20}43.62
& \cellcolor{blue!10}21.24 & \cellcolor{blue!10}71.09 & \cellcolor{blue!10}54.63 & \cellcolor{blue!20}39.00
\\
FedAdam
& \cellcolor{red!10}39.45 & \cellcolor{red!10}72.61 & \cellcolor{red!10}56.57 & \cellcolor{red!20}47.95
& \cellcolor{blue!10}42.49 & \cellcolor{blue!10}69.85 & \cellcolor{blue!10}39.83 & \cellcolor{blue!20}43.20
& \cellcolor{blue!10}21.55 & \cellcolor{blue!10}70.82 & \cellcolor{blue!10}55.19 & \cellcolor{blue!20}40.19
\\
FedAvgM
& \cellcolor{red!10}31.26 & \cellcolor{red!10}69.43 & \cellcolor{red!10}40.66 & \cellcolor{red!20}38.41
& \cellcolor{blue!10}35.81 & \cellcolor{blue!10}63.76 & \cellcolor{blue!10}25.73 & \cellcolor{blue!20}34.46
& \cellcolor{blue!10}16.54 & \cellcolor{blue!10}69.02 & \cellcolor{blue!10}40.25 & \cellcolor{blue!20}29.32
\\
FedAdagrad
& \cellcolor{red!10}32.47 & \cellcolor{red!10}68.46 & \cellcolor{red!10}48.82 & \cellcolor{red!20}42.01
& \cellcolor{blue!10}39.15 & \cellcolor{blue!10}67.50 & \cellcolor{blue!10}33.33 & \cellcolor{blue!20}39.86
& \cellcolor{blue!10}21.24 & \cellcolor{blue!10}69.85 & \cellcolor{blue!10}54.36 & \cellcolor{blue!20}36.49
\\
\bottomrule
\end{tabular}
\caption{
Performance comparison of FL algorithms on \textit{FedGUI-Full}.
The table reports Success Rate (SR, \%), as it is the most informative metric among the three evaluated. 
These results, in conjunction with those in Table \ref{tab:fedgui_platform}, reaffirm the significant performance variation of FL algorithms under different data distributions.
}
\label{tab:fedgui_full_alg}
\end{table*}

\begin{table*}[t]
\centering
\small
\setlength{\tabcolsep}{6pt} 

\begin{minipage}[t]{0.49\textwidth}
\centering
\begin{tabular}{l | p{1cm}p{1cm}p{1cm}p{1cm}}
\toprule
\multirow{2}{*}{\textbf{Algorithm}} & \multicolumn{4}{c}{\textbf{\faMobile\; Mobile Evaluation}}\\
\cmidrule(lr){2-5}
& AC & AitW &  GO &  Avg. \\
\midrule
\rowcolor{gray!10} Central & 55.08 & 58.02 & 69.78 & \cellcolor{gray!20}60.96 \\
\midrule
\rowcolor{red!10} FedAvg     & 42.19 & 52.19 & 55.29 & \cellcolor{red!20}49.89 \\
\rowcolor{red!10} FedProx    & 42.49 & 51.56 & 54.53 & \cellcolor{red!20}49.53 \\
\rowcolor{red!10} FedYogi    & 48.41 & 53.54 & 64.35 & \cellcolor{red!20}55.43 \\
\rowcolor{red!10} FedAdam    & 48.25 & 52.19 & 61.54 & \cellcolor{red!20}53.99 \\
\rowcolor{red!10} SCAFFOLD   & 41.88 & 52.19 & 55.56 & \cellcolor{red!20}49.88 \\
\rowcolor{red!10} FedAvgM    & 43.25 & 52.08 & 55.08 & \cellcolor{red!20}50.14 \\
\rowcolor{red!10} FedAdagrad & 46.89 & 47.92 & 59.13 & \cellcolor{red!20}51.31 \\
\midrule
\rowcolor{blue!10} FedAvg     & 23.98 & 47.60 & 60.10 & \cellcolor{blue!20}43.89 \\
\rowcolor{blue!10} FedProx    & 24.43 & 46.98 & 60.30 & \cellcolor{blue!20}43.90 \\
\rowcolor{blue!10} FedYogi    & 29.59 & 50.83 & 65.32 & \cellcolor{blue!20}48.58 \\
\rowcolor{blue!10} FedAdam    & 30.35 & 49.48 & 65.52 & \cellcolor{blue!20}48.45 \\
\rowcolor{blue!10} SCAFFOLD   & 24.43 & 61.30 & 59.89 & \cellcolor{blue!20}48.54 \\
\rowcolor{blue!10} FedAvgM    & 24.58 & 47.08 & 59.75 & \cellcolor{blue!20}43.80 \\
\rowcolor{blue!10} FedAdagrad & 27.92 & 50.17 & 70.77 & \cellcolor{blue!20}49.62 \\
\bottomrule
\end{tabular}
\end{minipage}
\hfill
\begin{minipage}[t]{0.49\textwidth}
\centering
\begin{tabular}{l | p{1cm}p{1cm}p{1cm}p{1cm}}
\toprule
 \multirow{2}{*}{\textbf{Algorithm}} & \multicolumn{4}{c}{\faGlobeEurope\; \textbf{Web Evaluation}} \\
 \cmidrule(lr){2-5}
 & M2W & GA-W & OA-W & Avg. \\
\midrule
\rowcolor{gray!10} Central & 17.73 & 52.72 & 67.74 & \cellcolor{gray!20}46.06 \\
\midrule
\rowcolor{red!10} FedAvg     & 5.18  & 32.79 & 53.77 & \cellcolor{red!20}30.58 \\
\rowcolor{red!10} FedProx    & 4.58  & 33.33 & 54.53 & \cellcolor{red!20}30.81 \\
\rowcolor{red!10} FedYogi    & 3.59  & 39.13 & 54.91 & \cellcolor{red!20}32.54 \\
\rowcolor{red!10} FedAdam    & 3.98  & 39.67 & 62.08 & \cellcolor{red!20}35.24 \\
\rowcolor{red!10} SCAFFOLD   & 4.38  & 33.88 & 54.15 & \cellcolor{red!20}30.80 \\
\rowcolor{red!10} FedAvgM    & 5.18  & 34.06 & 54.15 & \cellcolor{red!20}31.13 \\
\rowcolor{red!10} FedAdagrad & 5.38  & 35.33 & 60.19 & \cellcolor{red!20}33.63 \\
\midrule
\rowcolor{blue!10} FedAvg     & 6.77  & 13.04 & 29.81 & \cellcolor{blue!20}16.54 \\
\rowcolor{blue!10} FedProx    & 6.18  & 12.50 & 28.68 & \cellcolor{blue!20}15.79 \\
\rowcolor{blue!10} FedYogi    & 5.58  & 14.31 & 30.75 & \cellcolor{blue!20}16.88 \\
\rowcolor{blue!10} FedAdam    & 5.38  & 11.59 & 40.38 & \cellcolor{blue!20}19.12 \\
\rowcolor{blue!10} SCAFFOLD   & 5.38  & 12.68 & 30.57 & \cellcolor{blue!20}16.21 \\
\rowcolor{blue!10} FedAvgM    & 5.58  & 12.86 & 29.06 & \cellcolor{blue!20}15.83 \\
\rowcolor{blue!10} FedAdagrad & 7.57  & 17.21 & 31.51 & \cellcolor{blue!20}18.76 \\
\bottomrule
\end{tabular}
\end{minipage}

\caption{Supplementary experiments results on \textit{FedGUI-Mobile} (left) and \textit{FedGUI-Web} (right). 
Performance of each algorithm, measured as Success Rate (SR, \%), is evaluated under \colorbox{red!10}{IID} and \colorbox{blue!10}{Skewed} data distributions and compared against the \colorbox{gray!10}{Centralized} baseline. The table presents results on individual datasets within the mobile and web domains, along with their respective averages.
W
}
\label{tab:supplement_mobile_web}
\end{table*}

\section{Supplementary Experiments \& Results }
\label{sec:additional_experiment}

\subsection{Client Scalability under Heterogeneity.}
\paragraph{Setups.}
We conduct ablation studies on \textit{FedGUI-Full} to analyze the effect of scaling client number under different heterogeneity settings. By varying the number of participating clients (9, 18, 27, and 36) and applying multiple heterogeneous partitions (Full IID, Platform Skew, and Source Skew), we simulate federated training scenarios with different degrees of fragmentation and client participation.
\paragraph{Results.}
From the ablation results in Table~\ref{tab:scalability-results}, we draw the following conclusions: (1) As the client number scales from 9 to 36, performance consistently declines across most settings, indicating that finer partitioning leads to data sparsity and hampers effective GUI feature learning. (2) Regardless of client scales, Skew settings consistently underperform the IID setting, confirming that data heterogeneity remains the primary obstacle that cannot be mitigated by changing participation scale. (3) Increasing the number of participating clients does not compensate for performance loss under heterogeneity, suggesting that scaling alone is insufficient to address the challenges of non-IID data in federated GUI learning.
 
\subsection{Supplementary Comparison of FL Algorithms}

\paragraph{Setups.}
We further extend our evaluation to \textit{FedGUI-Mobile}, and \textit{FedGUI-Web} to make a more comprehensive view of result concluded in Section \ref{sec:experiment_fedgui_platform}.
For the mobile domain, we use AC, AitW, and GO each 2000 episodes.
For the web domain, we evaluate on M2W (500 episodes), GA-W (500 episodes), and OA-W (1,000 episodes).
In both settings, we simulate a federated environment with 15 clients and a participation rate of 3 clients per round, enabling a consistent comparison of federated algorithms under diverse UI distributions.

\paragraph{Results.}
Key observations from Tables~\ref{tab:supplement_mobile_web} include:
(1) Decentralized training is more feasible in the mobile domain, where IID performance reaches within 10--15\% of the centralized baseline, whereas the web domain exhibits a larger gap, particularly on M2W, due to higher variability in layouts, dynamic content, and interaction flows;
(2) data heterogeneity significantly degrades performance in both domains, with the web setting suffering nearly 50\% average degradation relative to IID, indicating that source skew poses a more severe challenge for cross-website GUI tasks than for mobile ones;
and (3) adaptive methods such as FedAdam, FedAdagrad, and FedYogi consistently outperform FedAvg across both domains, demonstrating that per-parameter adaptation and gradient scaling are crucial for handling high-variance GUI interaction sequences.

\begin{figure*}[t]
    \centering
    \begin{subfigure}[h]{0.24\linewidth}
        \centering
        \includegraphics[width=\linewidth]{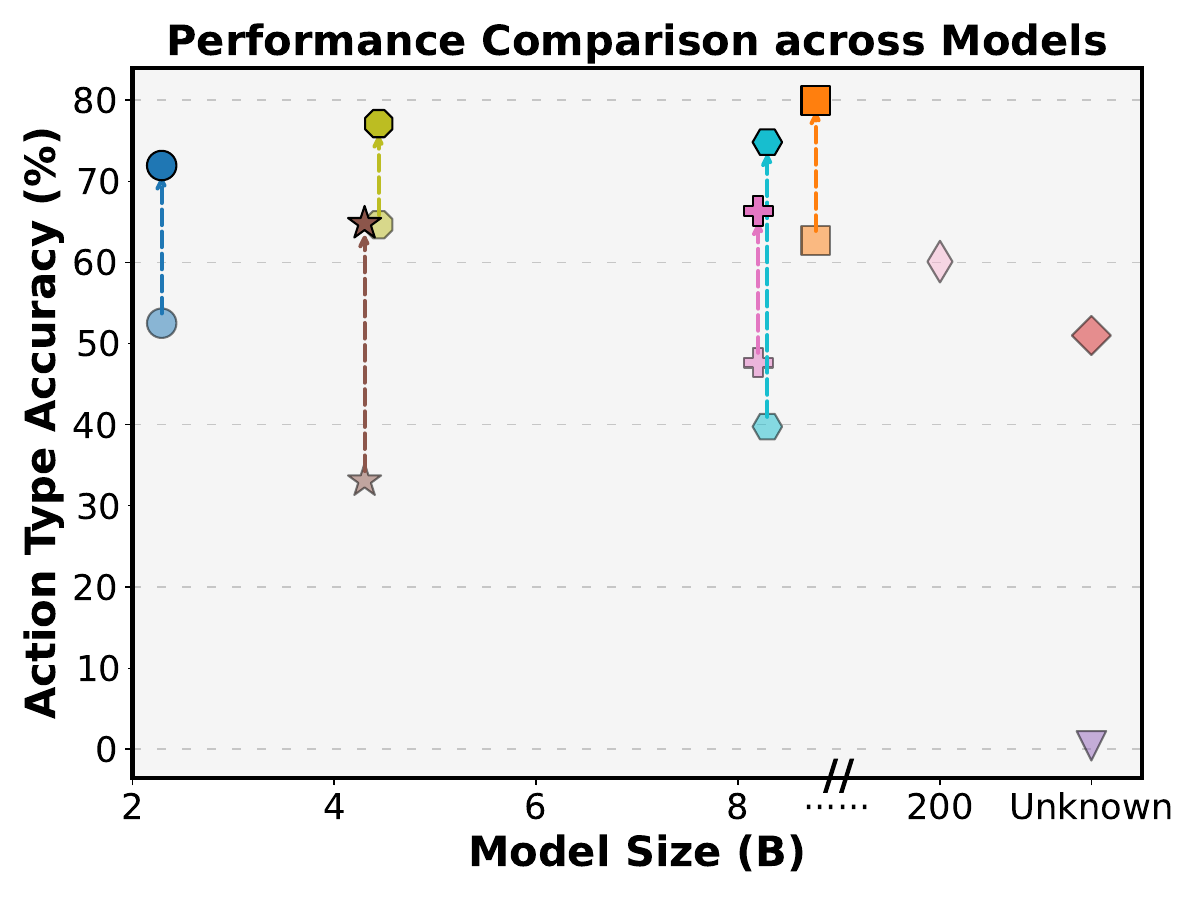}
        \caption{Mobile Domain}
    \end{subfigure}
    \hfill
    \begin{subfigure}[h]{0.24\linewidth}
        \centering
        \includegraphics[width=\linewidth]{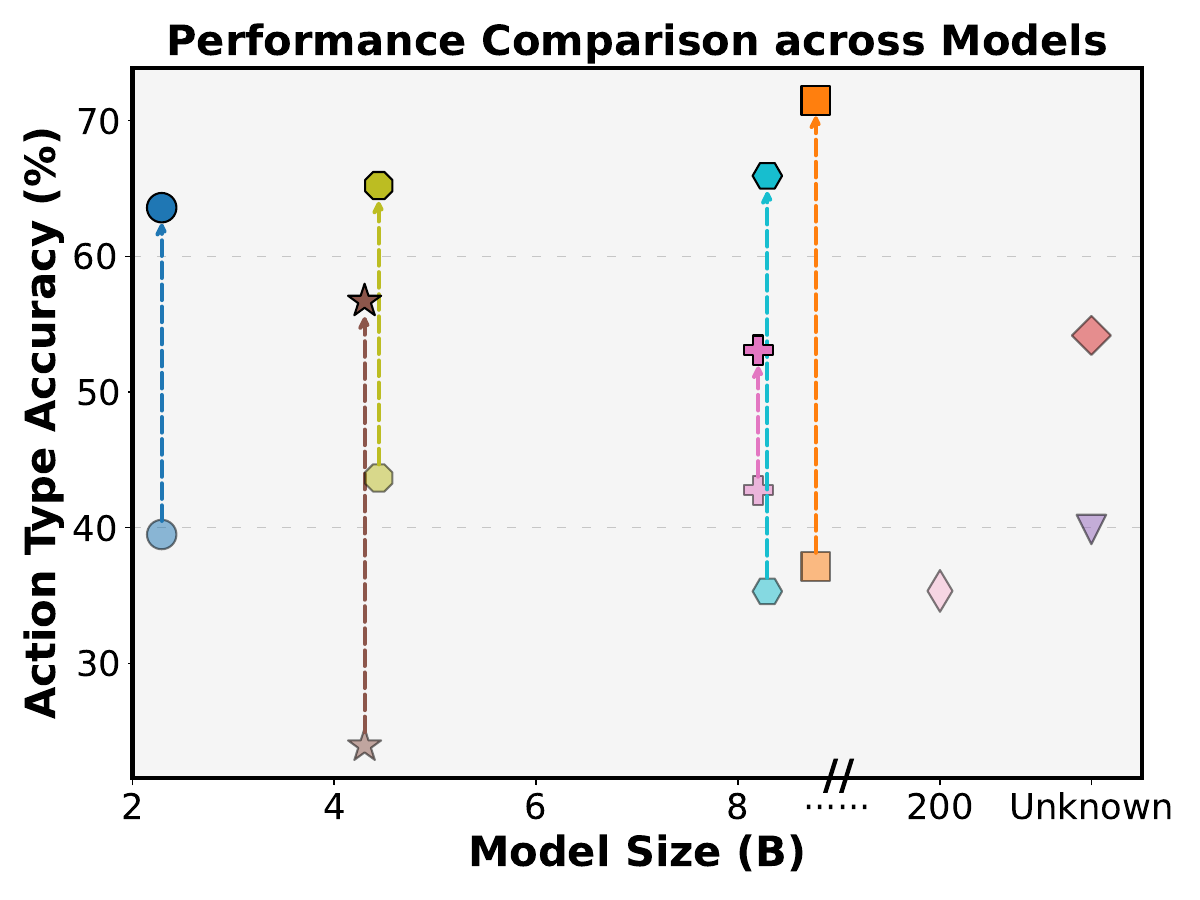}
        \caption{Web Domain}  
    \end{subfigure}
    \hfill 
    \begin{subfigure}[h]{0.24\linewidth}
        \centering
        \includegraphics[width=\linewidth]{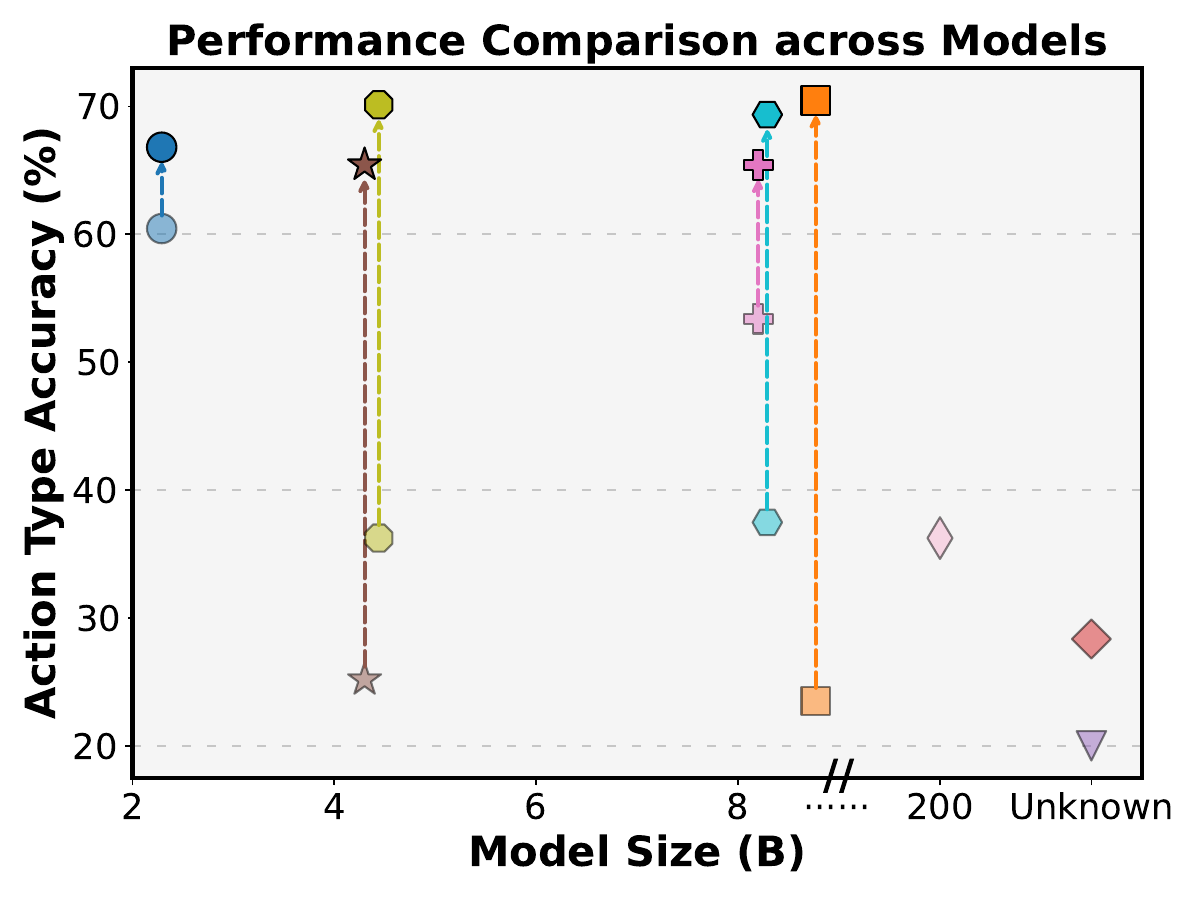}
        \caption{Desktop Domain}
    \end{subfigure}
    \hfill
    \begin{subfigure}[h]{0.24\linewidth}
        \centering
        \includegraphics[width=\linewidth]{figs/scatter_legend.pdf}
        \caption{Model Legend}
   
    \end{subfigure}


    \begin{subfigure}[h]{0.24\linewidth}
        \centering
        \includegraphics[width=\linewidth]{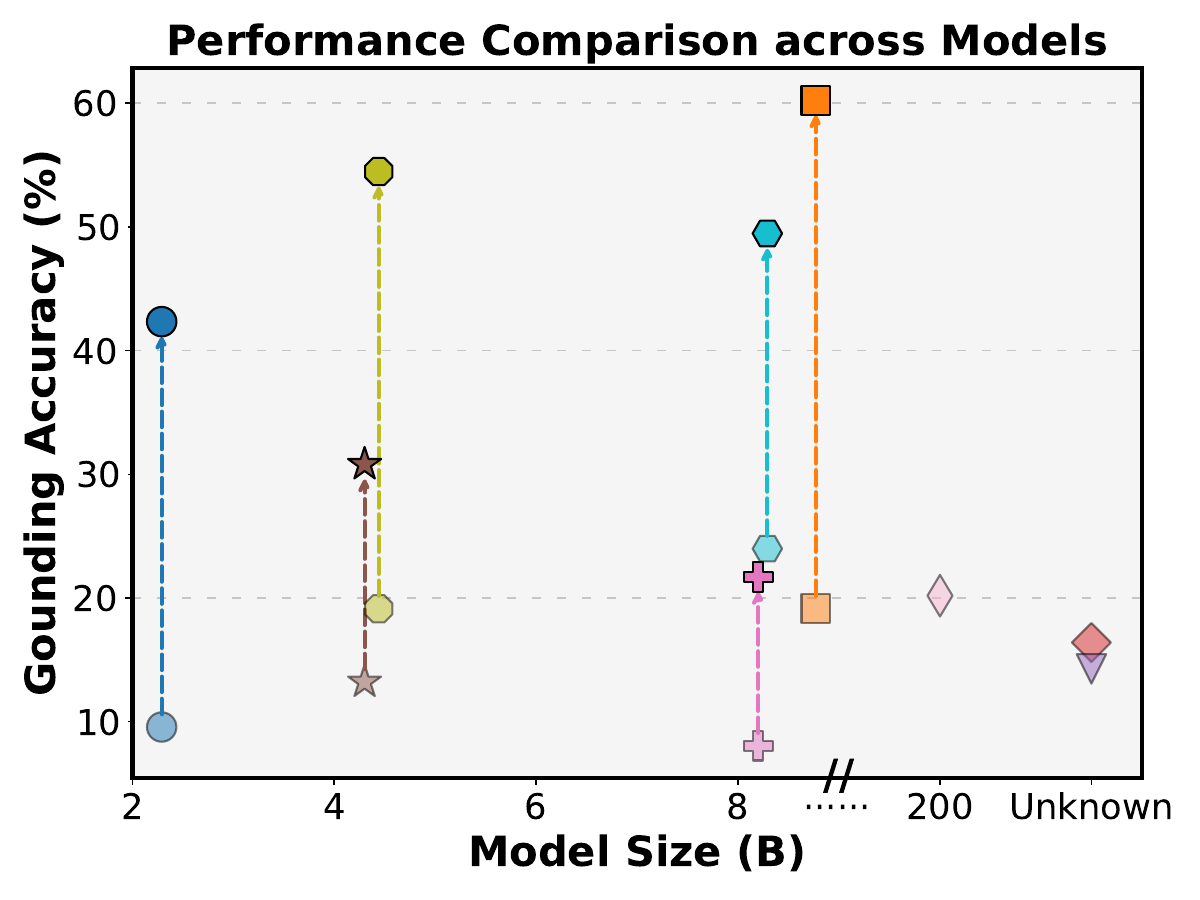}
        \caption{Mobile Domain}
    \end{subfigure}
    \hfill
    \begin{subfigure}[h]{0.24\linewidth}
        \centering
        \includegraphics[width=\linewidth]{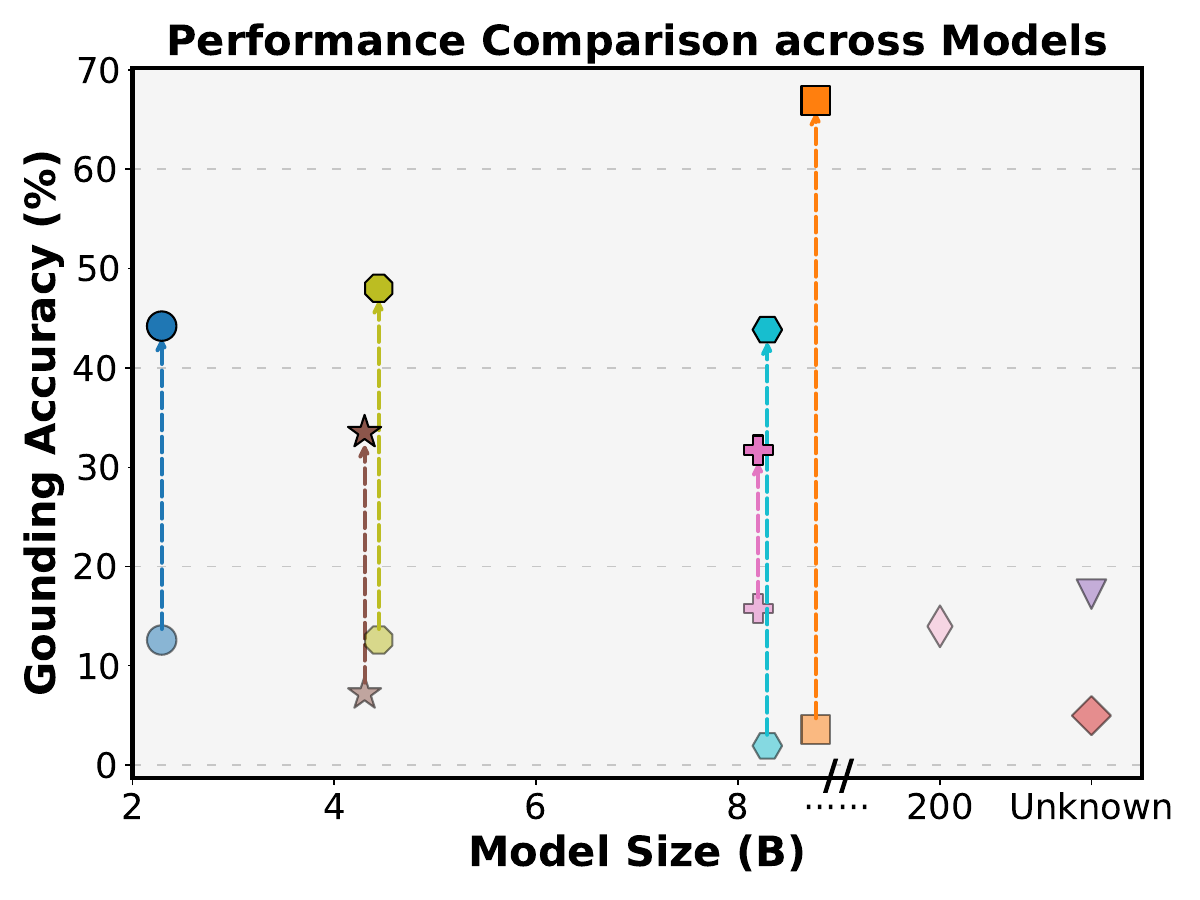}
        \caption{Web Domain}  
    \end{subfigure}
    \hfill 
    \begin{subfigure}[h]{0.24\linewidth}
        \centering
        \includegraphics[width=\linewidth]{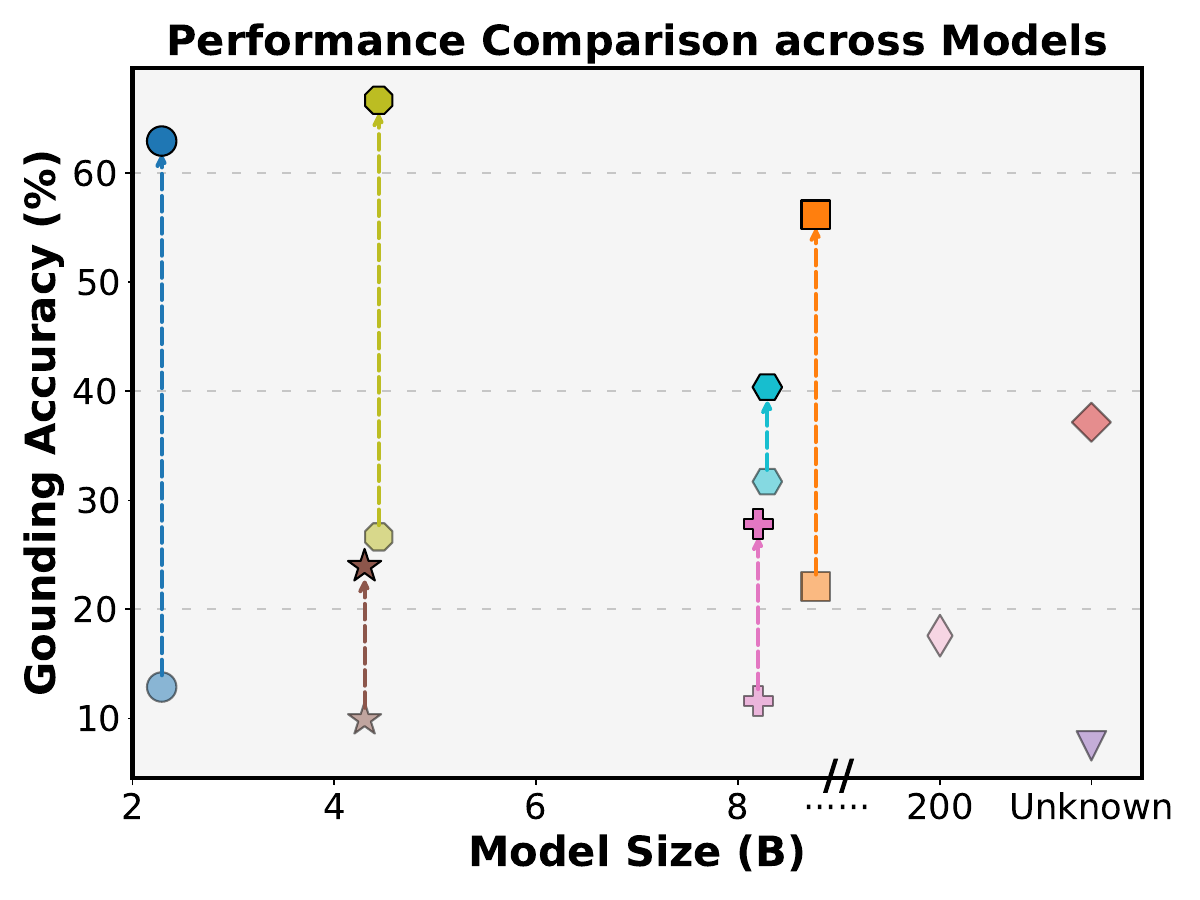}
        \caption{Desktop Domain}  
    \end{subfigure}
    \hfill
    \begin{minipage}[h]{0.24\linewidth}
        \centering

        \captionof{figure}{Model action type accuracy (top row) and grounding accuracy (bottom row) comparison across mobile, web, and Desktop domains. Base and FedAvg performances are shown with arrows indicating gains.}
        \label{fig:fedgui_model_supplementary}
        
    \end{minipage}

\end{figure*}

\section{Dataset Details}
\label{sec:dataset_details}

\subsection{Data Composition \& Example}
To illustrate the structure of our dataset and the composition of a data episode, we present representative examples from each domain in this section.

\begin{figure*}[t]
    \centering
    \begin{subfigure}{0.32\linewidth}
        \centering
        \includegraphics[width=\linewidth]{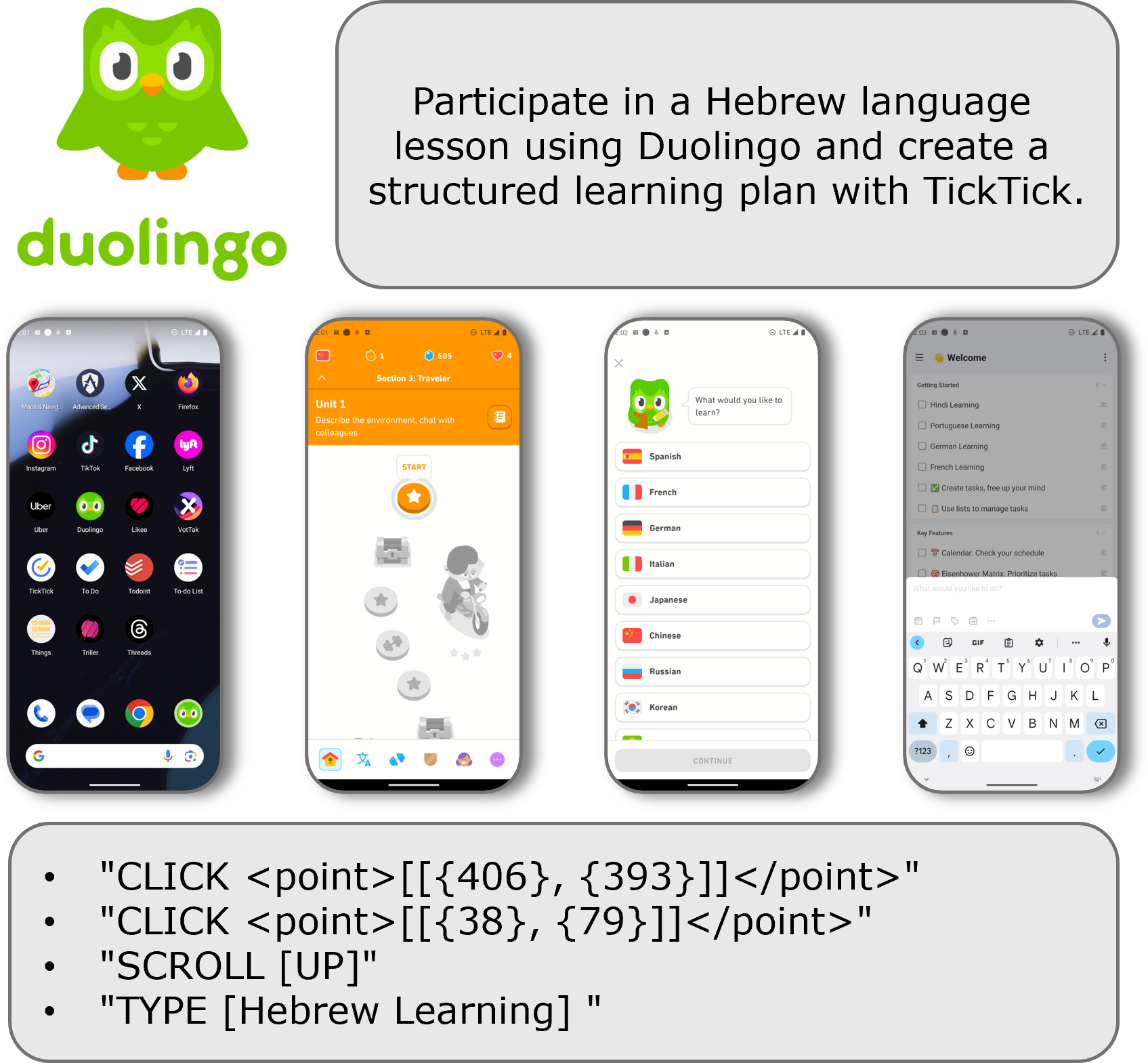}
        \caption{Mobile Data Example}
    \end{subfigure}
    \hfill
    \begin{subfigure}{0.32\linewidth}
        \centering
        \includegraphics[width=\linewidth]{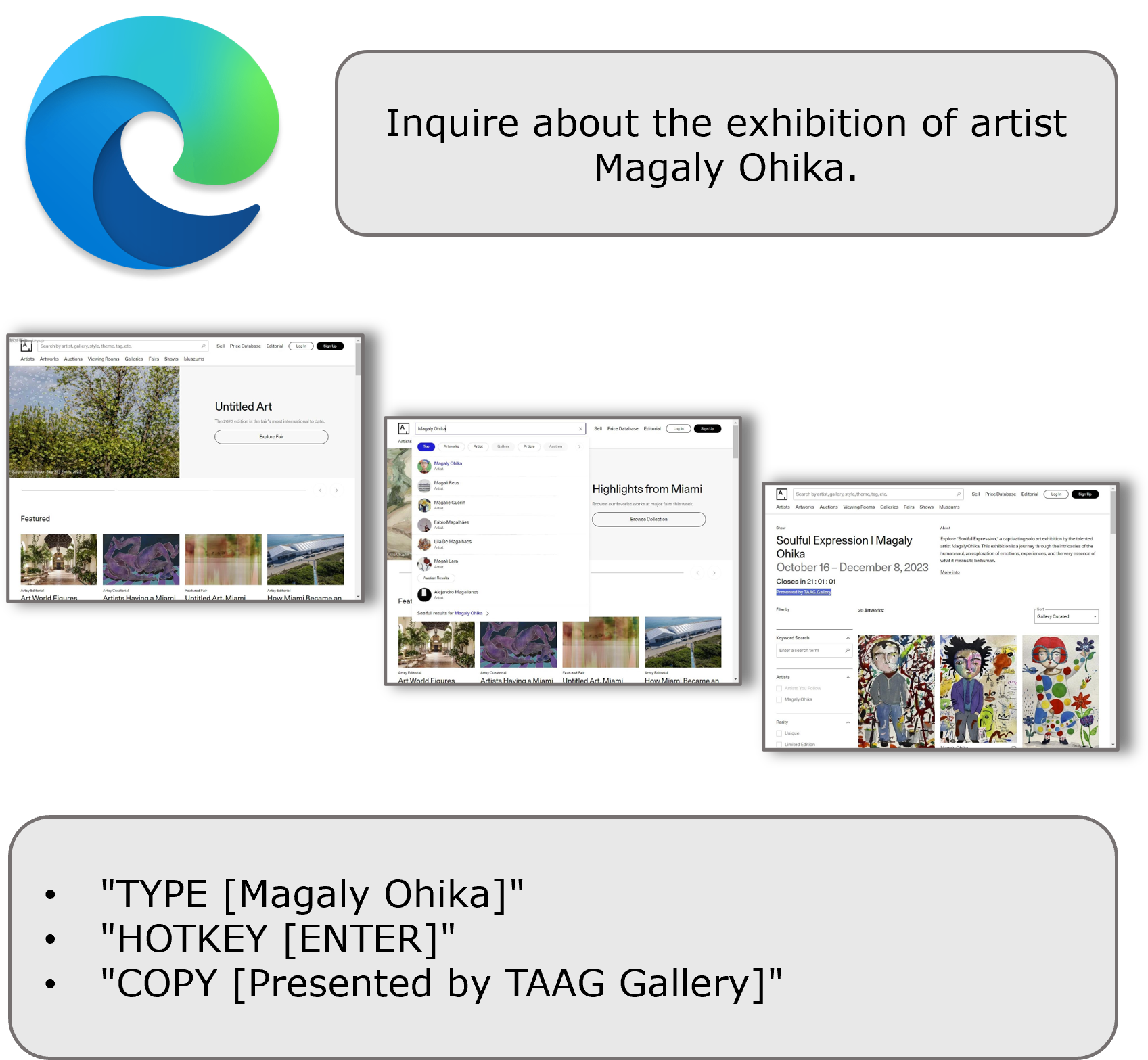}
        \caption{Web Data Example}
    \end{subfigure}
    \hfill
    \begin{subfigure}{0.32\linewidth}
        \centering
        \includegraphics[width=\linewidth]{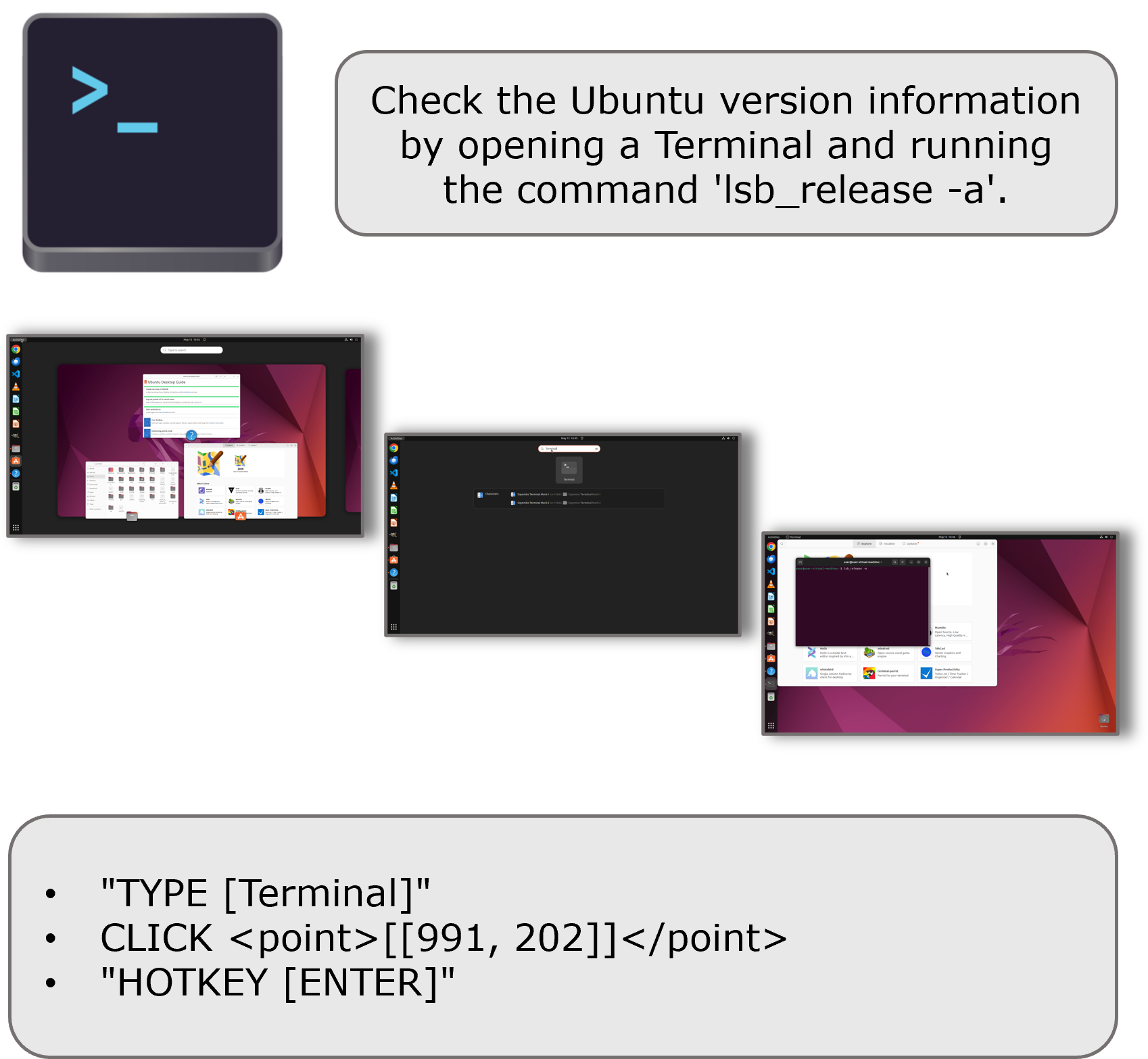}
        \caption{Desktop Data Example}
    \end{subfigure}

    \caption{Example data episodes for training GUI agents across mobile, web, and desktop platforms in FedGUI.}
    \label{fig:fedgui_data_example}
\end{figure*}

As shown in Figure~\ref{fig:fedgui_data_example}, each episode consists of three components, following typical GUI agent research \cite{chaiAMEXAndroidMultiannotation2024, mobilegpt, liu2025learnact,liu2026memguibenchbenchmarkingmemorymobile,zhao2025masbenchunifiedbenchmarkshortcutaugmented}:
(1) an instruction, i.e., a natural-language description of the task to be completed; (2) a sequence of screenshots captured from the beginning to the end of the task; and
(3) a corresponding sequence of actions, aligned one-to-one with the screenshots, specifying the user interactions that lead to the next state. Notably, an episode may involve cross-application or cross-website interactions, reflecting realistic multi-step workflows in GUI environments. All actions are defined within a unified action space of 17 action types, including both basic interactions and domain-specific custom actions, as detailed in Table~\ref{tab:action_space}. 

\subsection{Data Collection \&  Processing Details}
\label{data_process}

\begin{table*}[t]
\centering
\small

\begin{tabular}{c l l p{8.5cm}}
\toprule
\textbf{Index} & \textbf{Action Type} & \textbf{Attribute} & \textbf{Description} \\
\midrule
\multicolumn{4}{l}{\cellcolor{blue!10}\textbf{Basic Actions}} \\
1  & CLICK          & (x, y)           & Click at specific screen coordinates. \\
2  & TYPE           & input\_text      & Type the given input text into an active text field. \\
3  & SCROLL         & direction        & Scroll in a specific direction (\texttt{up}, \texttt{down}, \texttt{left}, or \texttt{right}). \\
4  & COMPLETE       & --               & Mark the task as completed. \\
5  & IMPOSSIBLE     & --               & Indicate that the current task cannot be completed. \\
6  & WAIT           & --               & Wait for a period (e.g., loading or animation). \\
\midrule
\multicolumn{4}{l}{\cellcolor{green!10}\textbf{Custom Actions for Mobile}} \\
7  & LONG\_PRESS    & (x, y)           & Long press at the given coordinates. \\
8  & OPEN\_APP      & app\_name        & Open a specific application by name. \\
9  & NAVIGATE\_BACK & --               & Go back to the previous page or screen. \\
10 & NAVIGATE\_HOME & --               & Return to the home screen. \\
11 & PRESS\_RECENT  & --               & Open the recent apps menu. \\
12 & PRESS\_ENTER   & --               & Simulate pressing the Enter key. \\
\midrule
\multicolumn{4}{l}{\cellcolor{red!10}\textbf{Custom Actions for Web \& Desktop}} \\
13 & DOUBLE\_CLICK  & (x, y)           & Perform a double click at the specified coordinates. \\
14 & RIGHT\_CLICK   & (x, y)           & Right click at the given position. \\
15 & MOVETO         & (x, y)           & Move the cursor to a target coordinate without clicking. \\
16 & HOTKEY         & keys             & Execute a keyboard shortcut (e.g., \texttt{ENTER}, \texttt{ESC}, \texttt{DOWN}, \texttt{SPACE}, \texttt{RIGHT}, \texttt{LEFT}, \texttt{TAB}). \\
17 & COPY           & text             & Copy the given text to clipboard. \\
\bottomrule
\end{tabular}
\caption{Action space used in FedGUI.
The table summarizes the unified set of 17 executable actions, consisting of platform-agnostic \colorbox{blue!10}{basic actions} and platform-specific custom actions for \colorbox{green!10}{mobile} and \colorbox{red!10}{web \& desktop} environments.Each action is defined with its required attributes and semantic description.}
\label{tab:action_space}
\end{table*}

To ensure the quality and consistency of the federated GUI datasets across heterogeneous platforms, we implement a rigorous data processing pipeline as follows:
We first perform systematic data cleaning to remove low-quality episodes, including samples with missing visual observations,
ambiguous or undefined action parameters, and redundant interaction steps \cite{wang2025mcpflowfacilitatingllmagents}.
Semantically unnecessary actions are filtered to ensure concise and meaningful interaction trajectories.

To improve cross-platform learnability, we normalize raw GUI actions into a compact and standardized representation. Platform-specific redundancies are removed by extracting core interaction primitives, e.g., collapsing compound operations such as $\mathtt{select\_from\_to}$ followed by $\mathtt{copy}$ into a single $\mathtt{COPY}$ action.Actions with equivalent semantics are further unified into a common format (e.g., $\mathtt{press\_enter}$ and $\mathtt{press\_button}$ are mapped to $\mathtt{HOTKEY}$), ensuring semantic consistency across mobile, web, and desktop environments.

We adopt a unified and extensible action space to support GUI interaction across mobile, web, and desktop platforms. As summarized in Table~\ref{tab:action_space}, the action space consists of 17 discrete action types, each optionally associated with structured attributes such as screen coordinates, text inputs, or key sequences.

The action space is organized into three categories: (1) {Basic Actions}, which capture platform-agnostic operations, including clicking, typing, scrolling, waiting, and task termination; (2) {Custom Actions for Mobile}, which extend the basic set with smartphone-specific interactions such as long-press gestures, application launching, and system-level navigation; and (3) {Custom Actions for Web \& Desktop}, which incorporate mouse- and keyboard-centric operations including double-clicking, right-clicking, cursor movement, hotkeys, and clipboard interactions. This unified design enables consistent modeling and evaluation of heterogeneous GUI tasks while preserving platform-specific interaction semantics, and all predicted actions are constrained to this predefined action space during both training and evaluation.

\begin{figure*}[t]
    \centering

    \begin{subfigure}[t]{0.32\textwidth}
        \centering
        \includegraphics[width=\textwidth]{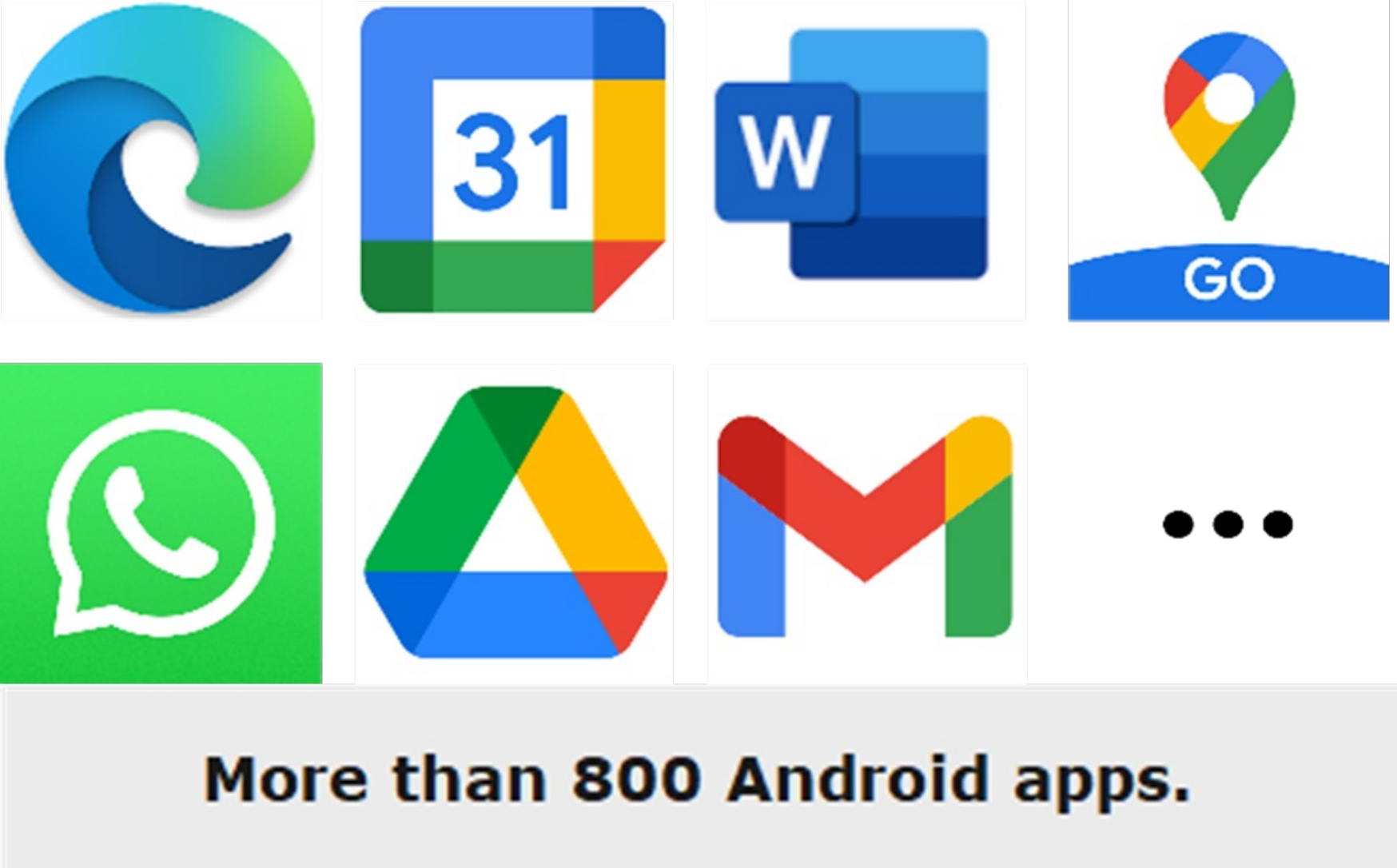}
        \caption{AndroidControl (AC) contains over 800 Android applications.}
    \end{subfigure}
    \hfill
    \begin{subfigure}[t]{0.32\textwidth}
        \centering
        \includegraphics[width=\textwidth]{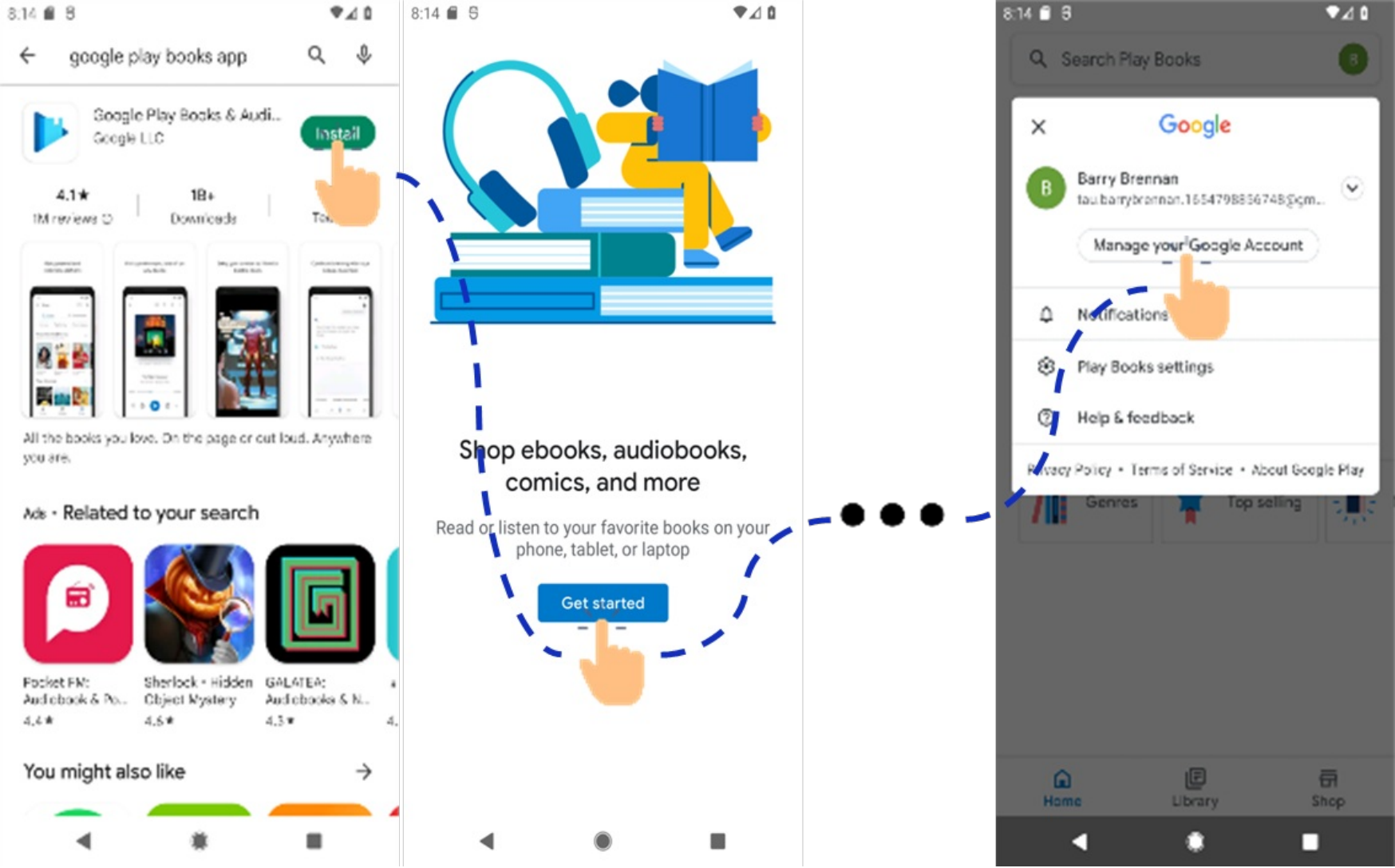}
        \caption{
        Android-in-the-Wild (AitW) is characterized by its procedural, multi-step tasks.}
    \end{subfigure}
    \hfill
    \begin{subfigure}[t]{0.32\textwidth}
        \centering
        \includegraphics[width=\textwidth]{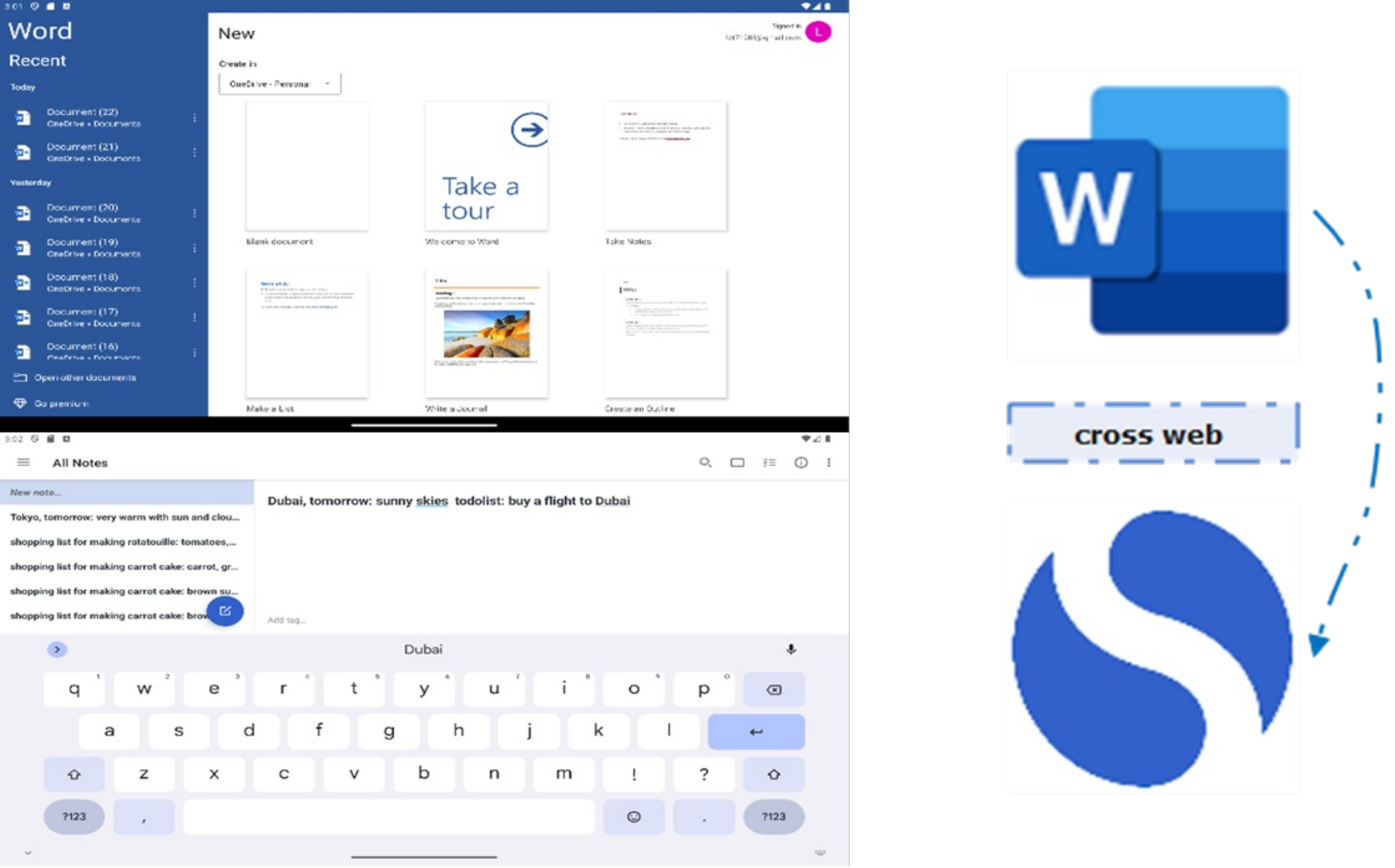}
        \caption{GUI Odyssey (GO) is a benchmark for cross-app GUI navigation.}
    \end{subfigure}
    
    \par\bigskip 

    \begin{subfigure}[t]{0.32\textwidth}
        \centering
        \includegraphics[width=\textwidth]{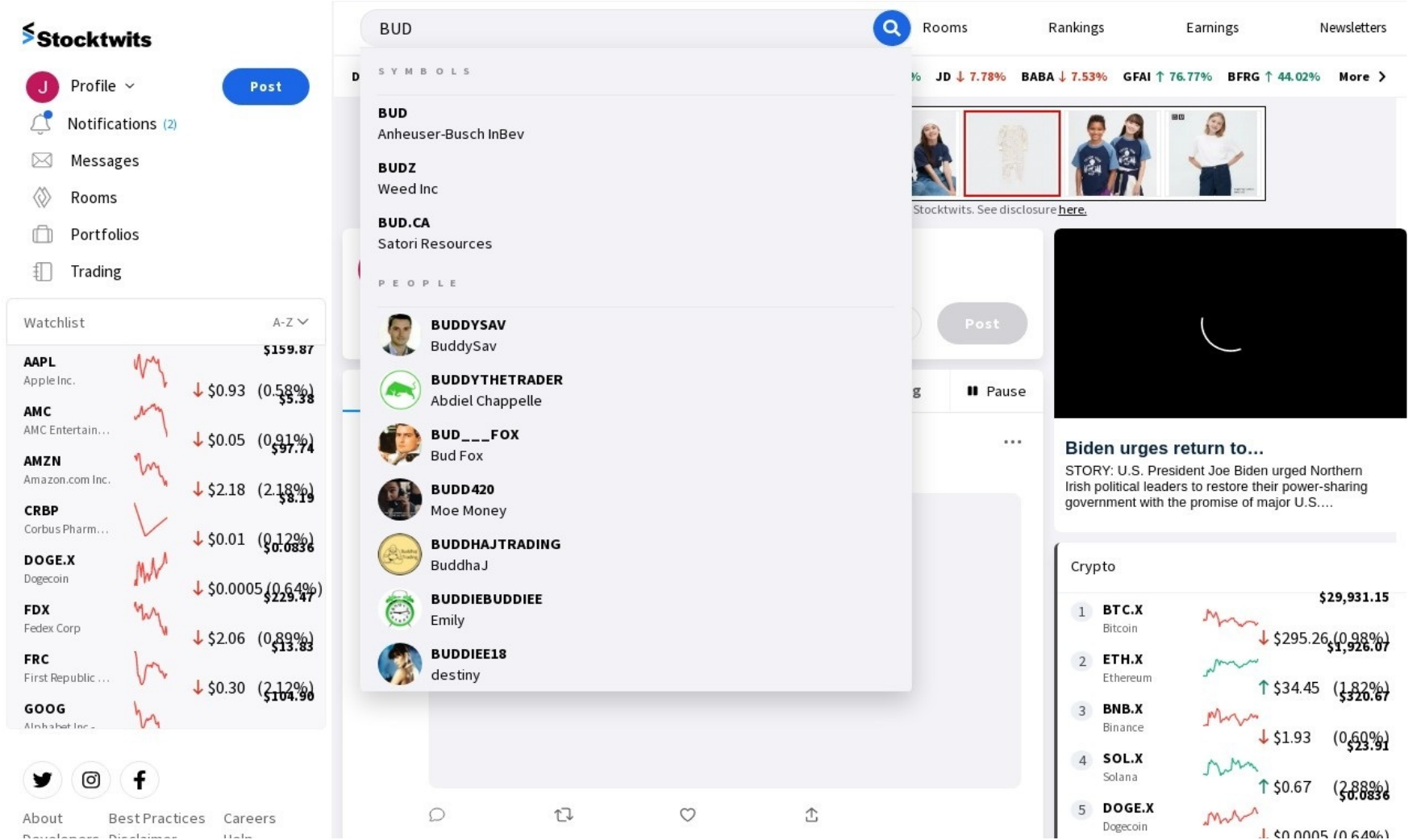}
        \caption{Mind2Web (M2W) covers diverse domains, websites, and user tasks.}
    \end{subfigure}
    \hfill
    \begin{subfigure}[t]{0.32\textwidth}
        \centering
        \includegraphics[width=\textwidth]{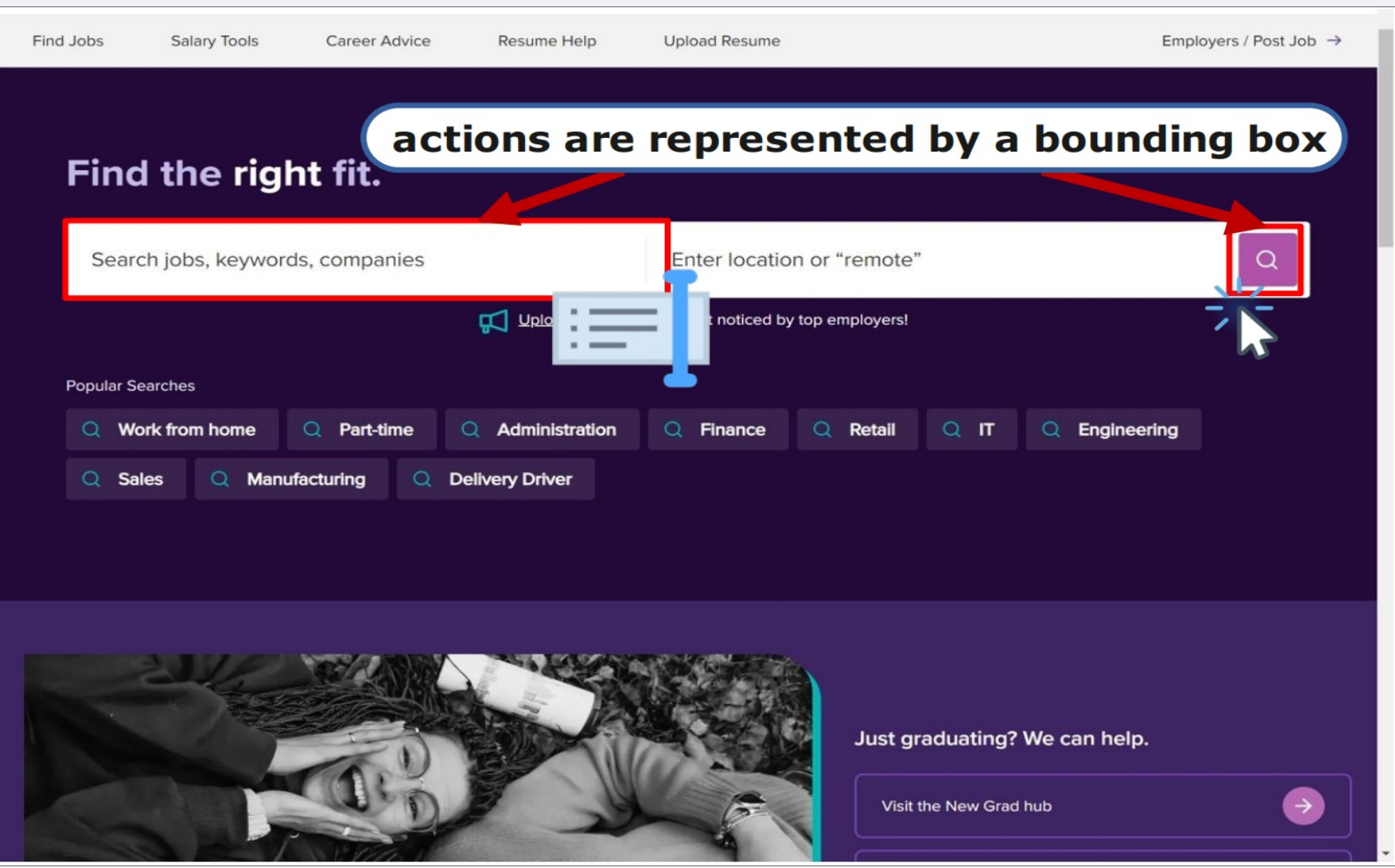}
        \caption{GUIAct-Web (GA-W), whose actions are represented by bounding boxes.}
    \end{subfigure}
    \hfill
    \begin{subfigure}[t]{0.32\textwidth}
        \centering
        \includegraphics[width=\textwidth]{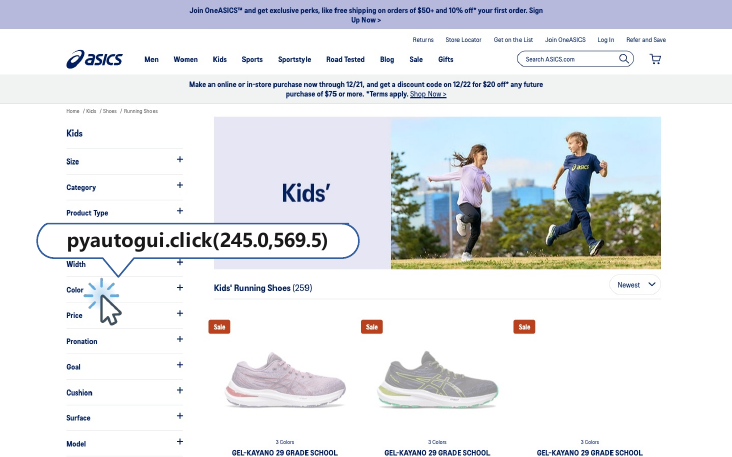}
        \caption{OmniAct-Web (OA-W) provides executable PyAutoGUI scripts.}
    \end{subfigure}

    \par\bigskip

    \begin{subfigure}[t]{0.32\textwidth}
        \centering
        \includegraphics[width=\textwidth]{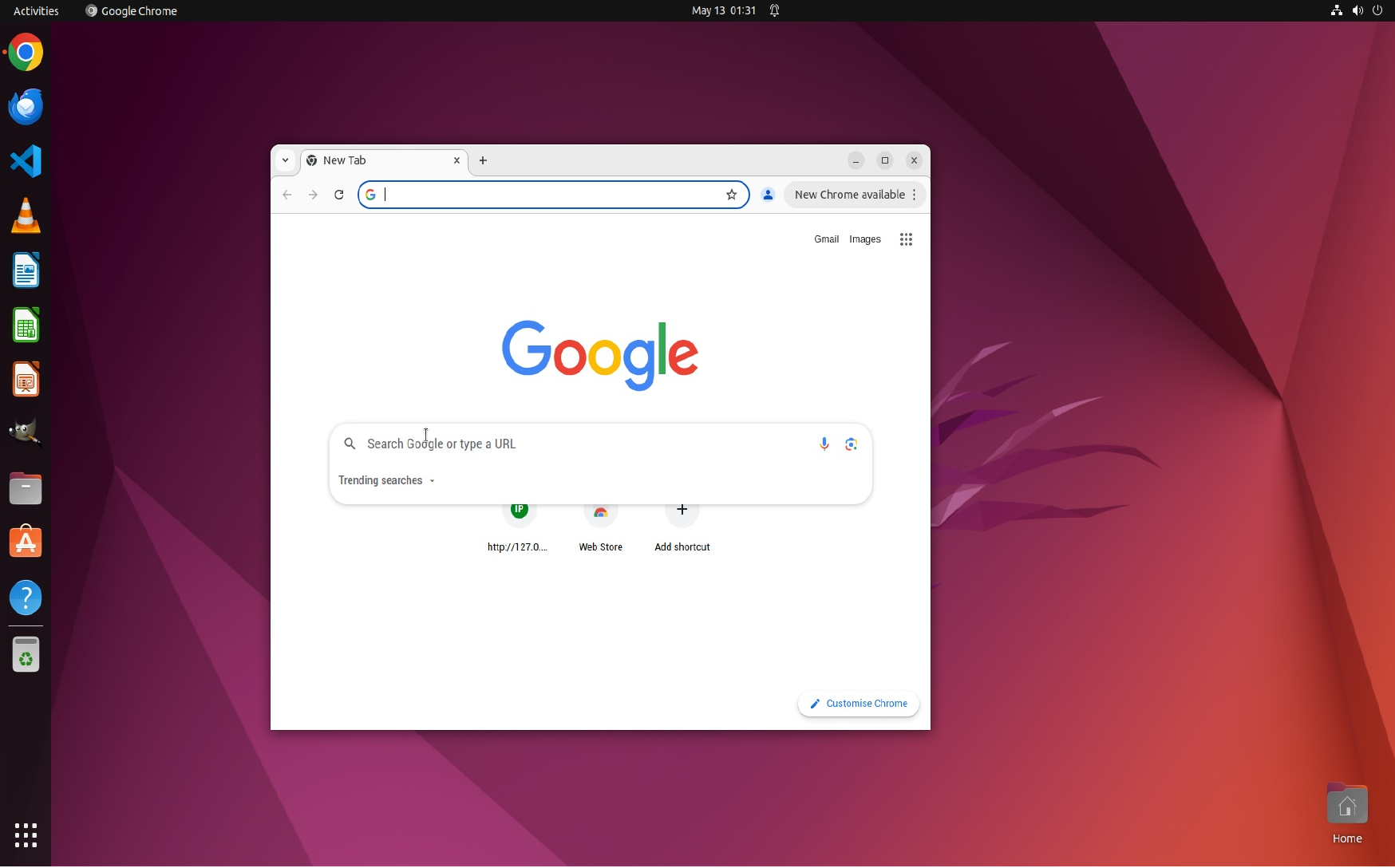}
        \caption{AgentSynth (AS) is collected on Linux Ubuntu environments.}
    \end{subfigure}
    \hfill
    \begin{subfigure}[t]{0.32\textwidth}
        \centering
        \includegraphics[width=\textwidth]{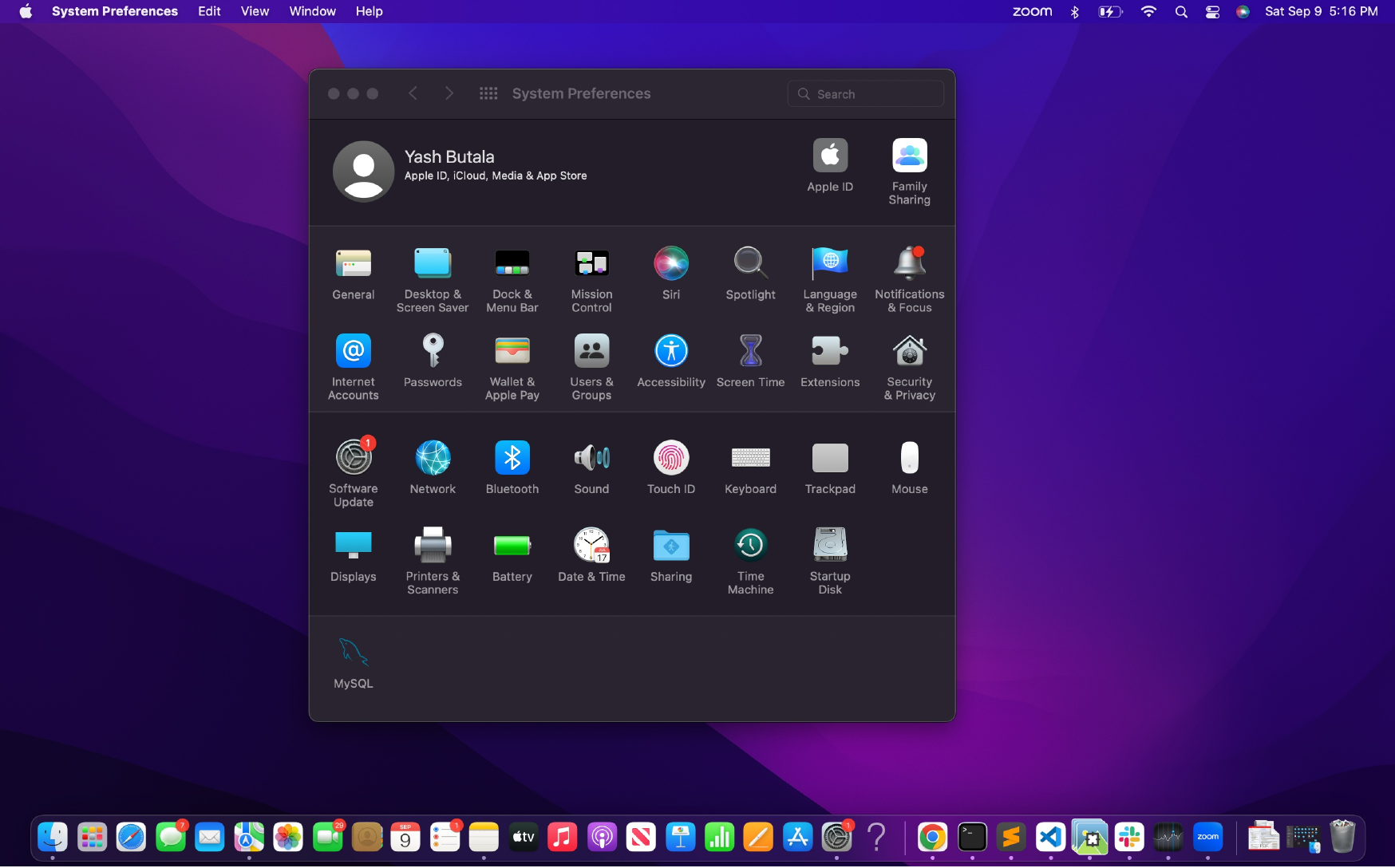}
        \caption{OmniAct-MacOS (OA-Mac) is collected on macOS environments.}
    \end{subfigure}
    \hfill
    \begin{subfigure}[t]{0.32\textwidth}
        \centering
        \includegraphics[width=\textwidth]{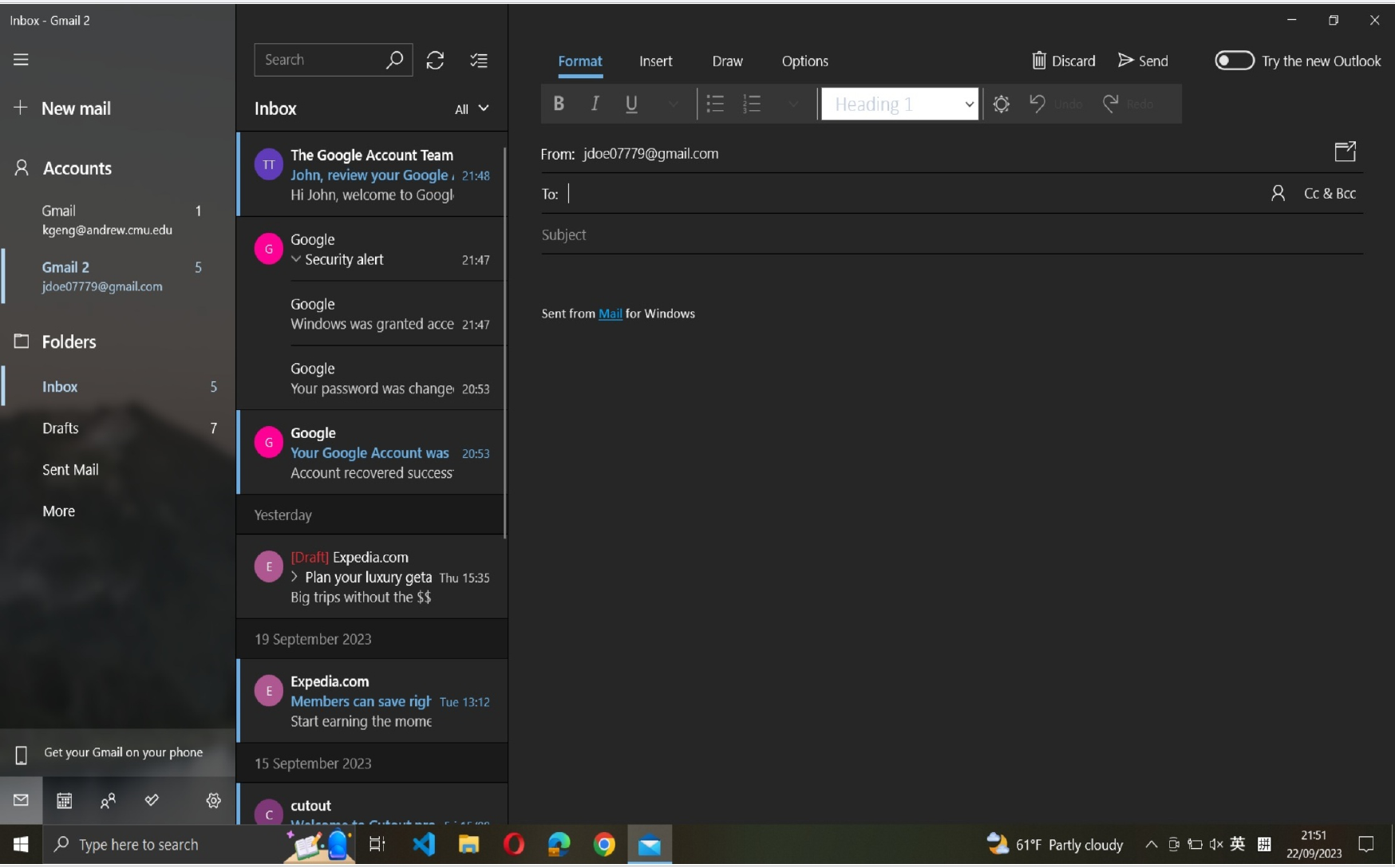}
        \caption{OmniAct-Windows (OA-Win) is collected on Windows environments.}
    \end{subfigure}

    \caption{Overview of the datasets used in our experiments across mobile (AC, AitW, GO), web (M2W, GA-W, OA-W), and desktop (AS, OA-Mac, OA-Win) environments. These datasets cover diverse task formulations, interface modalities, and interaction patterns, enabling evaluation of federated GUI agents under various heterogeneity.
}

    \label{fig:datasets}
\end{figure*}

\subsection{Source Datasets}
\label{app:dataset_details}

We consolidate multiple publicly available GUI interaction datasets spanning mobile, web, and desktop environments.
These datasets provide diverse task formulations, action spaces, and interface modalities, enabling comprehensive evaluation of federated GUI agents under heterogeneous data distributions. Figure~\ref{fig:datasets} provides a visual overview of the specific interface characteristics across these benchmarks.

AitW is a large-scale mobile device control dataset containing screenshots, human interaction trajectories, and natural language instructions collected across more than 800 Android devices and applications.
It supports multi-step interaction tasks and emphasizes generalization across apps and device configurations.

AndroidControl focuses on common Android application tasks with hierarchical task formulations, including both high-level goals and fine-grained low-level instructions.
The dataset contains approximately 15K episodes and is widely adopted for studying data scaling effects and instruction-following behaviors in mobile agents.

GUI Odyssey is a mobile GUI navigation dataset designed for cross-application reasoning.
It covers a wide range of device types and hundreds of mobile applications, with step-wise annotations that explicitly capture intermediate reasoning and navigation decisions.

\paragraph{GUIAct-Web (GA-W).}
GUIAct-Web is a web-based GUI action dataset comprising both single-step and multi-step instructions. It includes fundamental operations such as clicking, scrolling, and text input, and is designed to benchmark web interaction understanding and action prediction.

\paragraph{Mind2Web (M2W).}
Mind2Web targets generalist web agents and is collected from 137 websites across 31 domains.
The dataset contains complex multi-step tasks and places particular emphasis on cross-site and cross-domain generalization.

\paragraph{AgentSynth (AS).} 
AgentSynth is a synthetic dataset generated via a scalable task and trajectory synthesis pipeline, primarily collected in Linux environments. It enables low-cost construction of diverse GUI tasks and serves as a complementary data source for training and evaluating generalist agents.

\paragraph{OmniAct-Web (OA-W).}
OmniAct-Web is the web subset of the OmniAct benchmark.
It supports end-to-end evaluation of web interaction capabilities and facilitates systematic comparison between web and desktop GUI tasks under a unified action abstraction.

\paragraph{OmniAct-MacOS (OA-Mac).} OmniAct-MacOS is a desktop interaction dataset specifically collected within the macOS environment. It captures unique visual cues and interface paradigms of the macOS environment, enabling the evaluation of cross-platform agent adaptation and execution.

\paragraph{OmniAct-Windows (OA-Win).} OmniAct-Windows is the Windows-specific subset of OmniAct, focusing on productivity tasks in Microsoft Windows. It provides executable scripts paired with GUI states, enabling rigorous evaluation of task automation and correctness across Windows software.







\section{Experiment Details}
\label{app:ExperimentalDetails}

\subsection{Evaluation \& Metric Details}


\paragraph{Standardized Evaluation Sets.}
To ensure a robust and comprehensive assessment, we curate a standardized evaluation set by sampling 100 interaction episodes from the test partition of each source dataset. The sampling process is designed to capture diversity across multiple axes, including varying trajectory lengths and distinct task categories. This balanced selection ensures that the evaluation is not biased toward specific simple tasks and reflects the complexity of real-world GUI environments.

We implement a hierarchical evaluation protocol to dissect the agent's performance at different granularities. Following the implementation in our evaluation pipeline, we define three primary metrics:

\begin{itemize}[nosep]
    
     \item \textbf{Action Type Accuracy (Type):} This measures the consistency between the predicted and ground-truth interaction intent. A prediction is marked as a successful type match if the first token of the generated action string aligns with the target category in our unified action space.
    
    \item \textbf{Grounding Accuracy (Ground):} For coordinate-based actions such as \texttt{CLICK} and \texttt{DOUBLE\_CLICK}, we adopt a spatial proximity criterion. A grounding attempt is considered successful if the Euclidean distance between the predicted coordinates and the ground-truth coordinates is within a specific threshold. In our implementation, this threshold is set to 14\% of the screen diagonal. This relative ratio allows the benchmark to accommodate various UI element scales and screen resolutions across heterogeneous federated clients.

    \item \textbf{Success Rate (SR):} This metric represents the end-to-end execution accuracy, requiring both the action type and its associated parameters to be correct. For text-centric actions like \texttt{TYPE}, \texttt{OPEN\_APP}, or \texttt{COPY}, we evaluate semantic alignment using a \textbf{Similarity Score}. This score integrates F1-token matching and character-level intersection to support multilingual scenarios. As defined in our pipeline, a prediction is deemed successful if the similarity score exceeds 0.5, ensuring the agent captures the core semantic intent even with minor variations in the generated text.
\end{itemize}

\subsection{Training Details}
\label{sec:training_details}
This section summarizes the key configurations used in our experiments,
ensuring reproducibility and clarity.

The framework provides adaptive default hyperparameters for multiple federated optimization algorithms.
FedYogi~\cite{fedopt} uses momentum parameters $(\beta_1=0.9, \beta_2=0.999)$
with learning rate $\eta=10^{-3}$ and stabilization constant $\tau=10^{-6}$.
FedAvgM~\cite{fedavgm} applies a $0.9/0.1$ weighting between historical and current models.
FedProx~\cite{fedprox} incorporates proximal regularization with coefficient $\mu=0.2$
via a $||w - w^t||^2$ penalty term.
SCAFFOLD~\cite{scaffold} maintains a server learning rate of $\eta_s=1.0$
with client-side control variate compensation,
while FedAdam and FedAdagrad~\cite{fedopt} share base parameters $(\beta_1=0.9, \beta_2=0.999)$
with adaptive learning rate scaling.
All algorithm-specific coefficients are exposed through a unified parameter interface.

\paragraph{General Parameters.}
Our implementation is based on the $\mathtt{ms-swift}$ library~\cite{zhao2024swiftascalablelightweightinfrastructure}
and adopts parameter-efficient fine-tuning.
We employ Low-Rank Adaptation (LoRA) with a rank of 8 and an alpha scaling factor of 32,
together with a dropout rate of 0.05 to mitigate overfitting.
The maximum sequence length is set to 4,096 tokens.
We use a batch size of 1 with a gradient accumulation step of 4,
and the learning rate is fixed to $5\times10^{-5}$ across all experiments.

\paragraph{Hardware Configuration.}
Training is conducted on two NVIDIA GeForce RTX 3090 GPUs with CUDA version 12.4.
Under this configuration, training a single client for one communication round
with 10 episodes requires approximately 2 minutes.

\subsection{Model Details.}
\label{sec:model_details}
We employ two categories of models in our experiments: open-source vision-language models (VLMs) that support local training and federated optimization, and API-based closed-source models used exclusively for annotation and evaluation.

\paragraph{Open-Source VLMs.}
This category forms the backbone of our federated training framework.
We include a diverse set of state-of-the-art open-source VLMs spanning multiple model families and parameter scales, including
Qwen (Qwen3-VL-2B/4B/8B, Qwen2.5-VL-3B/7B, Qwen2-VL-2B/7B) \cite{Qwen2VL},
InternVL (InternVL2-1B/2B/4B/8B) \cite{chen2024internvl},
Gemma (Gemma-3-4B) \cite{gemma_2025},
DeepSeek (DeepSeekVL2, DeepSeekVL2-tiny/small) \cite{lu2024deepseekvl},
and Seed models (UI-TARS-1.5-7B) \cite{qin2025uitars}.
These models support supervised fine-tuning and parameter-efficient adaptation (e.g., LoRA), enabling systematic evaluation of federated learning algorithms on heterogeneous GUI interaction data across mobile, web, and desktop platforms.

We additionally incorporate commercial vision--language APIs, including GPT-5, GPT-4o, and Doubao-1.5-Vision.
Due to their closed-source and API-only nature, these models do not support supervised fine-tuning within our framework and are therefore exclusively used as annotation and reference models, providing strong off-the-shelf baselines for GUI understanding and action grounding.

\paragraph{Model \& Tokenization Configuration.}
Our primary experiments are conducted using the Qwen2-VL model.
The detailed model and tokenizer configuration is provided in Table~\ref{tab:config}.

\begin{table}[t]
\centering
\small
\setlength{\tabcolsep}{6pt}
\begin{tabular}{l l}
\toprule
\textbf{Configuration} & \textbf{Value} \\
\midrule
Model Architecture & \makecell{\texttt{Qwen2VLForConditional}\\\texttt{Generation}}\\
Model Type            & \texttt{qwen2\_vl} \\
Torch Dtype           & \texttt{bfloat16} \\
Tokenizer             & \texttt{Qwen2TokenizerFast} \\
Tokenization Strategy & Greedy search \\
\bottomrule
\end{tabular}
\caption{Model and tokenizer configurations.}
\label{tab:config}
\end{table}

\subsection{Prompt Format}
Following FedMobileAgent~\cite{wang2025fedmobileagenttrainingmobileagents}, we employ a structured prompt format, shown in Figure~\ref{fig:prompt_1} and \ref{fig:prompt_2}, to support executable decision-making.
The prompt integrates a high-level task goal, the current visual context (i.e., the screen screenshot), and an explicit action set comprising both basic and domain-specific actions.
This design grounds model decisions in the task objective while constraining outputs to valid, screen-aware GUI interactions.

\newpage

\begin{figure*}[t]
    \centering
    \begin{tcolorbox}[
        colback=white, colframe=black, sharp corners, boxrule=1pt,
        fontupper=\small\ttfamily, width=\linewidth,
        title=Full System Prompt (Part I),
        halign title=center, fonttitle=\bfseries,
        before upper={\setlength{\parskip}{0.6em}},
        left=10pt, right=10pt, top=8pt, bottom=8pt
    ]
        \raggedright
        You are a foundational action model capable of automating tasks across various digital environments, including desktop systems like Windows, macOS, and Linux, as well as mobile platforms such as Android and iOS. You also excel in web browser environments. You will interact with digital devices in a human-like manner: by reading screenshots, analyzing them, and taking appropriate actions. Your expertise covers two types of digital tasks:
    
    \begin{itemize}[leftmargin=2em, nosep] 
        \item \textbf{Grounding}: Given a screenshot and a description, you assist users in locating elements mentioned. Sometimes, you must infer which elements best fit the description when they aren't explicitly stated.
        \item \textbf{Executable Language Grounding}: With a screenshot and task instruction, your goal is to determine the executable actions needed to complete the task.
    \end{itemize}

    You are now operating in Executable Language Grounding mode. Your goal is to help users accomplish tasks by suggesting executable actions that best fit their needs. Your skill set includes both basic and custom actions:

    Basic actions are standardized and available across all platforms. They provide essential functionality and are defined with a specific format, ensuring consistency and reliability. 

    \begin{itemize}[leftmargin=1.5em, nosep]
        \item \textbf{Basic Action 1: CLICK} \\ - purpose: Click at the specified position. \\ - format: CLICK <point>[[x-axis, y-axis]]</point> \\ - example usage: CLICK <point>[[101, 872]]</point>
        \item \textbf{Basic Action 2: TYPE} \\ - purpose: Enter specified text at the designated location. \\ - format: TYPE [input text] \\ - example usage: TYPE [Shanghai shopping mall]
        \item \textbf{Basic Action 3: SCROLL} \\ - purpose: Scroll in the specified direction. \\ - format: SCROLL [direction (UP/DOWN/LEFT/RIGHT)] \\ - example usage: SCROLL [UP]
        \item \textbf{Basic Action 4: COMPLETE} \\ - purpose: Indicate the task is finished. \\ - format: COMPLETE \\ - example usage: COMPLETE
        \item \textbf{Basic Action 5: IMPOSSIBLE} \\ - purpose: Indicate the task is infeasible to reach. \\ - format: IMPOSSIBLE \\ - example usage: IMPOSSIBLE
        \item \textbf{Basic Action 6: WAIT} \\ - purpose: Wait for the screen to load. \\ - format: WAIT \\ - example usage: WAIT
    \end{itemize}
    
    \textbf{2. Custom Actions for Mobile Platforms} \\
    Custom actions are unique to each user's platform and environment. They allow for flexibility and adaptability, enabling the model to support new and unseen actions defined by users. 

    \begin{itemize}[leftmargin=1.5em, nosep]
        \item \textbf{Custom Action 7: LONG\_PRESS} \\ - purpose: Long press at the specified position. \\ - format: LONG\_PRESS <point>[[x-axis, y-axis]]</point> \\ - example usage: LONG\_PRESS <point>[[272, 341]]</point>
        \item \textbf{Custom Action 8: NAVIGATE\_BACK} \\ - purpose: Press a back button to navigate to the previous screen. \\ - format: NAVIGATE\_BACK
        \item \textbf{Custom Action 9: NAVIGATE\_HOME} \\ - purpose: Press a home button to navigate to the home page. \\ - format: NAVIGATE\_HOME
        \item \textbf{Custom Action 10: OPEN\_APP} \\ - purpose: Open the specified application. \\ - format: OPEN\_APP [app\_name] \\ - example usage: OPEN\_APP [Google Chrome]
        \item \textbf{Custom Action 11: PRESS\_RECENT} \\ - purpose: Press the recent button to view or switch between recently used applications. \\ - format: PRESS\_RECENT
    \end{itemize}
    \end{tcolorbox}
    \caption{The full system prompt for training and evaluation (Part I).}
    \label{fig:prompt_1}
\end{figure*}

\clearpage 

\begin{figure*}[!t]
    \centering
    \begin{tcolorbox}[
        colback=white, colframe=black, sharp corners, boxrule=1pt,
        fontupper=\small\ttfamily, width=\linewidth,
        title=Full System Prompt (Part II),
        halign title=center, fonttitle=\bfseries,
        before upper={\setlength{\parskip}{0.6em}},
        left=10pt, right=10pt, top=8pt, bottom=8pt
    ]
        \raggedright
      
    \textbf{3. Custom Actions for Web and Desktop Platforms}

    \begin{itemize}[leftmargin=1.5em, nosep]
        \item \textbf{Custom Action 12: DOUBLE\_CLICK} \\ - purpose: Double click at the specified position. \\ - format: DOUBLE\_CLICK <point>[[x-axis, y-axis]]</point>
        \item \textbf{Custom Action 13: RIGHT\_CLICK} \\ - purpose: Right click at the specified position. \\ - format: RIGHT\_CLICK <point>[[x-axis, y-axis]]</point>
        \item \textbf{Custom Action 14: MOVETO} \\ - purpose: Move the object to the specified position. \\ - format: MOVETO <point>[[x-axis, y-axis]]</point>
        \item \textbf{Custom Action 15: HOTKEY} \\ - purpose: Use the hot key. \\ - format: HOTKEY [keys] \\ - example usage: HOTKEY [CTRL+ALT]
        \item \textbf{Custom Action 16: COPY} \\ - purpose: Copy a sentence to answer user questions. \\ - format: COPY [text with answer] \\ - example usage: COPY [Wednesday]
    \end{itemize}

    In most cases, task instructions are high-level and abstract. Carefully read the instruction and action history, then perform reasoning to determine the most appropriate next action. Ensure you strictly generate one section: Actions.

    \textbf{Actions}: Specify the actual actions you will take based on your reasoning.
    \end{tcolorbox}
    \caption{The full system prompt for training and evaluation (Part II).}
    \label{fig:prompt_2}
\end{figure*}

\end{document}